\documentclass[useAMS,usenatbib]{mn2e}
\usepackage{natbib, graphicx, times}
\usepackage{amsmath}
\usepackage{mathtools}
\usepackage{array}
\usepackage{epstopdf}
\usepackage{amssymb}
\usepackage{extarrows}
\usepackage{color}
\usepackage{enumitem}
\usepackage{scrextend}
\usepackage{threeparttable}
\usepackage{scalerel}

\setlist{leftmargin=7.5mm}

\newcommand{\lsim}{\mathrel{\rlap{\lower4pt\hbox{\hskip1pt$\sim$}}
    \raise1pt\hbox{$<$}}}                
\newcommand{\gsim}{\mathrel{\rlap{\lower4pt\hbox{\hskip1pt$\sim$}}
    \raise1pt\hbox{$>$}}}                

\def\vecsign#1{\rule[1.388\LMex]{\dimexpr#1-2.5pt}{.36\LMpt}%
  \kern-6.0\LMpt\mathchar"017E}

\newcommand{\cbf}{} 

\long\def\mklr#1{}
\long\def\mklrc#1{}
\long\def\mklsrc#1{}

\def\arcsec{\hbox{$^{\hbox{\rlap{\hbox{\lower4pt\hbox{$\,\prime\prime$}}}\hbox{$\frown$}}}$}}

\title[Long-lived planet-induced vortices]{How to form compact \& other longer-lived planet-induced vortices: VSI, planet migration, or re-triggers, but not feedback}

\author[Hammer \& Lin]{Michael Hammer$^{1}$\thanks{E-mail: mhammer@email.arizona.edu},
Min-Kai Lin$^{1, 2}$ \\
$^{1}$ Institute of Astronomy and Astrophysics, Academia Sinica, Taipei 10617, Taiwan \\
$^{2}$ Physics Division, National Center for Theoretical Sciences, Taipei 10617, Taiwan  \\
}

\begin{document}

\date{Accepted XXX. Received YYY; in original form ZZZ}

\pagerange{\pageref{firstpage}--\pageref{lastpage}} \pubyear{2016}

\maketitle

\label{firstpage}

\begin{abstract}
Past computational studies of planet-induced vortices have shown that the dust asymmetries associated with these vortices can be long-lived enough that they should be much more common in mm/sub-mm observations of protoplanetary discs, even though they are quite rare. Observed asymmetries also have a range of azimuthal extents from compact to elongated even though computational studies have shown planet-induced vortices should be preferentially elongated. In this study, we use 2-D and 3-D hydrodynamic simulations to test whether those dust asymmetries should really be so long-lived or so elongated. With higher resolution (29 cells \cbf{radially} per scale height) than our previous work, we find that vortices can be more compact by developing compact cores when higher-mass planets cause them to re-form, or if they are seeded by tiny compact vortices from the vertical shear instability (VSI), but not through dust feedback in 3-D as was previously expected in general. Any case with a compact vortex or core(s)\mklrc{what is meant by cores here?} also has a longer lifetime. Even elongated vortices can have longer lifetimes with higher-mass planets or if the associated planet is allowed to migrate, the latter of which can cause the dust asymmetry to stop decaying as the planet migrates away from the vortex. These longer dust asymmetry lifetimes are even more inconsistent with observations, perhaps suggesting that discs still have an intermediate amount of effective viscosity.

\end{abstract}

\begin{keywords}
transition discs~\---~instability, hydrodynamics, methods:numerical, protoplanetary discs 
\end{keywords}



\section{Introduction} \label{sec:intro}

Even before there were any well-resolved observations of protoplanetary discs, it was already known that the Rossby Wave instability (RWI) could generate a long-lived vortex \citep{li00, li01} capable of trapping dust \citep{barge95} located in the immediate vicinity of a radial pressure bump in the disc \citep{lovelace99}. The activation of the Atacama Large Millimeter Array (ALMA) in 2011 quickly seemed to confirm that such vortices could actually occur in protoplanetary discs with the discovery of a large-scale asymmetric dust trap as the only prominent mm-dust feature in the disc around Oph IRS 48 \citep{vanDerMarel13}. 

There are various ways to create such dust asymmetries in protoplanetary discs. The mechanism that has been studied the most in-depth is gap-opening planets. More specifically, a planet in a low-viscosity disc can generate a vortex through the RWI at one or both of the pressure bumps that arise as the planet opens up a gap \citep[e.g.][]{koller03, li05, deValBorro07, mkl10, mkl14}. In discs with very low viscosity, it has been shown that the dust asymmetries corresponding to planet-induced vortices can be very long-lived, including for a wide range of low-mass planets, very high-mass planets more massive than Jupiter, and in thicker regions of a disc with high aspect ratios regardless of the planet mass \citep{fu14a, hammer21}. Despite that expected longevity, ALMA observations have found that dust asymmetries in protoplanetary discs are quite rare \citep{vanDerMarel21}. Various mechanisms for shortening vortex lifetimes have been suggested; however, the requirements for these mechanisms to take effect may either limit their applicability to a narrow range of conditions or even prevent them from having any effect at all (see Section~\ref{sec:dearth} for an in-depth discussion). As such, it remains an open question as to why there are so few dust asymmetries in observations of protoplanetary discs.

Beyond the low occurrence rate, another characteristic of the population of dust asymmetries is the wide degree to which their appearances can vary. They can be very compact like the original dust asymmetry in Oph IRS 48 or the two dust asymmetries in MWC 758 \citep{boehler18}. They can also be more elongated like in HD 135344 B \citep{vanDerMarel16b, cazzoletti18} or HD 142527 \citep{boehler17, boehler21}. These varying appearances may suggest that at least a few, if not many of these features are not planet-induced vortices, or possibly not even vortices at all. Computational studies on planet-induced vortices that took into account the planet's growth have shown that these vortices should preferentially be elongated, in terms of the vortex's azimuthal extent in both the gas and dust \citep{hammer17, hammer19}. This finding was a big departure from previous work that found planet-induced vortices to be compact when the introduction of the planet was not self-consistent with the evolution of the gap \citep[e.g.][]{fu14a}. The difference between elongated and compact vortices is not just their azimuthal extent, but also their minimum Rossby number being less negative or more negative respectively than the critical value of $\mathrm{Ro} = -0.15$ \citep{surville15, hammer21}, where $\mathrm{Ro}$ is the normalized local vorticity. Since actual observed asymmetries are not necessarily elongated, it remains of interest to see if planet-induced vortices can still be compact even if the planet's growth is taken into account.

In this work, we seek to improve on our previous studies by using a more realistic planetary accretion scheme and by incorporating several different physical effects into our simulations that we had previously neglected. Our two primary goals are to see if such effects may shorten dust asymmetry lifetimes to be more consistent with observations or if they can produce vortices that are more compact like what happens when the planet's growth is neglected. The effects we add separately are letting the planet migrate, the feedback of the dust on the gas, and extending the simulations from 2-D to 3-D -- which at high resolution allows the vertical shear instability (VSI) to manifest \citep{nelson13, flores20}. The VSI naturally occurs because of the expected radial temperature gradient in protoplanetary discs, which in turn results in vertical shear. We also improve the planet's accretion scheme compared to our previous work \citep[hereafter H21]{hammer21}  and have the planet grow to a range of different masses by varying the disc mass in our parameter studies instead of the planet's dimensionless accretion rate. Lastly, we conduct our simulations at higher resolution than before.

Each of these effects has significant potential to affect the evolution, properties, and lifetimes of the resulting planet-induced vortices. Planet migration can drastically alter the gap structure in low-viscosity discs in particular where gap-opening planets are capable of leaving the outer edge of the gap behind as they migrate inwards  \citep{lega21, kanagawa21}. Meanwhile, 3-D simulations have been shown to be able to counteract the major effects of dust feedback \citep{lyra18} that would otherwise greatly shorten vortex lifetimes in 2-D \citep{fu14b}, and they may do the same to stop cooling from killing the vortex as well \citep{rometsch21}. The VSI can generate tiny vortices of its own \citep{richard16, pfeil19, manger20} that may even grow to large-scale \citep{manger18}. It is unclear how  any of these effects will affect planet-induced vortices, particularly when incorporating the planet's growth time. \mklrc{nice motivation for this work}


We organize the paper as follows: In Section~\ref{sec:methods}, we describe the setups for our 2-D and 3-D hydrodynamical simulations in which we incorporate both the gas and dust components of the discs, and also outline the different parameter studies. In Section~\ref{sec:disc-mass}, we present the results from our high-resolution 2-D simulations, including the dependence of dust asymmetry lifetimes on disc mass and planet mass, and the dust morphology of the vortex in simulations and synthetic images. In Section~\ref{sec:results-migration}, we discuss the effects of migration. In Section~\ref{sec:results-3D}, we present our results from our 3-D simulations, both at low resolution with a comparison to low-resolution 2-D, and later at high resolution featuring the VSI. In Section~\ref{sec:feedback}, we present our results with dust feedback. In Section~\ref{sec:synthetic}, we show the synthetic images calculated from our simulations. In Section~\ref{sec:discussion}, we discuss the answers to our two main questions in connection to observational implications and applicable observed discs. In Section~\ref{sec:conclusions}, we conclude our results.

\section{Methods} \label{sec:methods}

We use the FARGO3D hydrodynamic code \citep{FARGO3D, FARGO3D-dust} to run 2-D and 3-D simulations of a planetary core \mklr{interacting gravitationally with its protoplanetary disc and}that interacts with the protoplanetary disc in which it is embedded by carving out a gap and accreting gas from its surroundings. Every simulation includes both a gas and a dust component, the latter of which is used to assess how the disc appears in observations of mm-sized dust and to determine vortex lifetimes.


\subsection{Hydrodynamic code}

\subsubsection{2-D Two-Fluid Hydrodynamics} \label{sec:hydro-2D}

FARGO3D \citep{FARGO3D} is a 3-D magnetohydrodynamic grid-based code. Like its predecessor FARGO \citep{FARGO}, one of its main functions is to conduct global simulations of protoplanetary discs. These codes speed up computational time by subtracting out the average azimuthal velocity when using the Courant-Friedrich-Levy (CFL) condition to calculate the timestep. This method, known as the FARGO (Fast Advection in Rotating Gaseous Objects) algorithm, makes the code well-suited for simulating various types of discs.

We begin our study with 2-D hydrodynamic simulations. In these simulations, FARGO3D solves the Navier-Stokes equations in cylindrical polar coordinates ($r$, $\phi$), namely the continuity equation and the momentum equations. The gas continuity equation is 
\begin{equation} \label{eqn:continuity}
\frac{\partial \Sigma}{\partial t} + \vec \nabla \cdot (\Sigma \vec{v}) = 0,
\end{equation}
where $\Sigma$ is the surface density, $\vec v$ is the velocity vector, and $t$ is time. The momentum equations are
\begin{align} \label{eqn:navier-stokes}
\Sigma \Big( \frac{\partial \vec{v}}{\partial t} + \vec{v} \cdot \vec  \nabla \vec{v} \Big) = &- \vec \nabla P - \Sigma \vec \nabla \Phi + \vec  \nabla \cdot \overset{\text{\tiny$\leftrightarrow$}} T \\
&-  \Sigma [2 \vec \Omega_\mathrm{f} \times \vec v + \vec \Omega_\mathrm{f} \times \big(\vec \Omega_\mathrm{f} \times \vec r \big) + \dot{ \vec \Omega}_\mathrm{f} \times \vec r ]
\nonumber
\end{align}
where $P$ is the pressure, $\Phi$ is the gravitational potential, $\vec \Omega_\mathrm{f}$ is the angular frequency vector of the reference frame\mklrc{need to define Omega, gravitational potential includes star and planet (is there indirect?). drag forces missing.}, and $\mathcal{T}$ \mklrc{notation conflict with temperature, maybe use bold for vectors and tensors}is the stress tensor. The gravitational potential $\Phi = \Phi_\mathrm{\bigstar} + \Phi_\mathrm{p} + \Phi_\mathrm{i}$ comprises the star's potential, the planet's potential, and the indirect potential arising from the non-inertial reference frame. The stress tensor itself is defined as
\begin{equation} \label{eqn:stress-tensor}
\overset{\text{\tiny$\leftrightarrow$}}{\cal{T}} = \Sigma \nu \Big[ \vec{v} + (\Sigma \vec{v})^T - \frac{2}{3}(\vec \nabla \cdot \vec{v}) \vec{I} \Big],
\end{equation}
where $\nu$ is the kinematic viscosity of the disk and $\vec{I}$ is the identity tensor. For simplicity, we use a locally isothermal setup to avoid needing to solve an energy equation\mklrc{need to define the EOS}. Neglecting cooling in this manner may be acceptable for the outer disc where cooling times are faster, and which is also the region where asymmetric features have been observed \citep{vanDerMarel21}. Assuming fast cooling times is also optimal for the few 3-D simulations we present with the VSI, which would otherwise be easily suppressed by slower cooling times \citep{mkl15, manger20}. \mklrc{"Rapid cooling is also needed to simulate the VSI in 3D."}

The dust is implemented as a fluid in a similar manner to the gas, as is appropriate for small dust grains\mklrc{fluid approx appropriate for small grains}. It does behave differently, however, because it does not experience pressure or viscous stress in the momentum equation. Unlike the gas, it does additionally experience radial and azimuthal drag forces resulting from the dust being coupled to the gas as well as turbulence from the gas viscosity that leads to diffusion. The drag forces added to the momentum equation are defined as
\begin{equation} \label{eqn:rad_drift}
\left.\frac{\partial v_\mathrm{r, d}}{\partial t}\right|_\mathrm{drag} = - \frac{v_\mathrm{r, d} - v_\mathrm{r}}{t_\mathrm{s}},
\end{equation}
\begin{equation} \label{eqn:az_drift}
\left.\frac{\partial v_\mathrm{\phi, d}}{\partial t}\right|_\mathrm{drag} = - \frac{v_\mathrm{\phi, d} - v_\mathrm{\phi}}{t_\mathrm{s}},
\end{equation}
where $v_\mathrm{r}$ and $v_\mathrm{\theta}$ are the radial and azimuthal velocity components and $v_\mathrm{d}$ and $v$ are the dust and gas components respectively. Since we focus on grains that around millimeter-sized, the stopping time falls in the Epstein regime \citep{weidenschilling77} and is defined in the midplane as
\begin{equation} \label{eqn:stopping}
t_\mathrm{s} = \frac{\mathrm{St}}{\Omega} = \left( \frac{\pi}{2} \frac{\rho_\mathrm{d} s}{\Sigma} \right) \frac{1}{\Omega},
\end{equation}
where $\rho_\mathrm{d}$ is the physical density of each dust grain, which we take to be 1 g / cm$^3$, and $s$ is the size of each grain. The dimensionless form of the stopping time is the Stokes number St, which is normalized to the orbital frequency $\Omega$. \mklrc{need to distinguish between Omega and Omega frame} The dust diffusion is added to the continuity equation and defined as
\begin{equation} \label{eqn:diffusion}
\left.\frac{\partial \Sigma_\mathrm{d}}{\partial t}\right|_\mathrm{diff} = \nabla \cdot \left( D \Sigma_\mathrm{tot} \nabla\left( \frac{\Sigma_\mathrm{d}}{\Sigma_\mathrm{tot}}\right) \right),
\end{equation}
where $\Sigma_\mathrm{d}$ is the dust surface density,  $\Sigma_\mathrm{tot} = \Sigma + \Sigma_\mathrm{d}$ is the total surface density, the diffusion coefficient is $D = \hat{D} r_\mathrm{p}^2 \Omega_\mathrm{p}$ and $\hat{D} \approx \hat{\nu}$ since the dust grains in this study are small enough such that St$~\ll~1$ \citep{youdin07}. With the low viscosity in our study, however, dust diffusion has little to no effect on our simulations. \mklrc{in the original code, the diffusive flux is prop to Sigmad/Sigmatotal. did we update it to Sigmad/Sigmag?}

\mklrc{just include this in describing the gas equations}
Lastly, in some of our simulations, we also incorporate the feedback of the dust on the gas. This feedback is added to the momentum equations and defined as:
\begin{equation} \label{eqn:rad_drift}
\left.\frac{\partial \vec v}{\partial t}\right|_\mathrm{feedback} = - \frac{\epsilon}{t_\mathrm{s}}\Big(\vec{v} - \vec v_\mathrm{d}\Big),
\end{equation}
where $\epsilon = \Sigma_\mathrm{d} / \Sigma$ is the dust-to-gas ratio.

\subsubsection{3-D Two-Fluid Hydrodynamics} \label{sec:hydro-3D}
\mklrc{yes, I would start with the 3D setup then put 2D as a reduced model}

We extend our simulations from 2-D to 3-D using spherical polar coordinates ($r$, $\phi$, $\theta$). The Navier-Stokes equations for these simulations are largely the same in 3-D as in 2-D. 
The only key differences are that the 2-D surface density $\Sigma$ is replaced by the 3-D spatial density $\rho$ and the velocities $\vec{v}$ include a vertical component $v_\mathrm{\theta}$. The vertically-integrated 2-D pressure $P$ is also replaced in 3-D by the actual pressure. \mklrc{there also vertical velocities in 3D} \mklrc{in 2D, P is the vertically-integrated pressure, while in 3D, P is the actual pressure} 

\cbf{When plotting the 2-D surface density from the 3-D simulations, we integrate the spatial density over the latitudinal direction following $\Sigma = \int dz \rho$. We note that this summation slightly underestimates $\Sigma$ in the outer disc because the coordinates are not perfectly cylindrical.}

\subsection{Planet setup}

We initialize each simulation with a $0.05~M_\mathrm{Jup}$ planetary core on a circular orbit at $r = r_\mathrm{p}$ accreting material from a disc of gas and dust around a star with mass $M_{\bigstar} = 1$. In our main parameter studies, the planet is on a fixed orbit at $r_\mathrm{p} = 1$. The planet's mass is introduced over $T_\mathrm{growth} = 5~T_\mathrm{p}$ following $m_\mathrm{p}(t) = \sin^2{\left(\pi t / 2T_\mathrm{growth}\right)}$. Its orbital period of $T_\mathrm{p} = 2\pi$ corresponds to its Keplerian angular frequency of $\Omega_\mathrm{p} = \sqrt{GM_\mathrm{\bigstar} / r_\mathrm{p}^3} = 1$. \mklrc{maybe describe units before results. MH --- not sure I understood what you meant???}

The planet is incorporated as a gravitational potential that follows
\begin{equation} \label{eqn:potential}
\Phi_\mathrm{p}(\mathbf{r}, t) = - \frac{Gm_\mathrm{p}(t)}{\sqrt{(\mathbf{r} - {\mathbf{r}_\mathrm{\mathbf{p}}})^2 + r_\mathrm{s}^2}},
\end{equation}
where $G = 1$ is the gravitational constant, $m_\mathrm{p}(t)$ is the planet's mass over time, $r_\mathrm{s}$ is the planet's smoothing length. The other terms in the  overall potential are the stellar potential $\Phi_\mathrm{\bigstar}(r)=-GM_\mathrm{\bigstar} / r$ and the indirect potential between the star and the planet that is given by $\Phi_\mathrm{i}(\mathbf{r}, t) = -Gm_\mathrm{p}(t)~[\mathbf{r}~\cdot~\mathbf{r_\mathrm{p}}] / r_\mathrm{p}^3$.

\subsubsection{Accretion scheme}

\mklrc{can describe all planet-related setups in one section}

We use the same general accretion scheme for the planet's growth as in our previous work (H21, where the scheme is described in more detail) in which we remove mass from the disc in the vicinity of the planet at each timestep and add it to the planet's mass.  Rather than varying the dimensionless accretion rate, we simulate different planet masses by using a range of disc masses as the main variable in our parameter studies. We note that the primary goal of our accretion scheme is to simulate a wide range of planet masses in a manner where the gap structure and planet growth develop self-consistently, and not to properly resolve the accretion process. Ultimately, we explore a similar range of final planet masses to our last study from about $0.1~M_\mathrm{Jup}$ to $1.5~M_\mathrm{Jup}$.

\mklrc{i found the following confusing. maybe it's clearer in previous papers?} 
Our accretion scheme is based on the one used in \cite{kley99} that was implemented for the original FARGO code (although see \cite{bergez20, bergez22} for a more detailed implementation). In this scheme, we remove a fraction of mass $f$ after each timestep \mklrc{at every time step? it's clearest to show this mathematically, i.e. Mij $\to$ Mij*(1-f) in Delta t, or dMij/dt = -f*Mij/tau} from any cell within a distance of $K$ from the planet. The operation we perform in each cell is 
\begin{align} \label{eqn:potential}
 \Sigma_\mathrm{ij}  \rightarrow (1 - f) \Sigma_\mathrm{ij} &\text{  in 2-D, and} \\
\rho_\mathrm{ijk} \rightarrow  (1 - f) \rho_\mathrm{ijk} &\text{  in 3-D,} 
\end{align}
where $i$, $j$, and $k$ are indices for the simulation domain. The amount of mass $\delta m$ corresponding to that reduced density, 
\begin{align} \label{eqn:potential}
\delta m_\mathrm{ij} = f\Sigma_\mathrm{ij} \times d\mathcal{A}_\mathrm{ij} &\text{  in 2-D, and} \\
\delta m_\mathrm{ij} = f \rho_\mathrm{ijk}  \times d\mathcal{V}_\mathrm{ijk} &\text{  in 3-D,} 
\end{align}
\mklrc{a summation over i,j is needed} is added to the planet, where $d\mathcal{A}$ is the cell area, $d\mathcal{V}$ is the cell volume, and the fraction $f$ is set to be $f = A \times \Omega_\mathrm{p} dt$. The parameter $A$ is the dimensionless accretion rate and the parameter $K$ is the accretion radius\mklrc{probably more intuitive to define it as Racc}. The inclusion of the timestep allows the fraction of accreted mass $A$ to be effectively removed from the disc over a time $\delta t = 1$ rather than instantaneously\mklrc{not sure i understand. don't we just reduce the cell mass?}, which also gives the disc time to fill the accretion zone where mass is being removed with more mass. \mklrc{what is A?} One key difference compared to that implementation and our previous work is that we only use one accretion zone instead of two with different accretion parameters. Although we suspect at least part of the reason two zones were originally used in \cite{kley99} to smooth the transition between the regions where material is or is not removed, we do not expect that to be necessary because the fraction $f$ is not removed instantaneously\mklrc{wording in this last sentence is awkward; shouldn't we know why we did what we did in our own work? MH: Clarified that I'm talking about Kley99 paper.}. 

We choose the fixed parameters to be $K = 0.6~R_\mathrm{H}$ and $A = 0.4$, which with the base surface density $\Sigma_\mathrm{base}$ corresponds to an accretion rate of a little more than $10^{-4}~M_\mathrm{Jup}$ per orbit over the first 1000 orbits.\mklrc{what is the typical (or range in) Mdot onto the planet, e.g. in Mjup/year? this might be useful to connect to physical disk models.} The accretion radius is larger than the expected value of about $0.25~R_\mathrm{H}$ from  \cite{lissauer09} because we wanted to use the same $K$ in all of our simulations and the latter value is small  enough to prevent the planet from accreting any material at all in our low-resolution parameter studies. \mklrc{we say in the first paragraph that we don't aim to resolve the accretion process, so why is it important to resolve the K region here?} Meanwhile, the value of $A$ is chosen to closely match the planet's growth track in our high-resolution simulation with the VSI, which uses the proper accretion radius of $K = 0.25~R_\mathrm{H}$. We discuss these choices in more detail in Appendix~\ref{sec:prescriptions}.


\subsection{Disc setup}

\subsubsection{2-D setup}
\mklrc{i would think the separate 2D/3D setups are for the disk. the planet setup doesn't depend on 2D/3D, does it?}

The gas disc starts out with a smooth power-law radial surface density profile following $\Sigma = \Sigma_\mathrm{0} (r / r_\mathrm{p})^{-1}$, where $\Sigma_0$ \mklrc{do you mean sigma 0?}is the main variable in our parameter studies. \cbf{We choose the disc to have a flat aspect ratio of $h \equiv H/r = 0.06$, where $H$ is the gas scale height, to avoid the possibility of our results depending on the local aspect ratio}. The dust density follows an identical profile to the gas except with the surface density reduced to $\Sigma_\mathrm{0,dust} = 0.01~\Sigma_\mathrm{0}$. This level of dust, however, has no effect on the gas except in the few simulations with dust feedback and only has a minor impact on the planet's growth. The size of the dust grains is $\mathrm{St} = 0.023$ to match what we used in H21. The disc extends across a domain of $r \in [0.4, 2.5]r_\mathrm{p}$ in radius and a full $2 \pi$ in azimuth $\phi$. In our main 2-D parameter study, this domain is resolved by $N_\mathrm{r} \times N_\mathrm{\phi} = 1024 \times 2048$ grid cells with arithmetic spacing in $r$ and uniform spacing in $\phi$. The radial resolution with an aspect ratio of $h = 0.06$ resolves a disc scale height $H$ at $r = 1$ by 29 grid cells, almost twice as many as the 16-grid-cell resolution used in H21. \cbf{The azimuthal resolution is a bit lower at a little less than 20 grid cells per scale height.}

Like all of our previous works, we use wave-killing zones, also called evanescent or Stockholm boundary conditions \citep[e.g.][]{deValBorro06}, at the inner edge ($0.4~r_\mathrm{p} < r < 0.5~r_\mathrm{p}$) and at the outer edge ($2.1~r_\mathrm{p} < r < 2.5~r_\mathrm{p}$) of the domain. These conditions weaken any waves\mklrc{they just damp waves to prevent reflecting, even if the waves have physical origin, i.e. planet} that may arise by damping changes to the density or velocity in these zones back towards the initial values. We use a stronger damping timescale of $\tau = \Omega^{-1} / 500$ at the outer boundary compared to $\tau = \Omega^{-1} / 3$ at the inner boundary. As is already standard in the code, these timescales are reduced by a parabolic ramp based on the distance a given location in the zone is away from its respective boundary.

We neglect self-gravity since none of the discs in our study are anywhere close to being massive enough to be affected by self-gravity\mklrc{justify this statement quantitatively: MH: sort of done below}. \cite{lovelace13} showed analytically that the RWI can be damped by self-gravity if $Q <(\pi / 2)(H / r)^{-1}$ in discs with a uniform background density, where $Q = c_\mathrm{s} \Omega / \pi G \Sigma$ is the Toomre Q \citep{toomre64}. Although our most massive simulations can get close to if not reach that critical analytic value at the location of the vortex in our simulations, our previous tests had shown that value to be an overestimate in actual simulations with a planet and a non-uniform background profile \citep{hammer21}.


\subsubsection{3-D setup}

\mklrc{i would describe physical setup before numerical. need to double check vertical BCs. i think they are equilibrium velocities, but density is extrapolated based on hydrostatic eqm.} 

We lower the resolution in the midplane for our 3-D simulations so that it is still feasible to simulate the disc evolution for thousands of orbits. The domain for these simulations is extended into 3-D through the colatitude $\theta \in [\pi - 2.5h, \pi + 2.5 h]$ and is resolved overall by $N_\mathrm{r} \times N_\mathrm{\phi} \times N_\mathrm{\theta} = 256 \times 256 \times 32$ grid cells. This grid resolves the radial direction with an aspect ratio of $h = 0.06$ by 7.3 cells at $r = 1$, while the colatitude is resolved by 6.4 grid cells. For comparison to 2-D, we also run a parameter study of low-resolution simulations with the same resolution of $N_\mathrm{r} \times N_\mathrm{\phi} = 256 \times 256$ in the midplane.

Our 3-D simulations have slightly different initial conditions than our 2-D simulations. We initialize the gas density to $\rho(r, z) = \rho_0(r) \exp \Big[\frac{\alpha}{\mathrm{St}}(\epsilon(z) - \epsilon_0) - \frac{1}{2}\frac{z^2}{H^2} \Big]$, where $\alpha = \nu / c_\mathrm{s} H$ is the dimensionless alpha viscosity \citep{alpha}. The dust-to-gas ratio is $\epsilon (z) = \epsilon_0 \exp \Big(-  \frac{\beta}{2 \alpha} \frac{z^2}{H^2} \Big)$, where $\beta = \frac{1}{2} \mathrm{St}^{-1} \Big(1 - \sqrt{1 - 4 \mathrm{St}^2} \Big)$. These initial conditions follow those from Equations 19 through 22 in \cite{mkl21}. We chose to use gas initial conditions that depend on the dust for the 3-D simulations with feedback, which we had issues running with initial conditions based on a simple hydrostatic equilibrium\mklrc{was the problem that we couldn't run the sim if we include feedback in the run but not in the initial conditions? MH: Yes}. Although the main motivation was to address this issue in the cases with feedback, we used such initial conditions for all of our 3-D simulations even though the majority did not have feedback. These initial conditions are only slightly different from the usual hydrostatic equilibrium, especially for the gas. For that reason, we do not expect our results to have a strong dependence on the precise initial conditions. We use fixed boundary conditions in the vertical direction. 
\mklrc{i think the feedback term in the gas density only slightly modifies it from Gaussian. MH: Added.}



\subsection{Suite of Simulations} \label{sec:simulations}

\begin{table}
\caption{2-D and 3-D parameter study varying the disc mass. The final masses $M_\mathrm{p}$ are in units of $M_\mathrm{J}$ and are recorded at the end of the vortex lifetime. The base density is $\Sigma_\mathrm{base} = 1.157 \times 10^{-4}$, indicated in \textbf{bold}.}
\begin{tabular}{ c c | c c c}
$\Sigma_0 / 10^{-4}$ & $ / \Sigma_\mathrm{base}$ & $M_\mathrm{p}$ [2-D] & [2-D low] & [3-D low]\\
   \hline
        \hline
  $5.787$ & 5.0 & 0.91 & 1.19 & 1.47 \\
  $4.629$ & 4.0 & 0.77 & 0.94 & 1.13 \\ 
  $3.472$ & 3.0 & 0.65 & 0.71 & 0.80 \\
  $2.315$ & 2.0 & 0.51 & 0.50 & 0.51 \\     
  $1.736$ & 1.5 & 0.42 & 0.39 & 0.40 \\
  \textbf{1.157} & \textbf{1.0} & 0.29 & 0.28 & 0.26 \\
  $.9259$ & 0.8 & 0.21 & 0.23 & 0.20 \\
  $.6944$ & 0.6 & 0.17 & 0.17 & 0.17 \\     
  $.5787$ & 0.5 & 0.15 & 0.15 & 0.15 \\
  $.3472$ & 0.3 & 0.10 & & \\
\end{tabular}
\label{table:2D-simulations}
\end{table}

\subsubsection{Disc mass parameter study}

We test the dependence of the lifetimes and morphologies of vortices on different disc masses with nine different disc surface densities presented in Table~\ref{table:2D-simulations}. These surface densities range from $0.3$ to $5.0~\Sigma_\mathrm{base}$, where $\Sigma_\mathrm{base} = 1.157 \times 10^{-4}$. That base surface density corresponds to a total disc mass of $1.6~M_\mathrm{Jup}$ and a Toomre $Q = 165$ at $r = r_\mathrm{0}$. \mklrc{toomre Q needs to be defined. MH: defined earlier} It yields a planet that grows to $0.29~M_\mathrm{Jup}$. As such, the total disc masses cover a range of 0.5 to $8.0~M_\mathrm{Jup}$ and correspond to a range of Toomre Q values from $550$ to $33$ at the location of the planet. The planets in these discs reach masses of 0.10 to $0.91~M_\mathrm{Jup}$.

\subsubsection{2-D and 3-D low resolution parameter studies}

For our 3-D parameter study, we run a 3-D set of simulations across a similar range of disc masses\mklrc{not sure what is meant by 3D parameter study with 2D and 3D sets of simulations}, only removing the case with the lowest surface density. To avoid comparing to 2-D only with different resolutions, we also run another 2-D set of simulations at the same low resolution that we use in 3-D. The planet masses in both low-resolution studies start at $0.15~M_\mathrm{Jup}$, extending to $1.19~M_\mathrm{Jup}$ in 2-D and to $1.47~M_\mathrm{Jup}$ in 3-D. The larger discrepancies for the higher-mass cases are discussed in Appendix~\ref{sec:growth-2D-3D}.


\begin{figure} 
\centering
\includegraphics[width=0.40\textwidth]{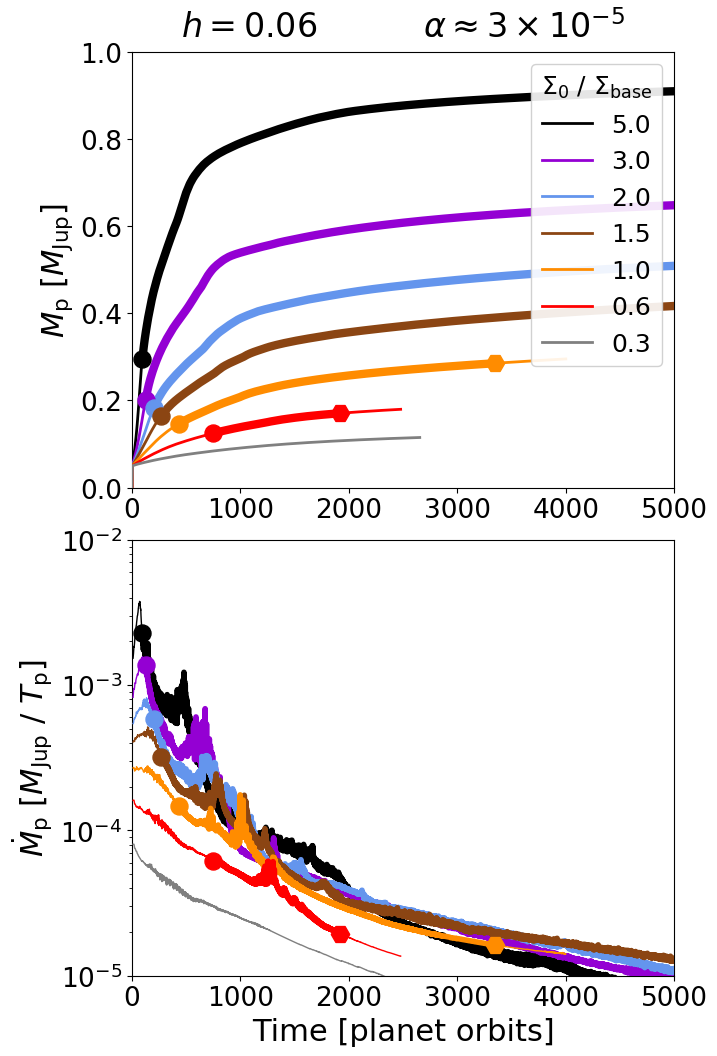}
\caption{Planet mass (\textit{top panels}) and planet mass accretion rate (\textit{bottom panels}) as a function of time over a range of disc masses. A circle denotes the start of the vortex lifetime; a hexagon denotes the end. The vortex lifetime is highlighted in bold. The start and end of the dust asymmetry lifetime are marked on each side of the bold line with a circle and a hexagon respectively. \textit{The accretion rates are measured as the increase in planet mass over each orbit divided by one orbit.}} 
\label{fig:mass}
\end{figure}


\section{Dependence on disc mass} \label{sec:disc-mass}

\begin{figure} 
\centering
\includegraphics[width=0.47\textwidth]{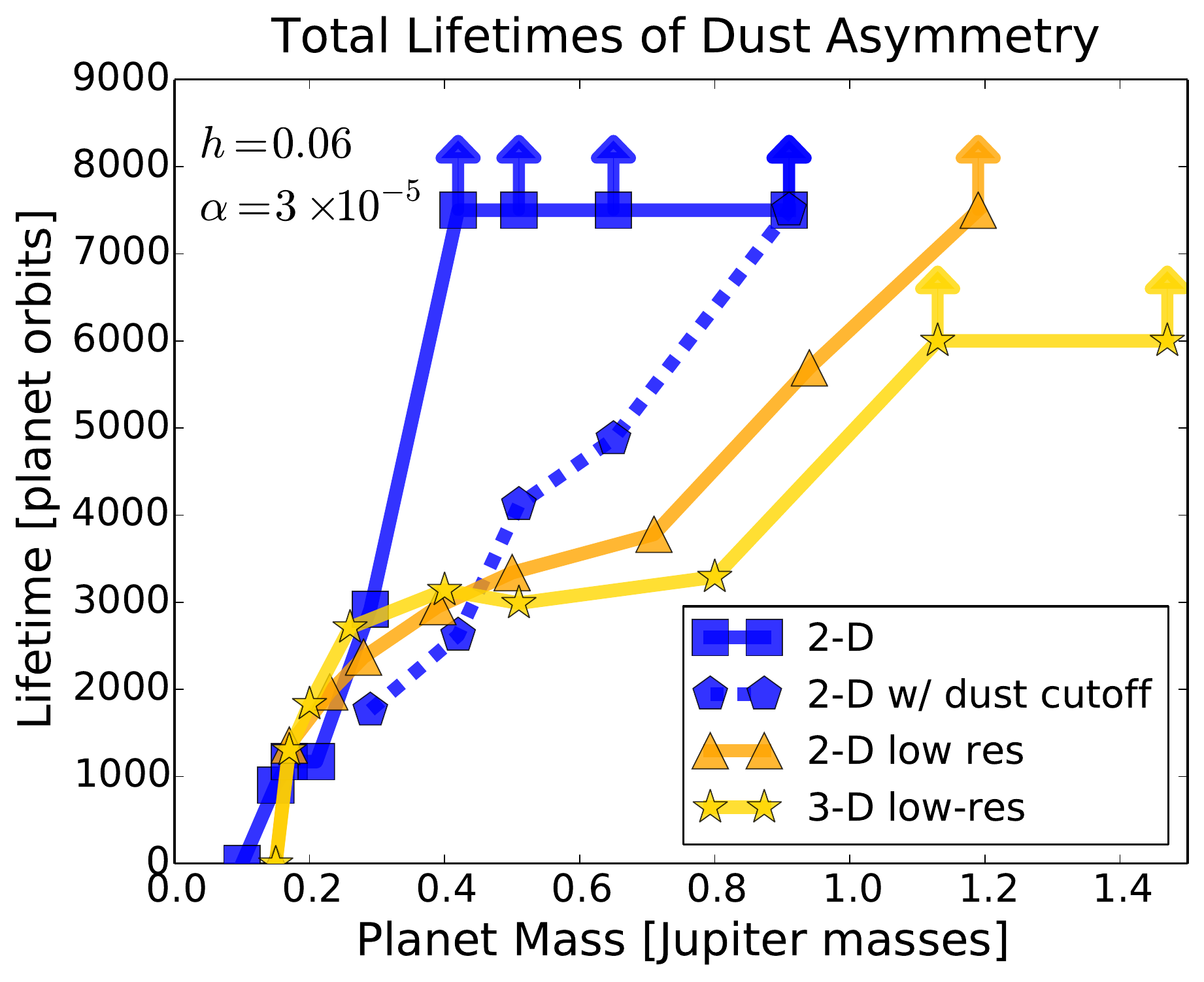}
\caption{Total dust asymmetry lifetimes as a function of planet mass for four different parameter studies. In our fiducial 2-D parameter study, the lifetimes increase monotonically with planet mass until they become indefinite and survive past the end of a simulation, as indicated by an up arrow. The lifetimes in the other three parameter studies -- 2-D w/ a dust cutoff, 2-D low-resolution, and 3-D low-resolution -- follow a similar dependence, except that the lifetimes do not become indefinite until a much higher planet mass. \textit{Like in H21, the end of a dust asymmetry is indicated by when the normalized azimuthal variance of the dust drops below $10\%$.}} 
\label{fig:lifetimes}
\end{figure}

When the final planet masses are set by varying the disc mass, we find that higher-mass planets (in higher-mass discs) generate longer-lived dust asymmetries than lower-mass planets (in lower-mass discs), as shown in Figure~\ref{fig:lifetimes}. We expect to see this largely-monotonic trend at any resolution. High resolution in particular, however, greatly augments the trend for higher-mass planets, facilitating them to re-trigger vortices that allow the dust asymmetries to survive more than twice as long as those generated by lower-mass planets, if they even decay at all. This is the complete opposite outcome to what we found with a fixed disc mass in H21, which instead resulted in the low-mass planets as the ones that could re-trigger vortices and in turn generate longer-lived dust asymmetries. 

Our results differ in this work because the higher resolution allows our simulations to better resolve the RWI critical function, the decay of the elongated vortices, and the flow of vorticity through the vortex. As a result, the initial vortices decay and re-form much earlier on compared to our previous work. And also unlike our previous work, the re-triggered vortices can develop compact cores. To show the results for high-mass planets in particular, we highlight the case with $\Sigma_0 / \Sigma_\mathrm{base} = 3$ as our featured simulation for many of the figures, typically to present facets of the results that were found across a range of simulations.

\subsection{Background} \label{sec:background}
\mklrc{should this be part of an extended intro, rather than part of the results section?}

Discs can become unstable to the non-axisymmetric Rossby Wave instability if they develop a maximum in the radial profile of the Rossby Wave instability critical function. In a locally isothermal disc, this function is given in \cite{mkl12a} as
\begin{equation} \label{eqn:maximum-iso}
L_\mathrm{iso} =  c_\mathrm{s}^2 \frac{\Sigma}{\omega},
\end{equation}
and represents the temperature-scaled inverse vortensity, that is, the density $\Sigma$ divided by the vorticity $\omega \equiv (\vec \nabla \times \vec{v})_\mathrm{z}$. A planet can create a maximum in this function if it opens up a gap, one maximum on each side of the gap. These maxima are located slightly offset from the pressure bump at each gap edge. When incorporating the planet's growth, the initial dominant azimuthal mode of the RWI that is triggered in an $h = 0.06$ disc is typically $m = 3$, albeit higher or lower-numbered modes are possible with higher or lower-mass planets respectively. A mode $m$ yields $m$ co-orbital vortices that continue to grow until the vortices become large enough to interact with each other. At that point, they then slowly merge into fewer vortices until just one big vortex is left. These vortices are typically elongated in shape with aspect ratios exceeding 20. Their minimum Rossby number, the normalized local vorticity of the vortex $\mathrm{Ro} \equiv \omega/2\Omega$, has a low amplitude that falls in the $\mathrm{Ro} > -0.15$ range expected for elongated vortices \citep{surville15}\mklrc{quote a typical aspect ratio for easier visualization. MH: done}, just like the minimum Rossby number of the smaller vortices that seeded it. Eventually, shocks from the planet's spiral waves disrupt that one big vortex -- which we label as the ``first-generation vortex" -- and cause it to spread into a ring in the gas.

After the first-generation vortex decays into a ring, the planet may induce another vortex if it can create another maximum in the RWI critical function. Such a maximum arises due to shocks from the planet's spiral waves as part of the gap-opening process \citep{cimerman21, cimerman23}. We found empirically in H21 that in order for the planet to re-trigger a vortex, the separation between the pressure bump and the maximum in the critical function must stay less than three scale heights. That does not necessarily happen because the pressure bump migrates outward over time as the planet opens up a gap, while the location where the maximum in the critical function develops stays in roughly the same location throughout the gap-opening process.\footnote{There is no analytic expression for the location of this maximum, but it could be estimated from Equation 40 of \cite{cimerman21}.} The pressure bump migrates away from the planet faster for higher-mass planets. In our previous work, the only planets that could re-trigger vortices were the lower-mass ones. The higher-mass planets could not because by the time the first-generation vortex decayed\mklrc{incomplete sentence?}, the pressure bump had already migrated too far away. After a later-generation vortex decays, the planet can re-trigger another later-generation vortex as long as the pressure bump is still close enough to the maximum in the critical function. Although the gas asymmetry decays and re-forms, the dust asymmetry persists throughout and survives even beyond when the final later-generation vortex decays.

\begin{figure*} 
\centering
\includegraphics[width=0.98\textwidth]{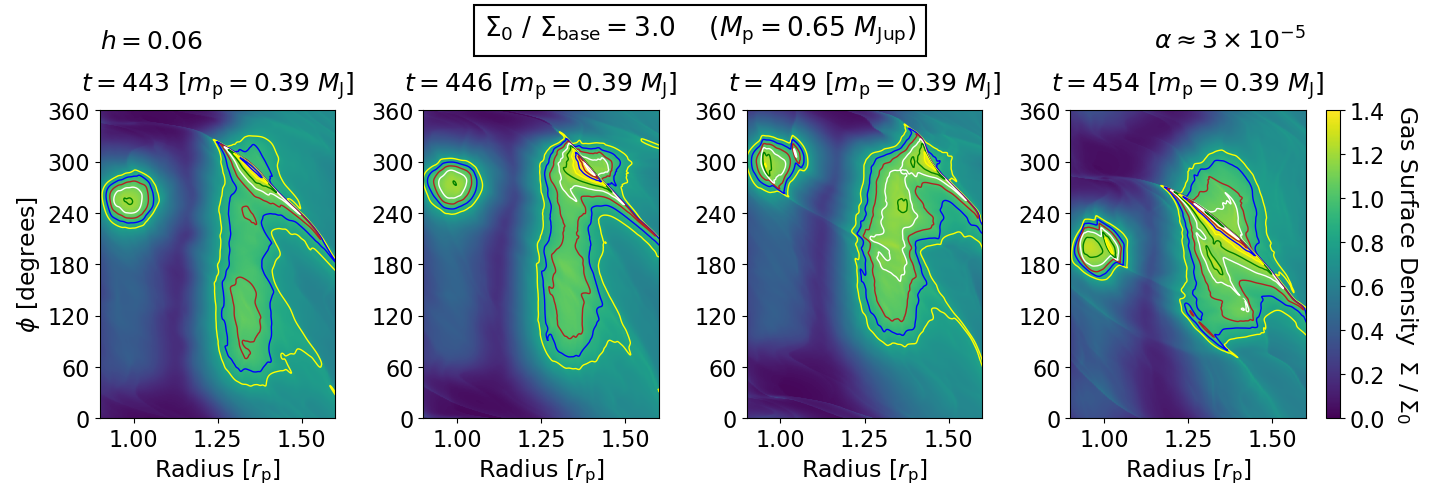} \\
\vspace*{0.5em}
\hspace*{1em}
\includegraphics[width=0.98\textwidth]{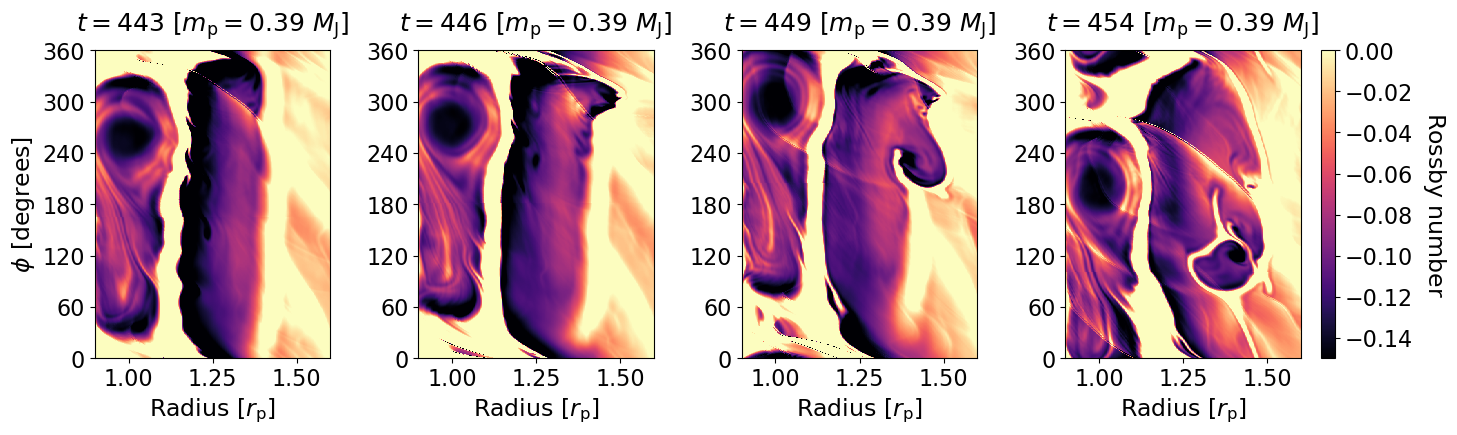}
\caption{High-cadence snapshots of the formation of a ``vortex-in-a-vortex" showing the evolution of the gas density (\textit{top panels}) and Rossby number (\textit{bottom panels}) in the presence of a super-thermal mass planet ($M_\mathrm{p} = 0.65~M_\mathrm{Jup}$) grown from a very massive disc with $\Sigma_0 / \Sigma_\mathrm{base} = 3.0$, the featured simulation. Density contours \cbf{(at $\Sigma / \Sigma_0 = 0.8, 0.9, 1.0,$ etc.)} are overlaid on the density panels. \textit{Column~1}: The elongated $m = 1$ vortex has just re-formed. At the front end of the vortex in azimuth, the lower vorticity from the inner side of the vortex is beginning to advect towards the far side of the vortex. \textit{Column 2}: As the lower vorticity reaches the far side of the vortex, it begins to partially detach from the rest of the vortex. \textit{Column 3}: As the lower vorticity partially detaches, it forms a compact core. \textit{Column~4}: The core spirals in to the interior of the elongated vortex, forming a ``vortex-within-a-vortex" while the larger elongated vortex also remains intact.} 
\label{fig:evolution_h06_s3472-2D-hi-rate}
\end{figure*}

\subsection{New high-resolution vortex evolutionary track} \label{sec:new-track}

\begin{figure*} 
\centering
\includegraphics[width=0.98\textwidth]{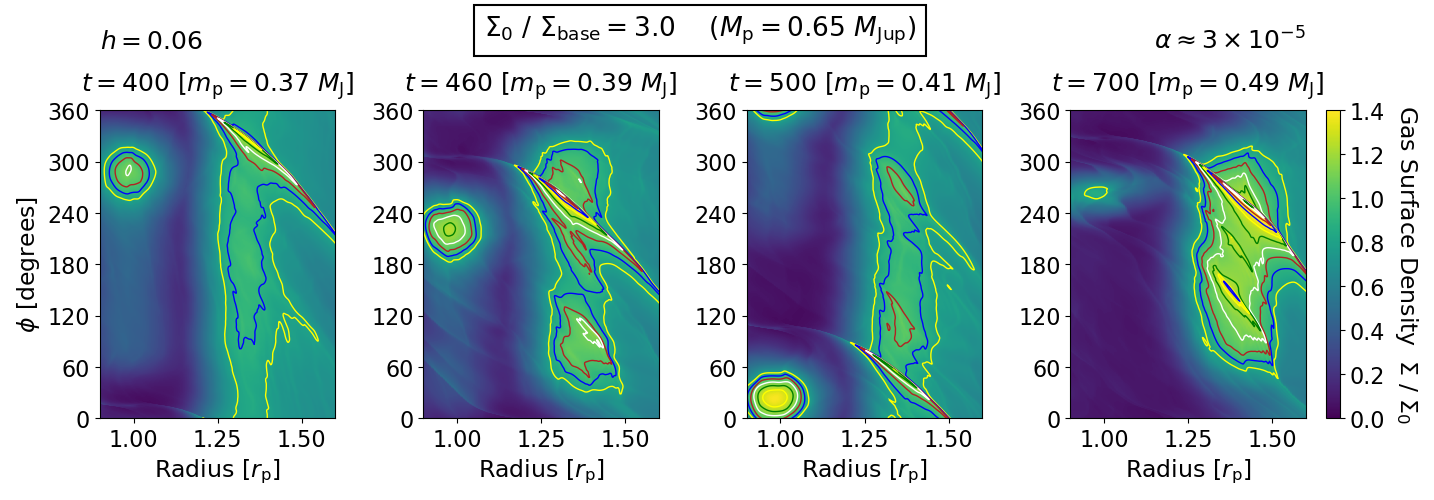} \\
\vspace*{0.5em}
\hspace*{1em}
\includegraphics[width=0.98\textwidth]{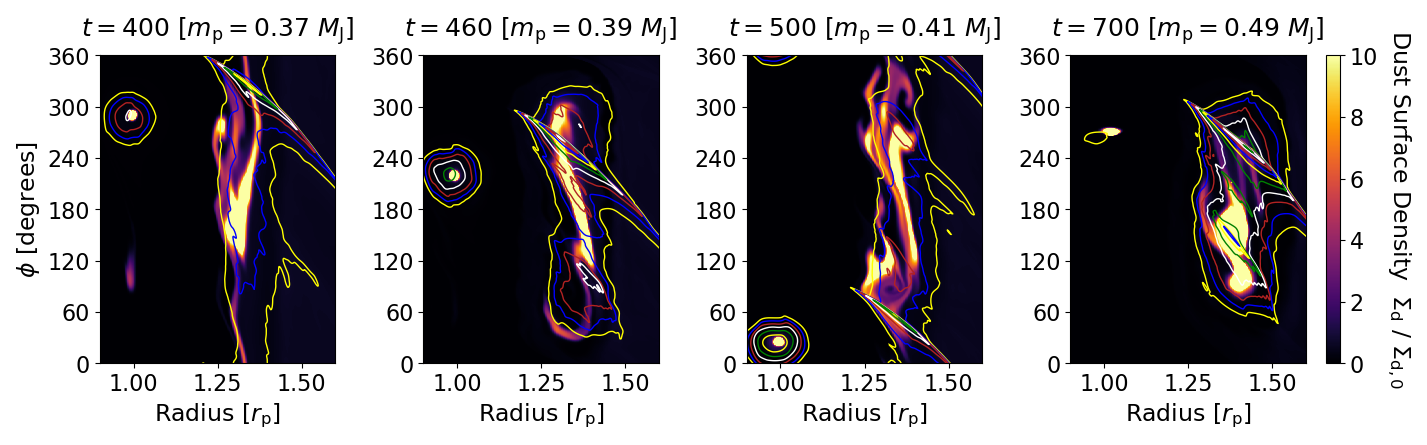} \\
\vspace*{0.5em}
\hspace*{1em}
\includegraphics[width=0.98\textwidth]{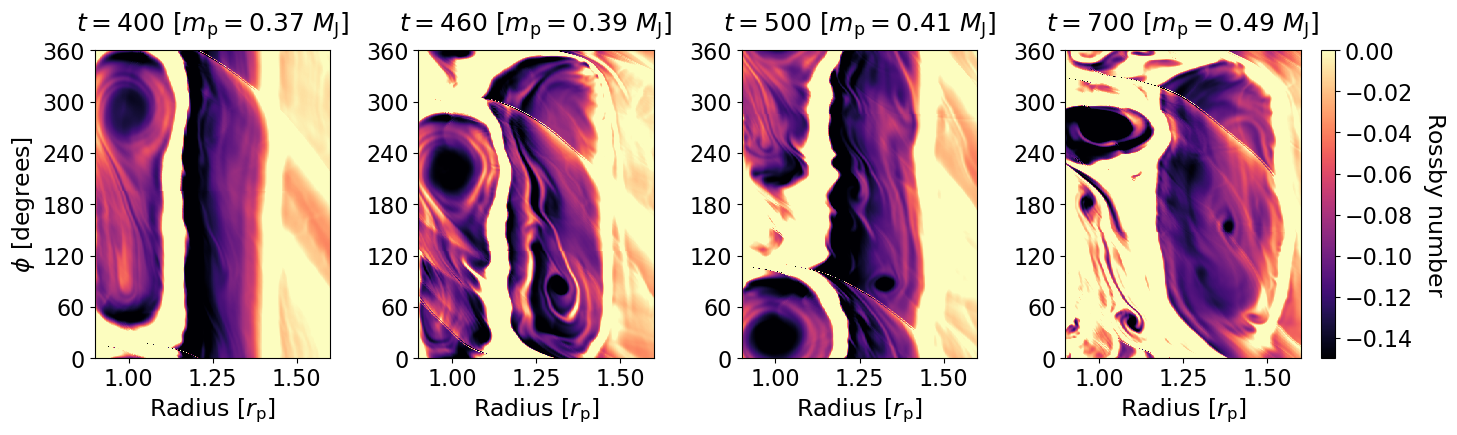}
\caption{Low-cadence snapshots of the formation of the second and third-generation vortices, each with a ``vortex-in-a-vortex", showing the evolution of the gas density (\textit{top panels}), dust density (\textit{middle panels}), and Rossby number (\textit{bottom panels}) for the featured simulation. Density contours \cbf{(at $\Sigma / \Sigma_0 = 0.8, 0.9, 1.0,$ etc.)} are overlaid on the density panels. \textit{Column~1}: The initial $m = 1$ vortex already decayed, leaving behind a ring in the gas while the asymmetry persists in the dust. \textit{Column~2}: The elongated $m = 1$ vortex has been re-triggered. During this process, a compact ``vortex-within-a-vortex" with Ro $< -0.15$ arose --- see Figure~\ref{fig:evolution_h06_s3472-2D-hi-rate} for how this compact core formed. \textit{Column~3}: Although the elongated vortex has again spread out into a ring in the gas, the compact core is still intact. \textit{Column~4}: A third-generation elongated $m = 1$ vortex has formed, and the compact core from the previous vortex is still present.} 
\label{fig:evolution_h06_s3472-2D}
\end{figure*}

\mklsrc{could some kind of flow chart be useful for illustration? e.g. an arrow on the bottom going from left to right, labeled first gen vortex, re-triggered vortex, next gen vortex, later gen vortex, final vortex, etc. with representative snapshots (maybe annotating what is a compact core, vortex-in-vortex, etc)?: MH: maybe in the revision}

With the higher resolution adopted in this work, we find that the first-generation vortices only survive a few hundred orbits in the gas for planets of any mass up to near that of Jupiter. That is about five times shorter than the 1000 to 1500 orbits they lasted in our previous work with lower resolution. Because those vortices are much shorter-lived, however, the pressure bump is still close to the location where the maximum in the critical function develops. As a result, the planet is capable of re-triggering a vortex across this entire range of planet masses, including the high-mass planets that in our previous work could not.

Not only are the planets that can re-trigger vortices different, but the morphology of the resulting re-triggered vortices can be different as well. When a later-generation vortex forms, the overall vortex is still elongated. Within the vortex, though, the structure can be more complex. The negative vorticity gas that accumulates inside the gap just interior to the vortex and near the location of the critical function maximum gets advected into the vortex and begins to circulate along its perimeter, as shown in Figure~\ref{fig:evolution_h06_s3472-2D-hi-rate} for our featured simulation. This excess vorticity has a Rossby number $\mathrm{Ro} < -0.15$, in the more negative range expected for a compact vortex. When this excess negative vorticity reaches the long exterior side of the vortex, it detaches itself outwards from the overall structure and forms a small compact vortex core with $\mathrm{Ro} < -0.15$, a ``vortex-in-a-vortex". Rather than settle at the center, this core continues to circulate around in the interior of the bigger elongated vortex.

Like the first-generation vortex, a second-generation vortex and subsequent later-generation vortices only last a few hundred orbits in the gas, or even less, as some do not even survive one hundred or even fifty orbits. With such short lifetimes, planets are prone to re-triggering more later-generation vortices within a few hundred orbits after the previous one decays, as depicted in Figure~\ref{fig:evolution_h06_s3472-2D}. These additional later-generation vortices may also form compact cores in the same manner as the second-generation vortex. It is possible for re-triggering to yield more than one compact core or none at all. The amount of time these compact cores survive varies. With higher-mass planets, one or more compact cores typically outlive the larger elongated vortex in which they were embedded. When a new later-generation vortex arises, it can harbor both new compact cores as well as those that remain from the previous generation, typically just one such core, if any at all. If the dominant core from a previous generation makes it to the next generation, it tends to dissipate after a new core arises, but not right away. \mklsrc{would a figure be helpful in visualizing these possibilities?: MH: maybe later}

Because the compact cores can greatly outlive the elongated vortices in which they reside, they can greatly prolong the lifetime of the asymmetry signature in the vorticity. The final later-generation vortex tends to decay at around $t = 1500~T_\mathrm{p}$ in the gas surface density. In the vorticity, however, one or more compact cores can still remain at this time. One by one, these cores get sheared by shocks from the planet's spiral waves and fade away. The strongest core by vorticity lasts much longer than the others. Although it only keeps its Rossby number in the compact range for a few hundred orbits, the core is still clearly visible with a Rossby number in the weaker elongated range for over a thousand orbits and remnants can still be seen until about $t = 4000~T_\mathrm{p}$, around 2500 orbits later. That lifetime of the vorticity signature of the last compact core is also more than twice as long as the elongated vortices survived in the gas from the start of the first to the end of the last. With no vortex in the gas during this late stage, the evolution of the compact cores in the vorticity play an even more dominant role in governing the appearance of the dust asymmetry. \mklrc{this paragraph can be shortened or omitted because vorticity/Rossby number isn't an intuitive quantity for most readers. dust asymmetry is more relevant for our target audience (as alluded to next)}

Even more important than extending the vorticity lifetime, the compact cores also extend the lifetime of the dust asymmetry, which survives long after the vorticity signature fades away. Towards the beginning when the first-generation vortex first arises, the dust has a very elongated extent typically between 90 to 180 degrees as it circulates around the vortex, as expected from our previous work. When the compact cores arise in the later-generation vortices, the core slowly traps a significant clump of dust that follows the circulating core even though the rest of the dust is still spread across the vortex. When that core starts to disappear after the next-generation vortex forms, the dust slowly moves to the new dominant compact core. Towards the end after the final later-generation vortex decays, the dust becomes very compact and is trapped in the compact cores that remain. 

Once each core's Rossby number weakens into the elongated range, its dust slowly becomes more spread out and no longer follows the compact shape of the core in its vorticity signature. But because the dust began this phase of evolution as very compact, the dust in the strongest core takes significantly longer to decay than the dust signatures left over by elongated vortices from our previous work. Eventually, after the dust has already spread out around the full 360$^{\circ}$ ring, we find that it stops decaying altogether, but still maintains a significant asymmetry in a steady state\mklrc{so dust decays into a ring, yet is still asymmetric}. 

\cbf{In Appendix~\ref{sec:steady}, we tested in various ways if this outcome of an asymmetric steady state was unphysical. We found that the dust asymmetries eventually decay if we cut off the dust supply to the vortices. Even with this decay, the resulting lifetimes are still among the longest in our study, as shown in Figure~\ref{fig:lifetimes}. With an unlimited dust supply though, we could not find any way to refute our finding that the steady state should still be asymmetric.} With this kind of a steady state, the dust asymmetry lifetimes are incomparably longer than those in cases without compact cores. \mklsrc{would a figure be helpful here? MH: maybe later}

\subsubsection{Planet mass dependence} \label{sec:mass-dependence}

The evolutionary track described in the preceding part of this section applies more for higher-mass planets ranging from a final mass of $M_\mathrm{p} = 0.42~M_\mathrm{Jup}$ to $0.77~M_\mathrm{Jup}$ ($1.5 \le \Sigma_0 / \Sigma_\mathrm{base} \le 4$). These higher-mass planets are needed to generate compact cores during the re-triggering process because those cores rely on the gas just interior to the elongated vortex having a large amount of negative vorticity. This gas interior to the vortex only has a very negative vorticity if it is supplied through shocks from the planet's spiral waves. Only higher-mass planets have strong enough shocks to generate a sufficient amount of negative vorticity. This pattern can be gleaned from the $M_\mathrm{p} = 0.45~M_\mathrm{Jup}$ case, which does not develop any compact cores in its second-generation vortex but still manages to produce a compact core by the final-generation vortex when the planet is much more massive.

Not only can the lower-mass planets not generate sustained compact cores, but the closer proximity of the pressure bump maximum to the location of the critical function maximum no longer helps them re-trigger vortices in a way that can extend the overall lifetimes either. When the lower planet masses are the result of lower disc masses, the planets carve out much deeper caps compared to our previous work. The gap depletion competes against the proximity between the two maxima, weakening the maximum in the critical function enough so that these planets can only re-trigger vortices early on just like the higher-mass planets. The lower-mass planets may still be massive enough to form compact cores, but they are not massive enough to sustain them.\mklrc{not sure what is meant here: almost massive enough to form compact cores also form compact cores, so i guess the mass is massive enough to form compact cores?} As a result, those cores are weaker and too short-lived to sustain any asymmetry in the vorticity or the dust much past the end of the last-generation gas vortex. The lowest-mass planet we simulate does not trigger the RWI at all, thereby setting $M_\mathrm{p} = 0.10~M_\mathrm{Jup}$, as about the minimum required mass for a static planet in an $h = 0.06$ disc to generate a vortex. Overall, the vortices with short finite lifetimes from 1000 to 3000 orbits are generated by planets with low final masses between $M_\mathrm{p} = 0.15~M_\mathrm{Jup}$ to $0.29~M_\mathrm{Jup}$ ($0.5 \le \Sigma_0 / \Sigma_\mathrm{base} \le 1.5$). Their short lifetimes are due to the gap being depleted and the compact cores in re-triggered vortices either never forming or disappearing too quickly.

The highest-mass planet we simulate that grows to $M_\mathrm{p} = 0.91~M_\mathrm{Jup}$ from $\Sigma_0 / \Sigma_\mathrm{base} = 5$ is the most different and does not follow the standard evolutionary track. Much more simple than that standard track, we find that the first elongated vortex survives indefinitely in the gas and therefore does not lead to any later-generation vortices or compact cores, nor does it need to survive that long. The gas and dust are both in the process of decaying, but very slowly. Although the vortex does appear elongated in both its azimuthal extent and minimum Rossby number, it still has a more compact structure. More specifically, it has much steeper density contours, a big difference from usual elongated vortices in the other cases that have a mostly flat density plateau across the nearly the entire vortex. We suspect this structure is responsible for the vortex's longevity.

\subsubsection{Role of RWI critical function} \label{sec:critical}

\begin{figure} 
\centering
\includegraphics[width=0.48\textwidth]{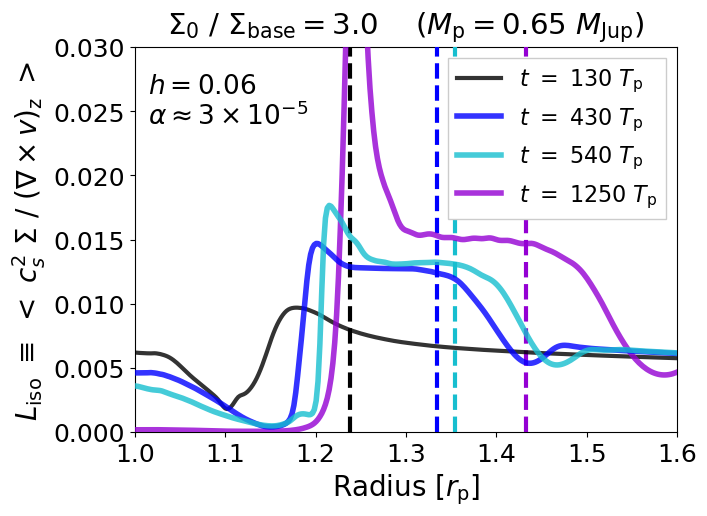} \\
\caption{For the featured simulation, the RWI critical function $L_\mathrm{iso}$ (\textit{solid lines}) and the associated pressure bump locations (\textit{dashed lines}) over time, in particular when the initial vortex is triggered (\textit{thin black line}) and when the later-generation vortices are triggered (\textit{thick coloured lines}). Compared to the lower-resolution studies in H21, the new spikes in $L_\mathrm{iso}$ interior to the pressure bump that result in the later-generation vortices are much more prominent.} 
\label{fig:maximum-condition_h06_s3472-2D-retriggers}
\end{figure}

\begin{figure} 
\centering
\includegraphics[width=0.48\textwidth]{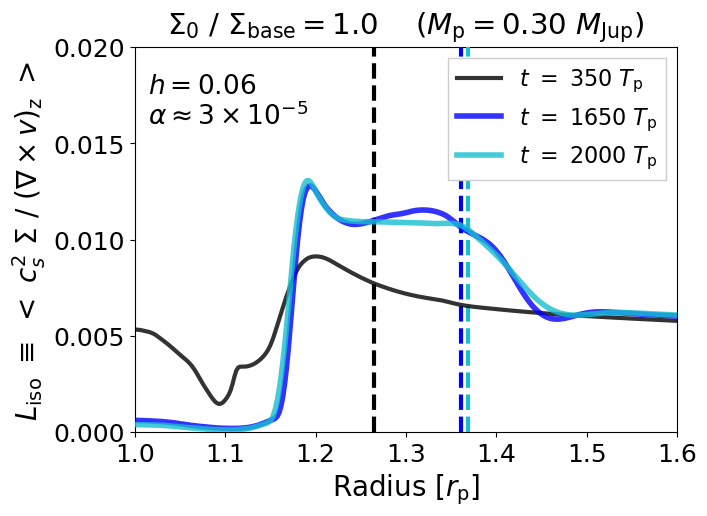} \\
\caption{With a low-mass planet, the RWI critical function $L_\mathrm{iso}$ (\textit{solid lines}) and the associated pressure bump locations (\textit{dashed lines}) at the beginning and end of a critical function bump that does not form a vortex (\textit{thick coloured lines}), in comparison to when the initial vortex is triggered (\textit{thin black line}). The critical function bump later on does not form a vortex because it is closer-in than usual due to the depletion of the gap.} 
\label{fig:maximum-condition_h06_s1157-2D-fail}
\end{figure}

With the higher resolution in this study, it is much clearer to see the RWI critical function set up the conditions for generating all of the later-generation vortices. Right before each re-triggered vortex arises, a new bump in the critical function appears, several examples of which are shown in Figure~\ref{fig:maximum-condition_h06_s3472-2D-retriggers}. When the same thing happened at lower resolution in H21, it was barely discernible in any isolated snapshot of the radial profile and could only easily be seen in a time evolution. The bumps were so difficult to discern that we did not include an example for any of the $h = 0.06$ cases and instead resorted to displaying an $h = 0.04$ case (see Figure 4 from that work). In this work, however, it is straightforward to see the bumps grow before each re-triggering, as would be expected from the conditions for triggering the RWI. This difference suggests that the low resolution in our previous work was not sufficient to resolve the development of the bumps in the critical function. Moreover, it is the additional negative vorticity that produces these stronger bumps that is also responsible for creating the compact cores that can ultimately control the dynamics of the dust asymmetry and extend its lifetime.

The bumps induced by low-mass planets in particular eventually do not yield any more later-generation vortices. Figure~\ref{fig:maximum-condition_h06_s1157-2D-fail} shows an example of such a bump that does not re-trigger the RWI. With a more depleted gap, the bump arises just interior to $r = 1.2~r_\mathrm{p}$. That is a bit closer-in than where it usually arises, just exterior to $r = 1.2~r_\mathrm{p}$ (see the thick coloured lines of Figure~\ref{fig:maximum-condition_h06_s3472-2D-retriggers}, or the orange line from the middle panel of Figure 5 in H21). We suspect the bump is closer inward because of the gap depletion. This difference in location is the main reason the planet stops re-triggering vortices much earlier than the similar-mass planets with less gap depletion from our previous work.

\subsubsection{Other aspect ratios} \label{sec:aspect-ratios}

Like in our fiducial $h = 0.06$ parameter study, we find that higher resolution also facilitates a disc with $h = 0.04$ to produce compact cores. We tested this with planets that grow to $M_\mathrm{p} = 0.36$ and $0.52~M_\mathrm{Jup}$ (from discs with $\Sigma_0 / \Sigma_\mathrm{base} = 3$ and $5$), using a higher resolution to match the radial grid cells per scale height in our fiducial $h = 0.06$ cases. Following the pattern in those cases, both of these simulations result in later-generation vortices with compact cores, but only the compact cores in the higher-mass case outlast the last-generation gas vortex. They extend the dust asymmetry lifetime to about 3700 orbits, longer than any of the lifetimes of the $h = 0.04$ cases with lower resolution from H21 and more than triple the lifetime of the case with a similar-mass planet. Unlike our fiducial parameter study though, the lingering compact core does not make the dust asymmetry survive indefinitely. We interpret this difference as further evidence that the gas structure at the outer pressure bump in the $h = 0.06$ cases is partially responsible for sustaining the dust asymmetry after the gas vortices have decayed.

The higher resolution in this study does not help planets in discs with $h = 0.08$ produce compact cores in later-generation vortices because the first-generation vortices already survive indefinitely, just like in our previous work.


\section{Effects of planet migration} \label{sec:results-migration}

\begin{figure} 
\centering
\includegraphics[width=0.40\textwidth]{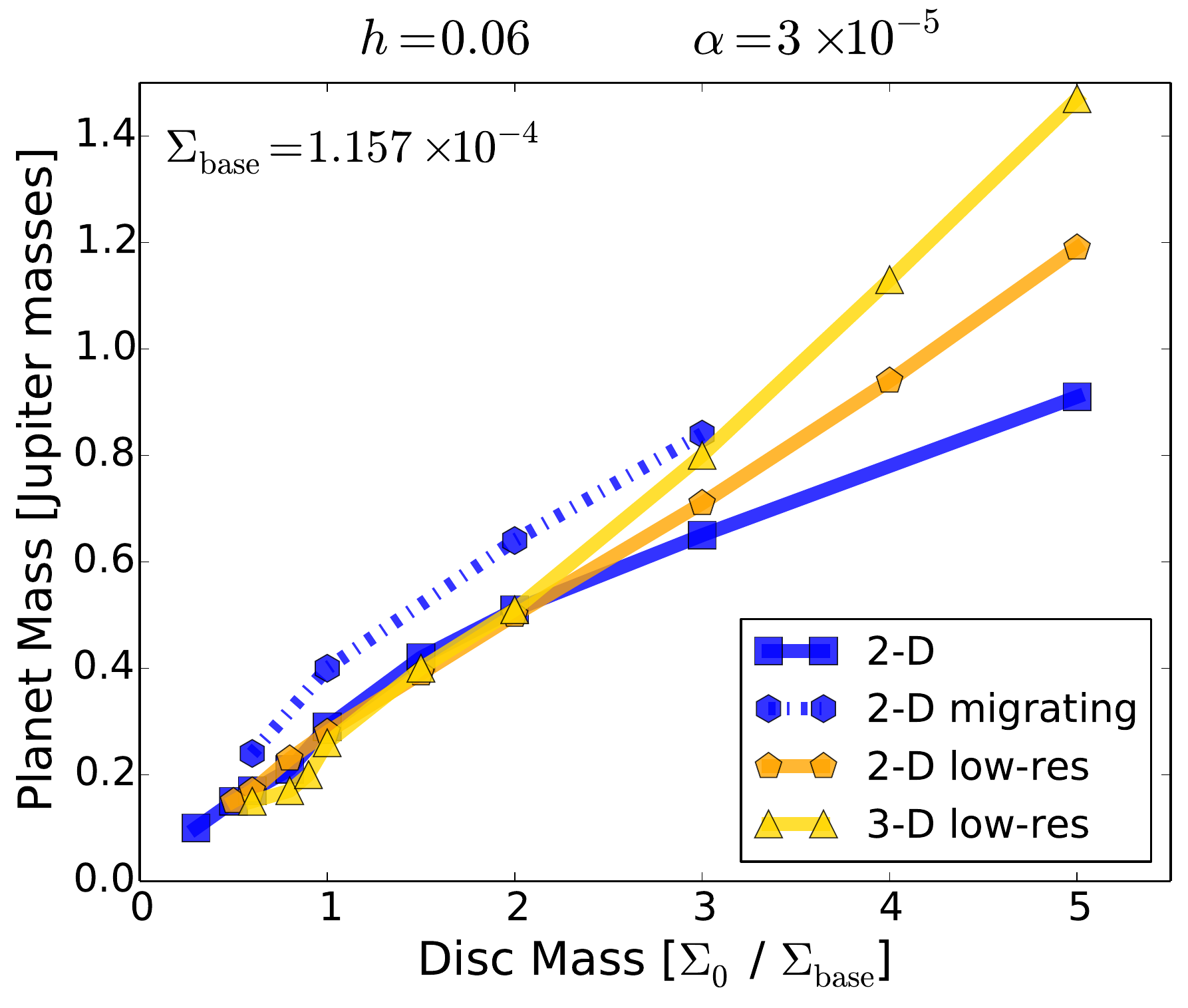}
\caption{Planet mass versus disc mass for four different parameter studies. In less massive discs ($\Sigma_\mathrm{base} / \Sigma_0 \le 2.0$), there is good agreement in all of the parameter studies except the migrating one, which is expected to yield more massive planets. In more massive discs, 3-D yields more massive planets than 2-D and low-resolution yields more massive planets than our higher fiducial resolution, also due to the different smoothing lengths used.} 
\label{fig:planet-vs-disc}
\end{figure}

\begin{figure} 
\centering
\includegraphics[width=0.40\textwidth]{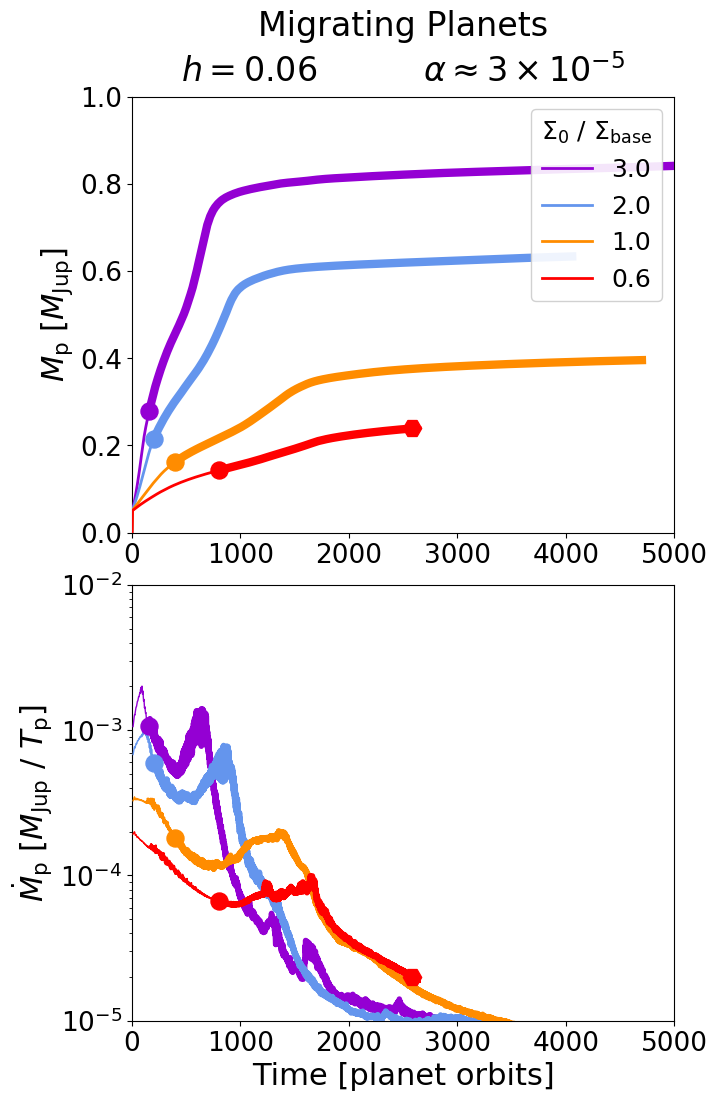} \\
\vspace*{1em}
\hspace*{1em}
\includegraphics[width=0.40\textwidth]{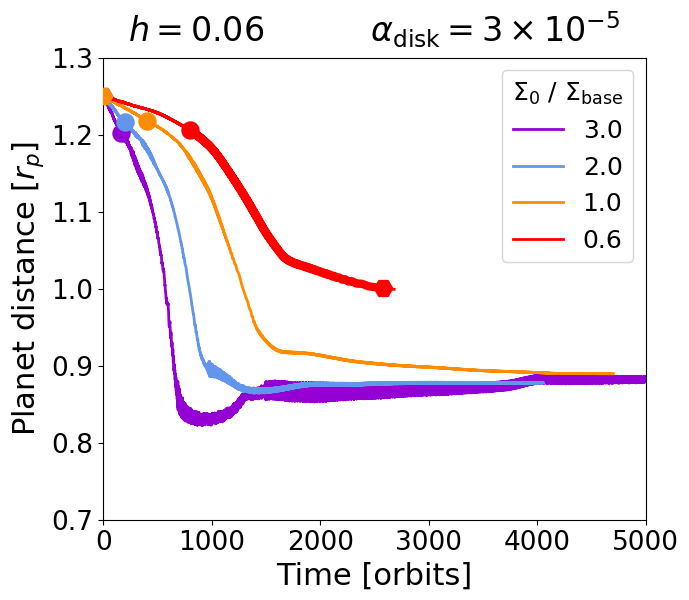}
\caption{Planet mass (\textit{top panels}), planet mass accretion rate (\textit{bottom panels}), and planet radii (\textit{bottom panels}) as a function of time for cases with a migrating planet. \textit{More details in caption for Figure~\ref{fig:mass}.}} 
\label{fig:mass-migrating}
\end{figure}

\mklrc{changing disk mass now affects both planet accretion and migration. MH: Added a sentence about the Type I migration rate.}

To build off our findings related to later-generation vortices, compact cores, and dust asymmetry lifetimes increasing with planet mass and disc mass, we now test if such results still hold if we allow the planet to migrate, as that would be more realistic. We ran a smaller parameter study for a migrating planet in which we repeated four different disc masses: $\Sigma_0 / \Sigma_\mathrm{base} = 0.5$, 1.0, 2.0, and 3.0. We initialised the planet further out at $r = 1.25~r_\mathrm{p}$ so that it would end up near its usual location of $r = 1.0~r_\mathrm{p}$ and the vortex would end up near $r = 1.5~r_\mathrm{p}$, about where it is located with a static planet. The more distant starting location of the planet also limits each planet's growth, preventing them from becoming much more massive than their static counterparts, as shown in Figure~\ref{fig:planet-vs-disc} and the top panel of Figure~\ref{fig:mass-migrating}. \footnote{\cbf{We do not include the torque correction from \cite{baruteau08} for simulations that neglect self-gravity because it primarily affects the Type I migration phase before there are vortices.}}

\subsection{Planet migration track} \label{sec:new-track}

\begin{figure*} 
\centering
\includegraphics[width=0.44\textwidth]{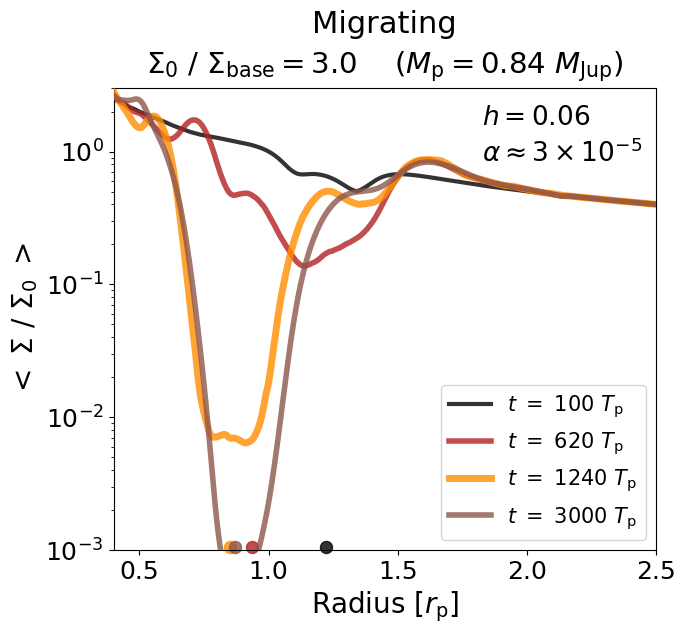}
\hspace*{3.5em}
\includegraphics[width=0.49\textwidth]{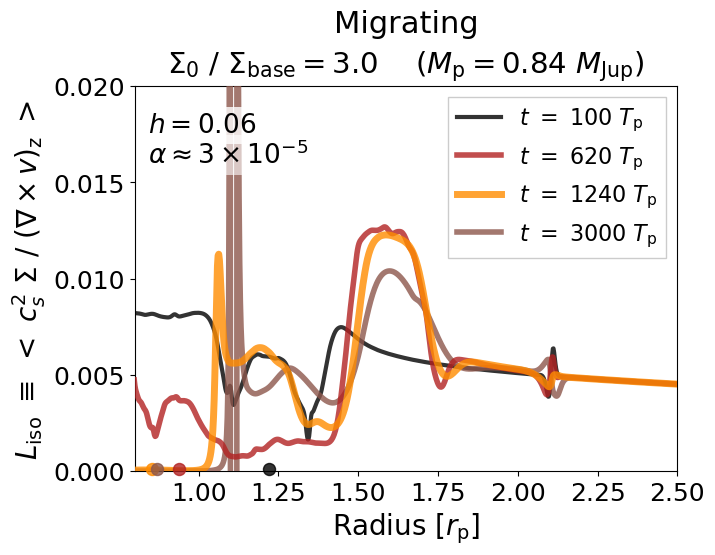}
\caption{Azimuthally-averaged gas density profiles (\textit{left panel}) and the RWI critical function $L_\mathrm{iso}$ (\textit{right panel}) over time for a migrating high-mass planet ($M_\mathrm{p} = 0.84~M_\mathrm{Jup}$) grown from a very massive disc with $\Sigma_0 / \Sigma_\mathrm{base} = 3.0$, the migrating analogue to the featured simulation. At $t=100~T_\mathrm{p}$ (\textit{thin black line}), the RWI is first triggered. At $t=620~T_\mathrm{p}$ (\textit{red line}), the planet has migrated inwards, leaving the pressure bump at the outer gap edge behind. At $t=1240~T_\mathrm{p}$ (\textit{thick orange line}) is the final instance of a bump in $L_\mathrm{iso}$ forming near $r = 1.1~r_\mathrm{p}$ interior to the original pressure bump, giving rise to the final later-generation vortex. At $t=3000~T_\mathrm{p}$ (\textit{brown line}), there is still another bump at about the same location, but it is too far interior to the inner pressure bump in the outer disc to form another vortex.} 
\label{fig:migration-profiles}
\end{figure*}

Overall, we find that the planets initially migrate inwards rapidly before slowly settling on a near-fixed and near-circular orbit. The rapid migration lasts until the planet opens a significant gap, as expected in the Type I regime\mklsrc{is the migration rate consistent with type i results?: MH: add the equation later}. Because the Type I migration rate increases proportionally to both the planet mass and disc mass \citep[e.g.][]{lin79}, each more massive planet migrates much faster than the next most massive one. When the planet migrates, it leaves the outer gap edge behind, a typical outcome found by \cite{kanagawa21} in their study of migrating planets in low-viscosity discs. Once the planet is sufficiently far from the outer gap edge, it creates a double gap exterior to the planet, as shown in Figure~\ref{fig:migration-profiles}. A new weaker pressure bump arises that is located interior to the original one. With the planet so far from the original outer gap edge, it mostly stops migrating, settling at around $r = 0.88~r_\mathrm{p}$ in all but the lowest-mass case, as shown in the bottom panel of Figure~\ref{fig:mass-migrating}.


\subsection{Vortex evolution with planet migration} \label{sec:new-migration-track}

\mklsrc{would it be useful to have side-by-side comparisons of the disk morphology between a migrating and a static planet? do you expect a migrating planet to populate the disk with vortices across a range of radii?: MH: not that different overall. No to the second part.}

Because the planet stops migrating and due to the low viscosity in the disc, the original gap edge is maintained and the general evolution of the generations of vortices is not that different than from the static planet cases. A new series of vortices also forms at the new weaker gap edge, but there is little to no dust to populate them because any dust from the outer disc that might drift inwards gets trapped at the original pressure bump. Only the case with the highest-mass planet can form compact cores. Even though they form in the vortices at both pressure bumps, the cores at the new pressure bump with no mm dust tend to be sustained longer while the cores at the original bump barely last at all. Eventually, the double gap merges into a single gap, leaving only the vortex at the original bump. In all but the lowest-mass case, we find that the dust asymmetry stops decaying and never goes away. Unlike with a fixed planet, these indefinite lifetimes do not directly arise from compact cores and even cutting off the dust supply does not shorten the dust asymmetry lifetime. 

Each gas vortex that forms tends to be even shorter-lived than those in from our parameter study with a static planet. Those short lifetimes have little effect on the overall dust asymmetry lifetime, however, because the gas vortex quickly re-forms and the dust never decays in-between, just like without planet migration enabled. The planet continues to re-trigger vortices at the original outer bump until about $t = 1000$ to $1500~T_\mathrm{p}$ even though the planet has migrated more than three scale heights away from the bump in the critical function\mklrc{how does the planet do that if it moves away from the bump? is the distance between Liso and Pbump still less than 3H?}. The higher-mass planets stop re-triggering vortices at this bump earlier than the low-mass ones because the higher-mass planets migrate inwards and away from the original bump much faster.

\subsubsection{Dust signature with planet migration} \label{sec:migration-dust-signature}

The dust signature can be more chaotic than with a static planet because the gas vortex keeps decaying and re-forming. It settles down after the last-generation vortex decays. Unlike with a static planet, only the dust in the highest-mass planet case ($\Sigma_0 / \Sigma_\mathrm{base} = 3$) becomes more compact at this point, albeit there is no clear sign of a compact core and the signature isn't nearly as compact as it is with a static planet. The dust asymmetries in the other cases do not even become more compact. Nevertheless, they still do not decay. Another reason this outcome is unexpected is that the vorticity signature at this pressure bump appears largely axisymmetric. 

We suspect the longevity of the dust asymmetry may be related to the planet migrating away from the original pressure bump. With a much bigger distance in-between as well an entirely new pressure bump forming in-between, the shocks from the planet's spiral waves may not affect the asymmetry as much as they do with a static planet.


\subsubsection{New pressure bump} \label{sec:new-pressure-bump}

\begin{figure} 
\centering
\includegraphics[width=0.48\textwidth]{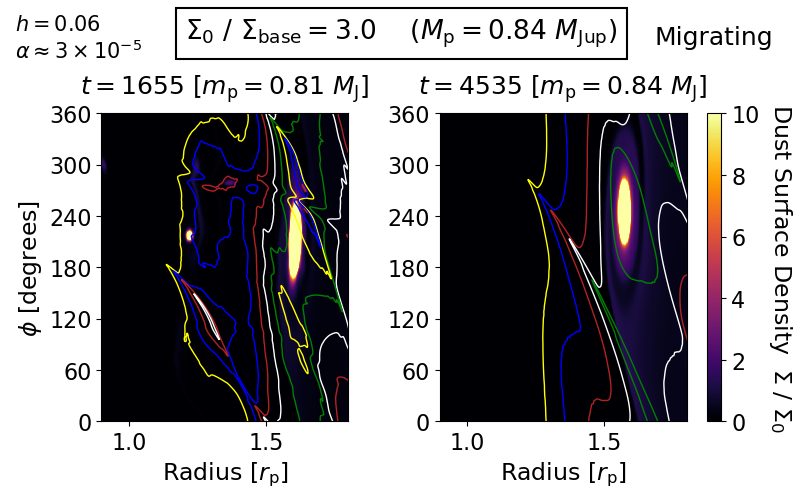} \\
\vspace*{0.5em}
\hspace*{0.5em}
\includegraphics[width=0.49\textwidth]{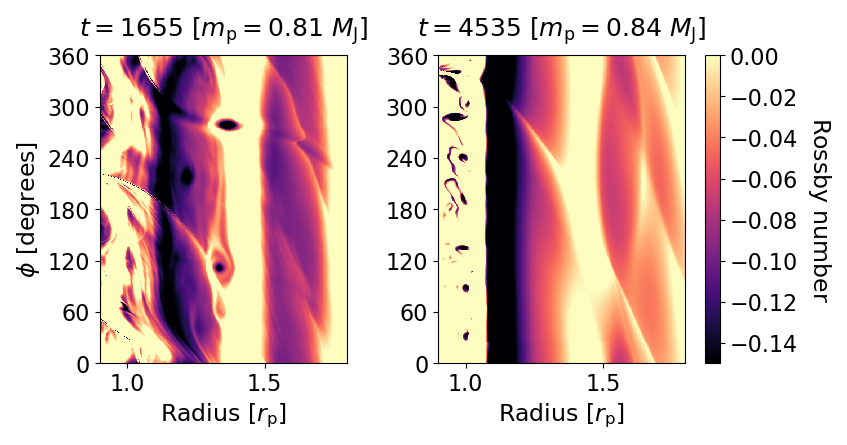}
\caption{Dust density (\textit{top panels}) and Rossby number evolution (\textit{bottom panels}) for the migrating analogue to the featured simulation. Contours begin at 0.4. \textit{Column~1}: At $t=1655~T_\mathrm{p}$, the gas vortex at the initial pressure bump has spread into a ring, albeit the dust there is still asymmetric. At the interior outer pressure bump, a later-generation vortex-in-a-vortex is present in the gas and it has trapped a small amount of dust at a concentrated point. The compact core from an earlier later-generation vortex has ended up in-between the two pressure bumps and its spiral waves affect both bumps. \textit{Column~2}: At $t=4535~T_\mathrm{p}$, even though vortices at both outer pressure bumps have long disappeared, the dust from the original vortex persists and has reached a steady state with no sign of decay. The vorticity signature at the corresponding location is not entirely symmetric.} 
\label{fig:migration-maps}
\end{figure}

Although the new inner bump is not populated by any mm-sized dust, it does yield a new series of later-generation vortices. The evolution of these vortices is also similar to those with a static planet. These vortices typically form after the last-generation vortex at the original bump decays, or even earlier in the case with the highest-mass planet.

That latter case is also the only one to develop compact cores. Although they have no direct impact on the prominent dust signature at the other pressure bump, the last compact core breaks away from the vortex, as depicted in the left panel of Figure~\ref{fig:migration-maps}. As expected for a vortex that is not at a pressure bump, it ends up migrating towards the original pressure bump \citep{paardekooper10}. In-between the two pressure bumps, the vortex develops more significant spiral waves that may affect both pressure bumps. At this point, the vorticity signature at the original bump becomes more asymmetric even though there is no sign the vortex there ever re-formed. 

Towards the end of the simulation, that asymmetry in the vorticity becomes stronger, as shown in the right panel of Figure~\ref{fig:migration-maps}. The dust that remains somewhat follows this asymmetric signature, albeit with a narrow signature. Although this behavior may help sustain the dust asymmetry in this case, the dust asymmetry in the other migrating cases\mklrc{do you mean lower mass cases?} can still be sustained even without any unusual behavior or compact cores at all. 

Lastly, even though the two pressure bumps merge into a single gap structure in the density, the vorticity signatures of the bumps remain separate beyond the end of the simulation.

\subsubsection{Lowest-mass migrating planet} \label{sec:low-mass-migrating-planet}

\begin{figure} 
\centering
\includegraphics[width=0.48\textwidth]{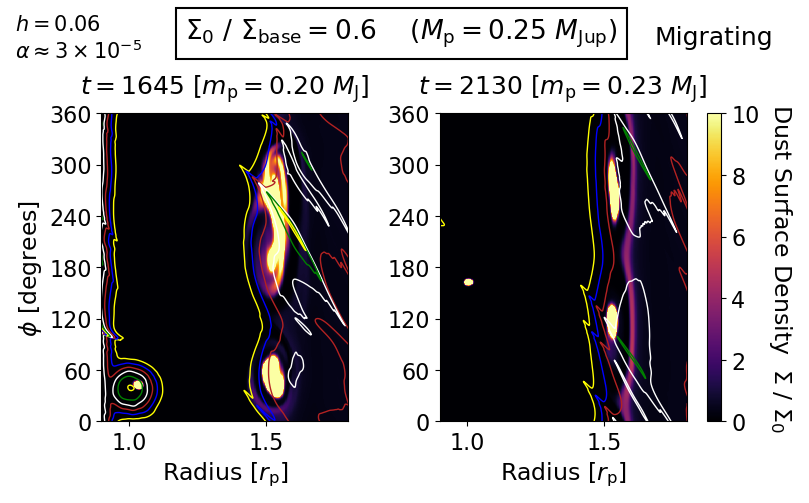} \\
\vspace*{0.5em}
\hspace*{0.5em}
\includegraphics[width=0.49\textwidth]{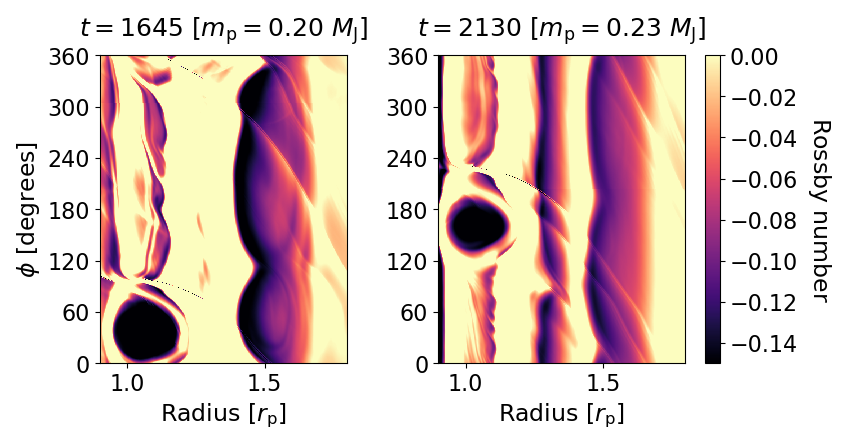}
\caption{Dust density (\textit{top panels}) and Rossby number evolution (\textit{bottom panels}) in the presence of a low-mass planet ($M_\mathrm{p} = 0.25~M_\mathrm{Jup}$) grown from a low-mass disc with $\Sigma_0 / \Sigma_\mathrm{base} = 0.6$. \textit{Column~1}: At $t=1645~T_\mathrm{p}$, the last later-generation set of vortices has yielded a long-lived $m = 2$ mode. \textit{Column~2}: At $t=2130~T_\mathrm{p}$, the $m = 2$ mode is decaying into a ring and the dust is collecting at the center of the pressure bump just exterior to the location of the vortices.} 
\label{fig:co-orbital}
\end{figure}

The lowest-mass migrating case ($\Sigma_0 / \Sigma_\mathrm{base} = 0.6$) differs the most because of how slowly the planet migrates. It is the only case in which the dust asymmetry does not survive indefinitely. Despite the behavior starting out similar to the static planet simulation with a similar planet mass, the dust develops a very different configuration around $t = 1500~T_\mathrm{p}$ when an $m = 2$ mode with two co-orbital vortices becomes dominant, as depicted in Figure~\ref{fig:co-orbital}. The dust stays separate in these vortices even though the vorticity signatures of these vortices have some overlap. This pattern eventually decays as dust escapes the vortices and settles at the actual peak of the pressure bump, a phenomenon we also see in the low-mass static planet simulations as well as in H21 with a ten-times higher viscosity. 

This low-mass simulation is also the only migrating case that doesn't have a clear double gap in the density and in turn no new pressure bump with another vortex. Even without another new pressure bump, the outer region of the gap still has two separate rings of lower vorticity just like all of the higher-mass planet cases with double pressure bumps. In observations, the resulting gap associated with this kind of low-mass migrating planet should still be much wider than that of a static planet of comparable mass, hosting an outer dust ring at $r \approx 1.6~r_\mathrm{p}$ compared to $r \approx 1.4~r_\mathrm{p}$ with the static planet.  \mklrc{any implications for this morphology? MH: Added note about gap}

\section{3-D simulations} \label{sec:results-3D}

Beyond testing if our 2-D results were applicable with a migrating planet, we also explore if the 2-D simulations are a good approximation for actual 3-D protoplanetary discs. In general, 3-D simulations are only used when necessary because 2-D simulations are much less computationally expensive. With isothermal simulations, however, an additional complication is that 3-D simulations with a high resolution yield the vertical shear instability \citep[VSI: e.g.][]{nelson13}, which affects any vortices that form through the RWI.

To avoid the VSI and do a more direct comparison to 2-D simulations, we instead looked for differences at lower resolution. We ran a suite of 3-D simulations with a low resolution of 256 $\times$ 256 $\times$ 32, which we compared to another new low-resolution 2-D parameter study with the same midplane resolution of 256 $\times$ 256 used in 3-D. Beyond comparing 2-D with 3-D, we also ran a few high-resolution simulations in 3-D primarily to investigate what happens when the VSI interacts with the RWI.

\subsection{Reproducing disc mass dependence}

\subsubsection{Low-resolution 2-D} \label{sec:low-2D}

\mklrc{going back to lower resolutions, do we just recover the 2021 results? MH: Yes, added a note about the appropriate appendix at the end of this section}


With a low resolution, we find the same general pattern for the dust asymmetry lifetimes as in our main parameter study\mklrc{are these equivalent. MH: What do you mean????}. As shown in Figure~\ref{fig:lifetimes}, the lifetimes just increase monotonically with planet mass. Even though the trend is the same, the vortex evolution is quite different. We do not see any later-generation vortices form, suggesting that this resolution is not sufficient for resolving how the shocks from the planet's spiral waves shape the vortensity profile during the re-triggering process. In spite of that limitation, the initial gas vortices survive much longer because this resolution also is not sufficient for the shocks to kill the vortex as quickly either. Since neither of the shock effects that lengthen or shorten the vortex lifetimes are properly resolved, the same trend emerges.

All of the vortices that form in these low-resolution simulations are elongated in terms of their Rossby number. Their shapes in the gas are also generally elongated, although that is not as clear as it is with higher resolution. The corresponding dust asymmetries have extents of about $120^{\circ}$ with no noticeable peak offset or dust circulation once the $m = 1$ vortex has formed. One major difference is that without the later-generation vortices, only the dust asymmetry in the highest-mass case survives 10000 orbits and is still alive at the end of the simulation. None of the other dust asymmetries reach 6000 orbits.

Beyond this resolution, we also tried some cases with a medium resolution of 512 $\times$ 512. We found that this doubled resolution still is not sufficient for allowing later-generation vortices to form in the $h = 0.06$ cases either. That same limitation along with such simulations being more computationally expensive in 3-D are why we did not use such higher resolution for the comparison between 2-D and 3-D. We also considered it valuable to be able to reproduce the trend with such a low resolution. This agreement shows the trend does not rely on re-triggered vortices and is more fundamentally related to the gap structure and the vortex's proximity to the planet. It also shows the usefulness of running low-resolution simulations that are not computationally expensive in order to look for trends.

Additionally, we tested if we could reproduce the ``U-shaped" lifetime trend from H21 in which low-mass planets also induced longer-lived vortices with a fixed-value very massive disc. As presented in Appendix~\ref{sec:planet-mass-3D}, we did indeed find the same trend at low resolution in both 2-D and 3-D.


\subsubsection{Low-resolution 3-D} \label{sec:disc-mass-3D}

\mklrc{i guess similarity can be expected because RWI vortices are fundamentally 2D}

Like our main 2-D parameter study and the low-resolution 2-D study, we find that the dust asymmetry lifetimes for the most part monotonically increase with the disc mass and planet mass in 3-D, as shown in Figure~\ref{fig:lifetimes}. In particular, the lifetimes are rather short at under 2000 orbits for planets with masses less than $0.25~M_\mathrm{Jup}$. The gas and dust asymmetries in the highest-mass case above $1.0~M_\mathrm{Jup}$ are still not close to decaying by the end \mklrc{phrase confusion. MH: Fixed!!}of our simulations at $5000~T_\mathrm{p}$. For intermediate-mass planets between $0.25~M_\mathrm{Jup}$ and $0.80~M_\mathrm{Jup}$, there is a slight difference in the trend in that all of these cases have close to the same lifetime at about 3000 orbits. Near $0.50~M_\mathrm{Jup}$, there is even a slight drop in the lifetimes. Overall though, we consider the results with 3-D low resolution to be very similar to those with 2-D low resolution, including the lack of re-triggered vortices and the general vortex evolution in the gas and dust.

\subsection{High-resolution 3-D}
\begin{figure*} 
\centering
\includegraphics[width=0.98\textwidth]{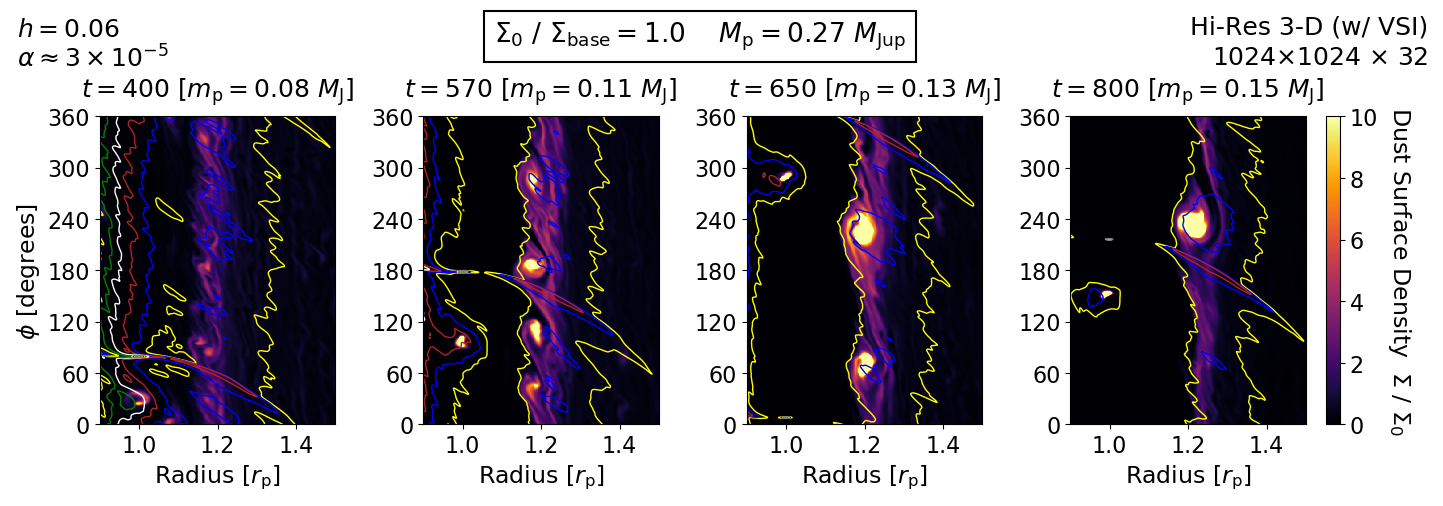} \\
\vspace*{0.5em}
\hspace*{1em}
\includegraphics[width=0.98\textwidth]{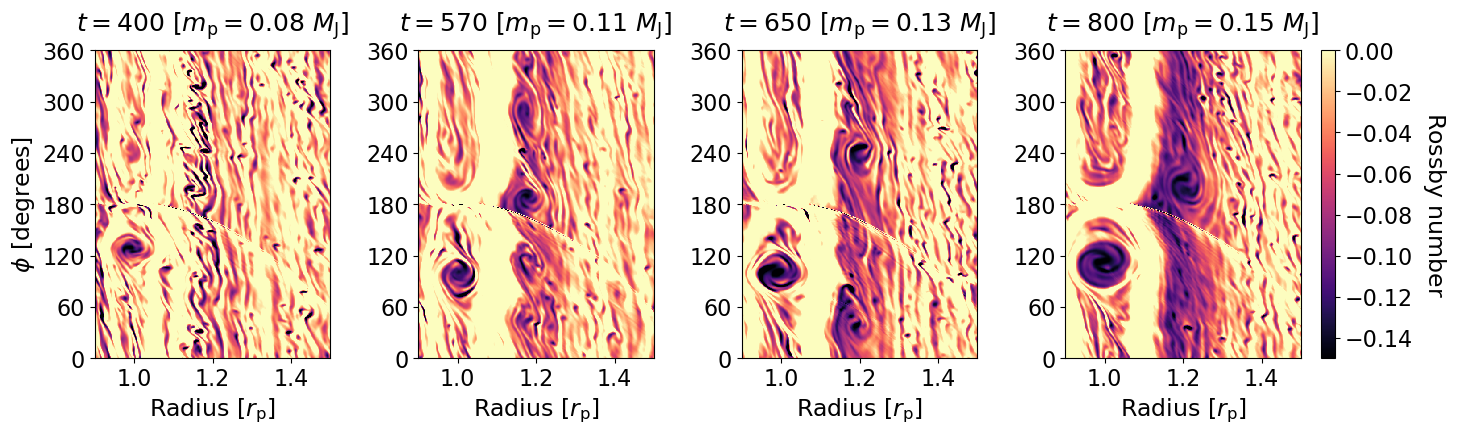}
\caption{Gas density (\textit{top panels}) and Rossby number evolution (\textit{bottom panels}) from a high-resolution 3-D simulation with an embedded low-mass planet ($M_\mathrm{p} = 0.27~M_\mathrm{Jup}$) grown from an intermediate-mass disc with $\Sigma_0 / \Sigma_\mathrm{base} = 1.0$ and the VSI in operation. Density contours \cbf{(at $\Sigma / \Sigma_0 = 0.6, 0.7, 0.8,$ etc.)} are overlaid on the density panels. \textit{Column~1}: With just the VSI and no RWI, many tiny compact dust-trapping vortices arise across a wide range of the disc, in particular near the pressure bump at the outer gap edge. \textit{Column 2}: With the VSI vortices already present, the RWI also arises, resulting an $m = 4$ mode series of larger co-orbital vortices that also have compact Rossby numbers. \textit{Column 3}: Less than 100 orbits later, mergers have turned the $m = 4$ mode into an $m = 2$ mode. \textit{Column~4}: The last two large vortices have merged, forming a single compact vortex larger than the VSI vortices, but not too much larger than the initial RWI series of vortices. Some dust is still present outside of the vortex, albeit it will be consumed within a few hundred orbits, leaving behind only the vortex as the predominant signature at the pressure bump.} 
\label{fig:evolution_vsi}
\end{figure*}

Beyond the low resolution 3-D parameter study, we also present one 3-D simulation with a very high resolution in the midplane of $1024 \times 1024$, which is the same radial resolution we used in our fiducial 2-D parameter study and half the azimuthal resolution. \cbf{Despite the low vertical resolution of 6 cells per scale height, we resolve the VSI enough to slightly exceed the levels of vertical kinetic energy and accretion stress found by \cite{ziampras23} with 16 cells.} With the higher \cbf{radial} resolution, we are able to use a smaller accretion radius of $K = 0.25~R_\mathrm{H}$ while setting $A = 1$. This case has the base surface density ($\Sigma_0 / \Sigma_\mathrm{base} = 1$) and the planet grows to $0.27~M_\mathrm{Jup}$ by the end of the simulation at $t = 2900~T_\mathrm{p}$. Although we had found that planets with this low of a mass do not typically induce long-lived dust asymmetries in low-mass discs, we find that they can in this case because the VSI is present.\mklrc{do we need a no-planet sim to show "full" vsi? MH: Yes. I'll mention it later in VSI behavior section!!!}

\subsubsection{VSI background} \label{sec:VSI-background}

The interaction between planets and the VSI and any associated vortices has previously been explored by \cite{stoll17b}. They found that the resulting RWI vortex appears stronger and has a much longer lifetime in the presence of the VSI than without, albeit they are comparing the VSI case to a viscous case with a ten times higher viscosity than in our study. In their simulations, the VSI is suppressed in the outer disc from $1.2$ to $2.0~r_\mathrm{p}$, which they attribute to the vortex.

We posit that the region of suppression may be connected to the nature of the VSI itself. A much more recent work by \cite{svanberg22} demonstrated that even when the VSI operates throughout the vast majority of the simulation domain, they can identify discrete modes of the VSI. More specifically, they show analytically that each mode of the VSI is expected to operate across a finite range of radii beginning from the radius at which the mode's fastest growth should occur (see their Equation 18) and extending to the radius at which the mode frequency $\omega$ matches the near-Keplerian orbital frequency $\Omega$\mklrc{might be helpful to cite some of svanberg's vsi equations to better explain this, and prep for the mode analysis below}. They isolate the modes of the VSI in their simulations by taking a Fourier transform of the azimuthally-averaged vertical velocity over time across the entire range of radii in the simulation\mklrc{i think the FT is only done for the time axis}. With a much larger simulation than our study, they found four discrete VSI modes operate and there are no more than two at any given location in the disc.

We build on these past works by determining which type of vortex forms when taking into account the planet's growth track, incorporating dust, testing whether the vortex is indeed responsible for the suppression of the VSI, and investigating how that suppression relates to the discrete modes of the VSI.

\subsubsection{Vortex evolutionary track with the VSI} \label{sec:VSI-vortex}

With the VSI active, we find that the vortex that arises from the RWI is compact throughout, a striking difference between this case and all our other simulations without the VSI. The RWI vortex becomes compact because the VSI produces bunches of really small micro-vortices across wide ranges of the disc before the RWI vortex forms, as shown in the left column of Figure~\ref{fig:evolution_vsi}. With such a low-mass planet, the RWI develops rather slowly at the gap edge once the Rossby waves become unstable. As a result, the very small vortices produced by the VSI are able to seed the much larger vortices that develop through the RWI, as shown in the middle columns of Figure~\ref{fig:evolution_vsi}. Since the VSI vortices start out as compact with Rossby numbers Ro $< -0.15$, the resulting RWI vortices also end up compact. In turn, after those vortices merge, the final vortex is also compact, as shown in the final column of Figure~\ref{fig:evolution_vsi}. When there is no VSI and the growth of the planet has been incorporated, that pathway to a compact vortex does not typically occur because the initial small vortices that seed the final vortex are almost always elongated.

Because the disc already has vortices from the VSI, it is difficult to tell when exactly the RWI starts to develop. Although the planet is not very massive, the first clear mode that emerges is $m = 4$ at about $t = 550~T_\mathrm{p}$, a stark contrast from most low-mass planets that typically only trigger $m = 1$ or $m = 2$ modes of the RWI at the onset. The $m = 4$ mode reduces to $m = 3$ by $t = 600~T_\mathrm{p}$ and then $m = 2$ by $t = 650~T_\mathrm{p}$. Finally at $t = 750~T_\mathrm{p}$, it has settled into a single compact vortex with a width of only about $30^{\circ}$. The 250 orbits it takes the initial set of vortices to merge into a single vortex is much longer than the less than 100 or so orbits it normally takes. By about $t = 1000~T_\mathrm{p}$, all of the rest of the dust at the pressure bump settles in the vortex. From there, the vortex remains relatively unchanged until the end of the simulation around $t = 2900~T_\mathrm{p}$ and shows little to no signs of decay.

\subsubsection{Where the VSI operates} \label{sec:VSI-behavior}

\begin{figure} 
\centering
\includegraphics[width=0.47\textwidth]{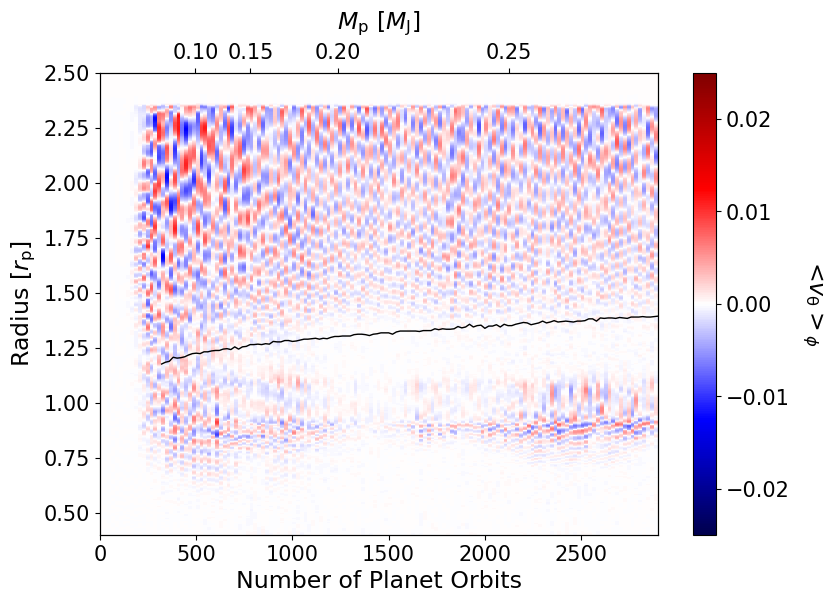}
\caption{Azimuthally-averaged vertical velocity ($v_\mathrm{\theta}$) over time. The VSI takes about 200 orbits to develop. It operates everywhere (from a bit exterior the inner boundary to the outer Stockholm boundary) until the planet becomes massive enough to suppress it. The region of suppression for the VSI is roughly the outer edge of the horseshoe region to the location of the pressure bump (black line).} 
\label{fig:vertical-velocity-over-time}
\end{figure}

\begin{figure*} 
\centering
\includegraphics[width=0.49\textwidth]{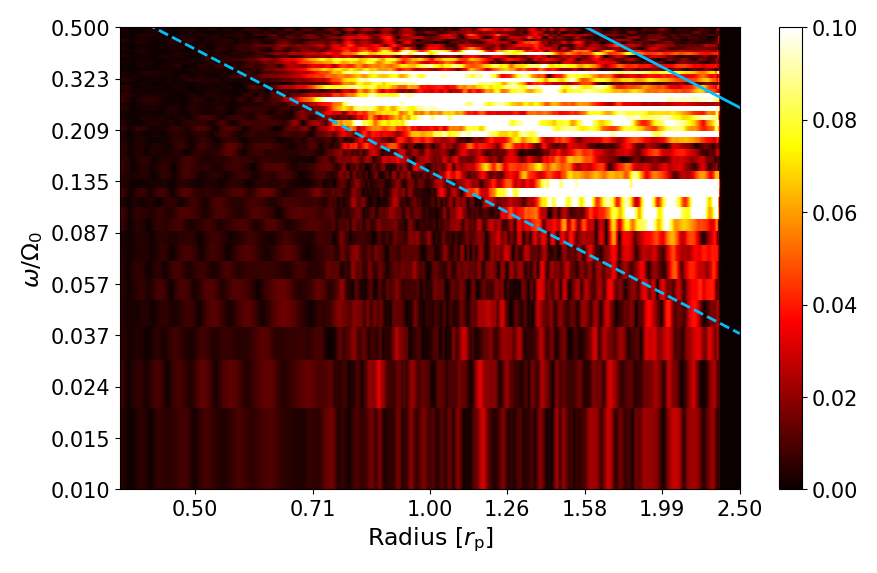}
\hspace*{\fill}
\includegraphics[width=0.49\textwidth]{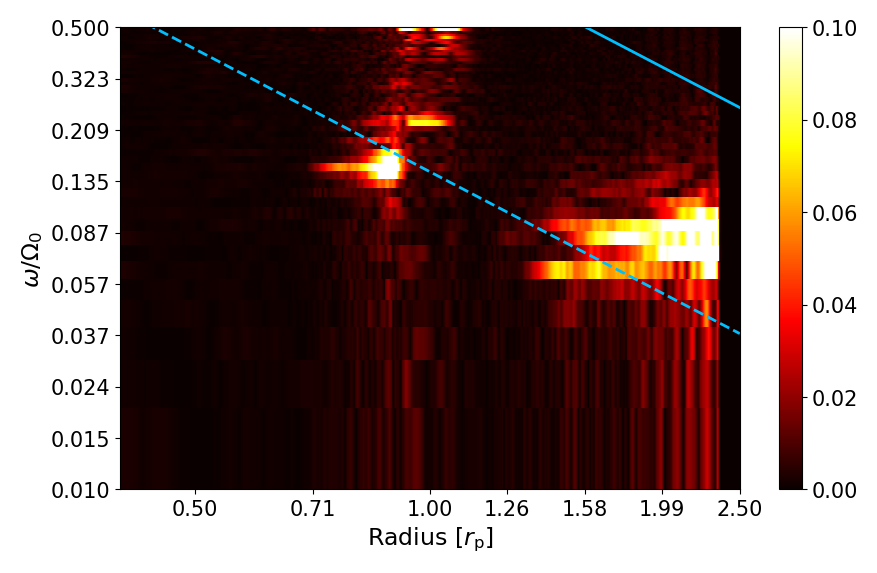}
\caption{Fourier amplitudes of the azimuthally-averaged vertical velocity with no planet (\textit{left panel}) and with a planet (\textit{right panel}). Without a planet, there are two distinct modes. The inner mode occupies a thicker range of frequencies. Both modes approximately propagate from $\omega = (\sqrt{109} - 10) \Omega / 3$,  the location with the fastest growth  (\textit{dashed line}), to the location resonant to the Keplerian orbital frequency (\textit{solid line}). The former line follows from Eq. 17 in \citealp{svanberg22} with the values adapted to our parameters. With a planet, the outer mode is still present, but further out and with a lower frequency. It is less clear if the inner mode is still present, but there is some signature present. Both modes are cut off in the region where the VSI is observed to be suppressed in the outer disc near $r \sim 1.2~r_\mathrm{p}$.} 
\label{fig:vsi-modes}
\end{figure*}

Besides the VSI affecting the RWI, the planet also affects the VSI. The effect of the VSI on the disc can be seen in the vertical velocity, which only attains significant non-zero values after the VSI is present. Figure~\ref{fig:vertical-velocity-over-time} shows the evolution of the azimuthally-averaged vertical velocity over time across the entire disc. It takes about 200 orbits for the VSI to develop, at which point it operates everywhere except both boundaries where the damping condition is applied and some of the inner region of the disc adjacent to the inner boundary. 

Not until a bit after the $m = 1$ vortex arises when the planet has already grown past $0.15~M_\mathrm{Jup}$ does the VSI start to fade in part of the region exterior to the planet. Specifically, we find that the VSI is suppressed from the exterior edge of the planet's horseshoe region at around $r = 1.15~r_\mathrm{p}$ to the peak of the pressure bump, which migrates outwards from around $r = 1.25$ to $1.40~r_\mathrm{p}$ over the course of the simulation. That inner edge is about the same as what was found by \cite{stoll17b}, but the outer edge near the pressure bump is noticeably closer-in than their work. One major difference is that our compact vortex is much smaller than their vortex, which appears elongated and much larger (see the ``$100M_{\oplus}$ VSI" panel of their Figure 4). That larger vortex also coincides better with their region of suppression (see the same panel of their Figure 2). In contrast, our much smaller vortex only barely extends into the region of suppression in radius and covers a much, much smaller azimuth even though the VSI is affected across the entire azimuth throughout the region of suppression. As a result of these differences, we do not suspect the vortex is responsible for the VSI suppression in our work like in theirs. Because the VSI can still be suppressed even with a much smaller vortex, we suspect the planet itself also plays a role in suppressing the VSI beyond just through the RWI vortex.\cbf{\footnote{A parallel study by \cite{ziampras23} likewise found that the planet was directly suppressing the VSI and attributed that suppression to the shocks from the planet's spiral waves.}}

Exterior to the pressure bump, the VSI operates largely the same as it does before the growth of the planet. Interior to the planet and inside the horseshoe region; however, the VSI shows more significant changes. It does fade somewhat at times, in particular for a few hundred orbits around $t = 1500~T_\mathrm{p}$. Overall, the signature of the VSI is affected interior to the outer pressure bump only when a planet is present, but not always.

\subsubsection{VSI modes} \label{sec:VSI-modes}

To test the planet's effect on the modes of the VSI, we also ran a comparison 3-D run for a few hundred orbits that was identical to the other VSI simulation, except without a planet.\mklrc{is this a no-planet sim or do you mean the evolution before the planet is introduced. MH: A new simulation with no planet.} As illustrated in the left panel of Figure~\ref{fig:vsi-modes}, we found two separate modes, both of which operate roughly across the expected analytic range derived by \cite{svanberg22}, albeit across a bit wider range of frequencies compared to the narrower modes in their work. \mklrc{is it necessary to list the equations for the VSI-zones derived by svanberg? MH: I put it in the figure caption.}

With the planet, however, the VSI modes display a noticeable change, as shown in the right panel of Figure~\ref{fig:vsi-modes}. Whereas the outer mode begins at about $r = 1.2~r_\mathrm{p}$ with a corresponding frequency near $\omega = 0.12~\Omega_0$ without the planet, the same mode moves further out to beyond $r = 1.3~r_\mathrm{p}$ and consequently drops in frequency to around $\omega = 0.08~\Omega_0$. As a result of the frequency drop, the outer mode still begins approximately at the expected radius for the fastest growth. On the other hand, the inner mode is much less recognizable. The strongest amplitude for the inner mode is located at about $r = 0.7$ to $0.9~r_\mathrm{p}$, which is not too much further out than where the mode normally begins without a planet. The main difference, however, is that its frequency dropped all the way down to $\omega = 0.15~\Omega_0$. It is much narrower than the wide frequency range of $\omega = 0.20$ to $0.40~\Omega_0$  it had without a planet and and twice as slow as the average across that range\mklrc{i wouldn't say 0.15 is far from 0.2}. Moreover, this peak amplitude only appears just a bit too far interior to the expected radial range. The only strong part of the inner mode that does appear in the expected range is a less prominent at about $\omega = 0.20~\Omega_0$ located more co-orbital with the planet. 

Beyond the two strongest signatures with no planet, there is a non-zero amplitude across the entire frequency range at these radii. With a planet, on the other hand, there is almost no signature of any mode interior to the start of the inner mode and the signature in the small radial range in-between the inner and outer modes is much fainter. Both of those regions had signatures of the VSI when there was no planet in the disc. Overall, adding the planet to the simulations weakens and constricts the spatial range of the inner mode while having little effect on the outer mode except for reducing its frequency so it can begin further out in the disc.


\section{Planet-induced vortices with dust feedback} \label{sec:feedback}

With the high concentrations\mklrc{how much?} of dust easily greater than order unity that can naturally be trapped in vortices, we were interested in whether this amount of dust would affect the lifetimes or appearances of the encompassing vortices through feedback, an effect we neglected in our other simulations. With dust feedback, we find that the dust configurations and dust asymmetry lifetimes can be qualitatively different depending on whether the simulation is 2-D or 3-D, and also the resolution in the vertical direction. 

We ran two sets of simulations: a fixed-mass case with $M_\mathrm{p} = 0.30~M_\mathrm{Jup}$, and a growing case that reaches about $M_\mathrm{p} = 0.50~M_\mathrm{Jup}$ (from $\Sigma_0 / \Sigma_\mathrm{base} = 1.5$ and $A = 0.6$). The fixed-mass case reaches its final planet mass after $T_\mathrm{growth} = 200~T_\mathrm{p}$ and follows a growth track of $m_\mathrm{p}(t) = M_\mathrm{p} \sin^2{\left(\pi t / 2T_\mathrm{growth}\right)}$. We needed the fixed-mass case to compare the lifetimes in 2-D and 3-D because the growing case already has different dust asymmetry lifetimes in 2-D and 3-D even without feedback due to the planet's slightly different growth tracks. We also increase the dust size from $\mathrm{St} = 0.023$ to $0.07$ to induce more dust to drift into the vortex and thereby strengthen the effects of feedback. The 2-D and 3-D simulations have a low midplane resolution of $256 \times 256$ like our 3-D simulations without feedback or the VSI, and the 3-D simulations have either 32 or 64 grid cells in \cbf{the vertical direction}. \mklrc{strength of feedback is reflected by epsilon, st represents coupling. in fact, for large st there is no feedback since the two fluids are decoupled.}

\subsection{Base cases in 2-D} \label{sec:feedback-2D}

\begin{figure} 
\centering
\includegraphics[width=0.48\textwidth]{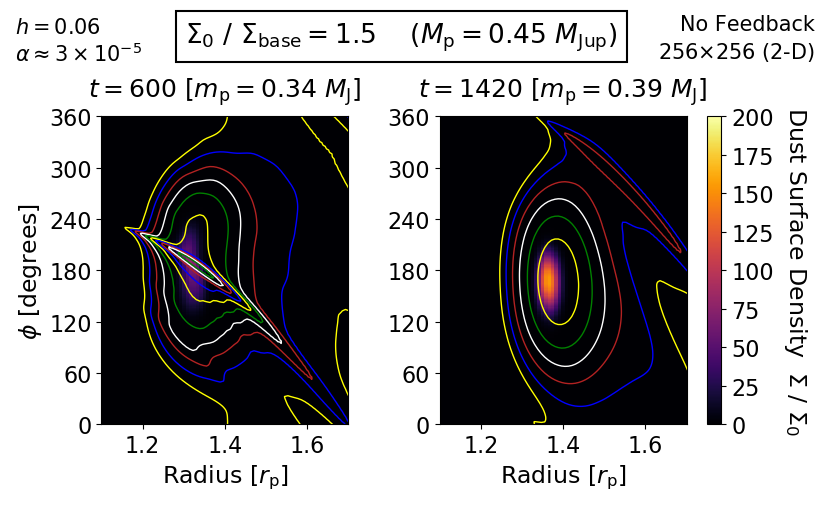} \\
\vspace*{0.5em}
\includegraphics[width=0.48\textwidth]{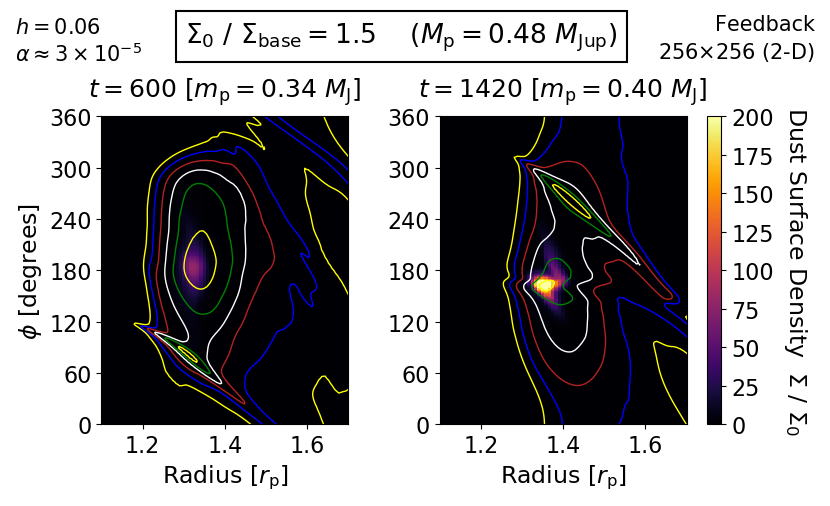}
\caption{Low-resolution 2-D comparison of the dust density distribution in a vortex without (\textit{top panel}) and with dust feedback (\textit{bottom panel}). Without dust feedback, the dust is more spread out and circulates around the vortex. With dust feedback, the dust forms a very compact clump that stays in a fixed Keplerian orbit and is unaffected by the gas motion.} 
\label{fig:2D-feedback-comparison}
\end{figure}

In 2-D, feedback causes nearly all of the dust in the vortex to clump together across a very narrow azimuthal extent $<30^{\circ}$, consistent with expectations from previous work \citep{fu14b}. Figure~\ref{fig:2D-feedback-comparison} compares the dust configurations with and without feedback. The clumping has the opposite effects of shortening the gas lifetime of the vortex while preventing the dust asymmetry from decaying. The clump also generates a pair of waves on both sides that are easiest to see in the vorticity (\citealp[see Figure 2 from][]{fu14b}) and seem like they could be a numerical artifact. 

The initial clump can break up into multiple clumps, but the feedback prevents these individual clumps from ever spreading out towards a ring. The clumps never decay because they are massive enough to be unaffected by the gas motion, leaving them on fixed Keplerian orbits. The gas vortex only survives until about 1000 orbits with feedback compared to 2100 orbits without feedback, also as expected from \cite{fu14b}. Despite that shorter lifetime in the gas, the lifetime of the dust asymmetry is still indefinite and thus longer with feedback compared to a lifetime of just 2500 orbits without feedback.\mklrc{is the shortened gas vortex lifetimes consistent with previous works? MH: Yes, added!}

\subsection{3-D behavior with feedback} \label{sec:feedback-2D}

\begin{figure} 
\centering
\includegraphics[width=0.48\textwidth]{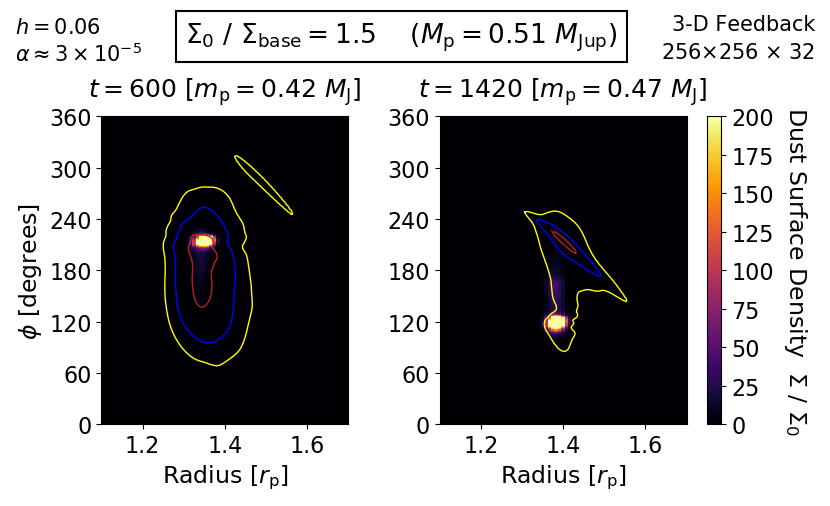} \\
\vspace*{0.5em}
\includegraphics[width=0.48\textwidth]{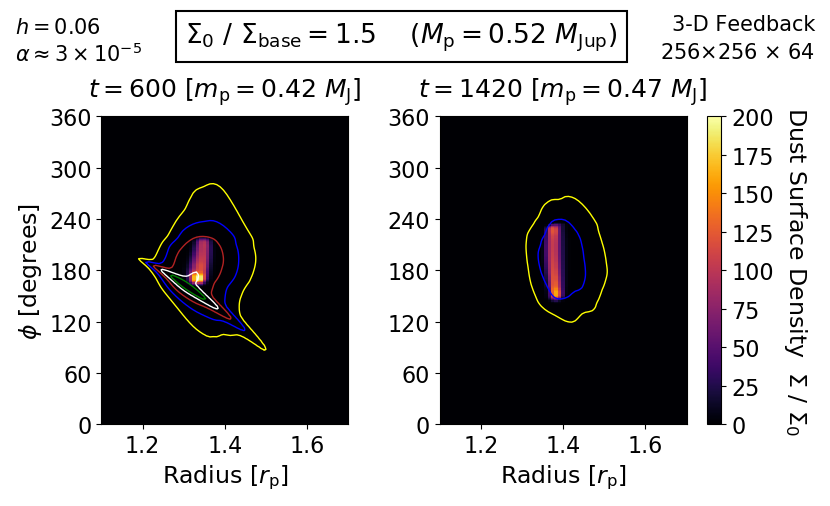}
\caption{3-D comparison of the dust density distribution at lower vertical resolution ($N_\mathrm{\theta} = 32$; \textit{top panel}) and higher vertical resolution ($N_\mathrm{\theta} = 64$; \textit{bottom panel}). At low resolution, the dust all ends up in a single clump like in 2-D. At high resolution, some of the dust ends up in a clump, but the overall structure is still elongated like when there is no feedback.} 
\label{fig:3D-feedback-comparison}
\end{figure}

\begin{figure} 
\centering
\includegraphics[width=0.48\textwidth]{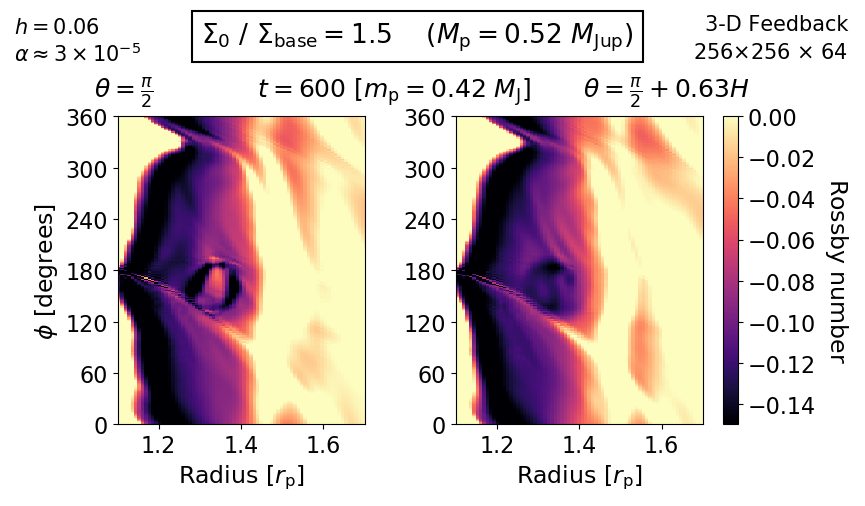}
\caption{Comparison of the Rossby number in the midplane ($\theta = \pi / 2$) and out of the midplane ($0.63H$ above). The vortex signature is visible at both latitudes, albeit with a more positive signature from the dust in the midplane and a more negative signature from the dust at altitude.} 
\label{fig:3D-feedback-vorticity}
\end{figure}

In 3-D, dust feedback can have a similar effect compared to 2-D if the vertical resolution is low. Like in 2-D, the vortex still forms a singular clump that itself does not decay and may eventually split into multiple clumps that also do no decay. Meanwhile, feedback likewise shortens the gas vortex lifetime, albeit not by as much as in 2-D\mklrc{is this because feedback is only important in the midplane? MH: Probably yes. Added!}. Although the gas vortex lifetimes are nearly identical in 2-D and 3-D without feedback \cbf{regardless of the vertical resolution}, the gas vortex only survives until about 1500 orbits with feedback. That is about twice as long as it survives with feedback in 2-D, but still a $30\%$ reduction from the gas vortex lifetime without feedback. We suspect feedback may have less of an effect in 3-D because feedback should only be important in the midplane where the dust has settled.\mklrc{tbc: with feedback, dust clump stays intact in 2D but splits in 3D?. MH: No, that happens in both. Clarified.}

With double the vertical resolution, however, we find that the dust no longer completely gathers into a single clump and feedback no longer reduces the lifetime of the gas vortex at all. Although there is still typically one dominant clump, the rest of the dust in the vortex is more spread out in such a way that that the vortex still has a significant azimuthal extent much larger than the clump itself. The clump is rarely if ever centered, and instead is typically located at the very back of the dust configuration. Figure~\ref{fig:3D-feedback-comparison} compares how the dust configurations typically appear in 3-D simulations with low and high vertical resolution. Both ends of the vortex can have peaks, and occasionally there are two significant peaks. The front peak can also be more dominant than the back peak. 

The effect of dust feedback also imprints itself on the vorticity, as shown in Figure~\ref{fig:3D-feedback-vorticity}, even if the clumps do not produce their own waves. There is a small region of more positive vorticity near where the dust clumps in the midplane, and a more negative region of vorticity at the same location away from the midplane. Both of these results differ from that of \cite{lyra18}, who instead found that the vortex was unrecognizable in the midplane vorticity and unaffected at altitude, neither of which we see in our simulations. We suspect this difference is due to our vortices having both gas density and vorticity signatures because they were produced by the RWI while their vortices lack a gas density signature because they were produced by a different instability, the convective overstability \citep[COV:][]{klahr14, lyra14, latter16}. \mklrc{citation for COV}




\section{Synthetic images} \label{sec:synthetic}

\begin{figure*} 
\centering
\includegraphics[width=0.812\textwidth]{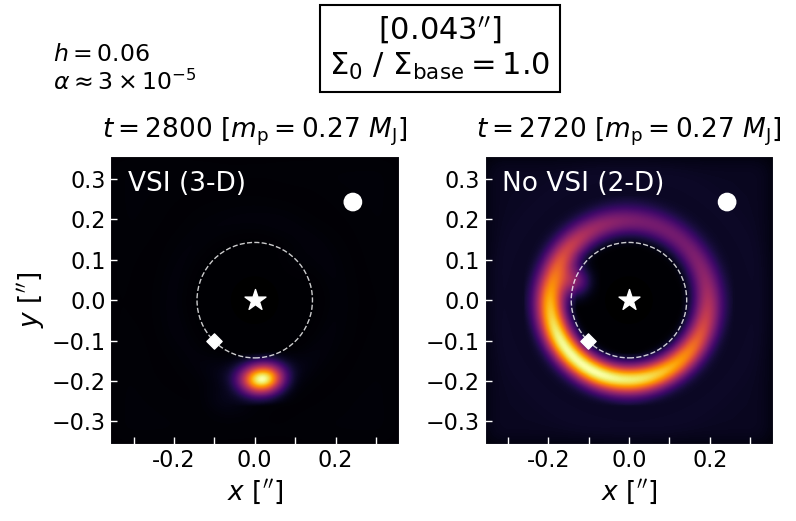}
\includegraphics[width=0.130\textwidth]{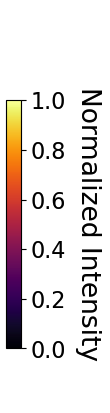}
\caption{Synthetic image comparison between the fiducial high-resolution 3-D simulation with VSI (\textit{left panel}) and the corresponding case with the same disc mass (\textit{right panel}) at about the same time and planet mass ($t \approx 2750~T_\mathrm{p}$, $m_\mathrm{p} = 0.27~M_\mathrm{Jup}$). The vortex in the case is compact and in a steady state, whereas the vortex in the case with no VSI has already decayed in the gas and the dust is in the midst of spreading into a ring.} 
\label{fig:images_vsi}
\end{figure*}

We calculate synthetic images to more directly show how the new behavior of the dust should influence the types of asymmetries observed in actual protoplanetary discs. The most dramatic difference from our previous results is how compact a compact vortex appears with the VSI compared to an elongated vortex without the VSI in a case with a planet of similarly-low mass. We also find that before the last gas vortex decays, it is much more likely to have an asymmetry with multiple peaks in the azimuthal direction, a feature that was quite rare in H21. Relatedly, the offset between the peak and the azimuthal center of the asymmetry can also be much larger in that early chaotic stage. Later on after the last gas vortex, the azimuthal peak of the dust asymmetry is barely off-center at all.

\mklrc{maybe some intro? do we only work with 2D simulations? what happens if we input 3D data? should we give surface densities in physical units from the beginning because we will be producing synthetic images (which use real units)? MH: No on the physical units, because all of the images are normalized anyway. I added the point about 3-D to the methods. Maybe I should still add a results-based intro. MH: Done!}

\subsection{Method} \label{sec:synthetic-method}

The synthetic images are derived primarily using the surface density distribution as input and a beam convolution for the final output. We largely follow the same method for generating the synthetic images as in H21. The flux in the image is calculated from the intensity as 
\begin{equation} \label{eqn:flux}
F_\mathrm{\lambda}(r, \phi) = I_\mathrm{\lambda}(r, \phi) \frac{r \delta r \delta \phi}{d^2},
\end{equation}
where $\lambda$ is the wavelength and $\delta r$ and $\delta \phi$ are the radial and azimuthal dimensions of the cell such that $d\mathcal{A} = r \delta r \delta \phi$\mklrc{this is just dA defined earlier}. The intensity $I_\mathrm{\lambda}$ itself is calculated from the optical depth as
\begin{equation} \label{eqn:intensity}
I_\mathrm{\lambda}(r, \phi) = B_{\lambda}(T)[1 - e^{-\tau_{\lambda}}],
\end{equation}
where $B_{\lambda}(T)$ is Planck's law for the spectral radiance. All of the images are presented normalized to the maximum intensity. The optical depth itself is calculated from the density as
\begin{equation} \label{eqn:tau}
\tau_{\lambda} = \Sigma_\mathrm{dust}(r, \phi; s) \kappa_{\lambda},
\end{equation}
where $\kappa_{\lambda}$ is the dust opacity at the observed wavelength based on the grain size $s$.\mklrc{should we quote typical opacities? MH: maybe in the revision, I'm not sure how useful it is.} We still use $\Sigma_\mathrm{dust}$ when calculating images from the 3-D simulations because the dust mostly all settles in the midplane. The opacity values are retrieved from the Jena database\footnote{http://www.astro.uni-jena.de/Laboratory/Database/databases.html}, assuming the grain size is similar enough to the wavelength for Mie theory to apply \citep[$2 \pi s \approx \lambda$, see][]{bohren83} and a dust composition of magnesium-iron silicates \citep{jaeger94, dorschner95}.

Like in H21, we calculate all of the synthetic images at $\lambda = 0.87$ mm, which corresponds to Band 7 of ALMA, albeit with this method the images would appear largely the same regardless of the band choice. We set a physical scale by defining the planet's separation of $r_\mathrm{p} = 1.0$ to be 20 AU and assuming a distance to the system of $d = 140$ pc. With the physical scale defined, the base surface density of $\Sigma_\mathrm{base} = 1.157 \times 10^{-4}$ corresponds to $2.57$~g/cm$^2$ and \cbf{the Stokes number} $\mathrm{St} = 0.023$ corresponds to a grain size of $s = (2 / \pi)(\Sigma / \rho_\mathrm{d}) \times \mathrm{St} = 1.64$~cm at the location of the planet with that surface density. The temperature profile of $T = T_0 (r / r_\mathrm{p})^{-1}$ is consistent with the lack of flaring in our simulations. The star's temperature is assumed to be the Sun's temperature of $T_\mathrm{\odot} = 5770$ K, albeit normalizing the final images takes away any effect of this parameter. We use the simplest viewing inclination angle of $i = 0^{\circ}$ so that it is easiest to identify any features in the image. \mklrc{what is the surface density in physical units?}

We focus on a beam diameter of $0.043^{\prime \prime}$, which corresponds to a scale of 6 AU, to convolve the flux image because we see that much larger beam sizes may not resolve features within the vortex. We filter out the dust from the inner disc ($r < 0.90~r_\mathrm{p}$), but unlike in H21, we leave in the dust co-orbital to the planet, which primarily resides at the Lagrange points. Keeping that dust in the image allows it to overlap with dust from the vortex when a larger beam diameter is used.

\subsection{Background} \label{sec:synthetic-background}

In our previous work, we had found that vortices in observations should appear rather simple, characterized by just two measurable properties: their azimuthal extents and their peak offset, the latter of which is the angle between the azimuthal location of an asymmetry's peak intensity and its perceived azimuthal center.\cbf{\footnote{Like in \cite{hammer19}, the center itself is determined using an appropriate threshold intensity $I_\mathrm{cut} / I_\mathrm{max}$ to identify the edges and only weakly depends on the choice of threshold.}} Azimuthal extents were usually elongated and could vary greatly, particularly if the dust supply was cut off, but typically did not exceed $180^{\circ}$. Another much rarer feature we only found with the dust cut off was that the vortex peak could split into two as it circulated around the vortex.

\subsection{High-resolution 3-D with VSI} \label{sec:synthetic-vsi}

\mklrc{this result on vsi vs non-vsi is significant, maybe it's worth highlighting?} The striking impact of the VSI on the vortex morphology may be easiest to see in the synthetic images. Figure~\ref{fig:images_vsi} compares our high-resolution 3-D simulation with the VSI to the high-resolution 2-D simulation with the same disc mass. Around the selected timestep, the gas vortex with the VSI is still very compact in both the gas and dust. Meanwhile, the vortex without the VSI has already decayed in the gas and the dust is already spread out. It is several hundred orbits away from completely spreading into a ring with no remaining asymmetry. This snapshot is characteristic of how the vortex with the VSI appears for a very long timescale on the order of thousands of orbits, whereas the vortex without the VSI changes its appearance on the timescale of a few hundred orbits, if not less or much less early on.

\begin{figure} 
\centering
\includegraphics[width=0.48\textwidth]{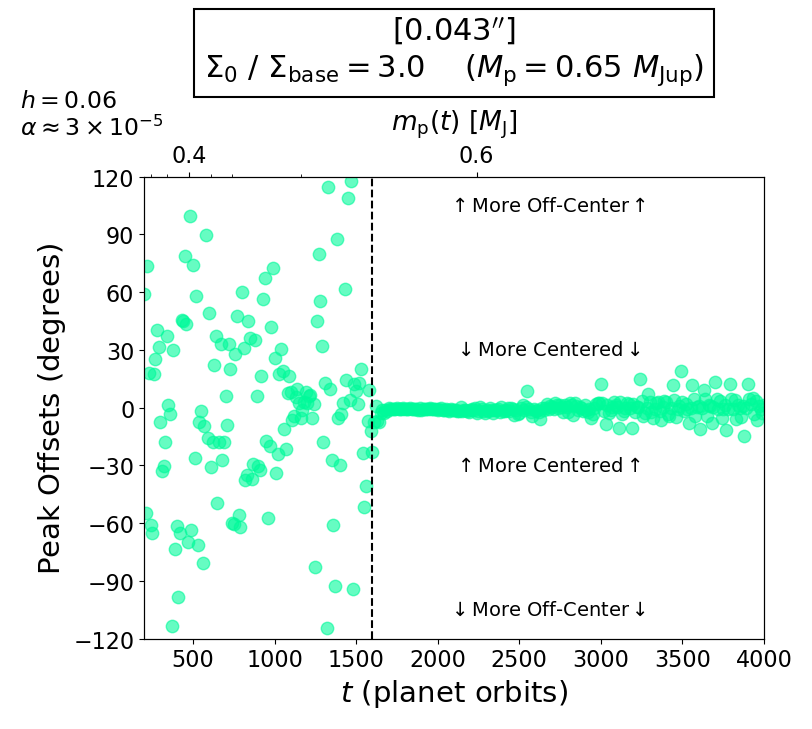} \\
\caption{Peak shifts for the featured simulation. In the early stage, the peak shifts are not preferentially centered or off-center are largely random in the early stage, and typically range between $-60^{\circ}$ to $60^{\circ}$ off-center most of the time. In the later stage when the vortex becomes compact, the peak is much more centered, albeit the shift slowly grows over time as the dust spreads out and becomes more elongated.} 
\label{fig:offsets_h06_s3472}
\end{figure}

\begin{figure} 
\centering
\includegraphics[width=0.45\textwidth]{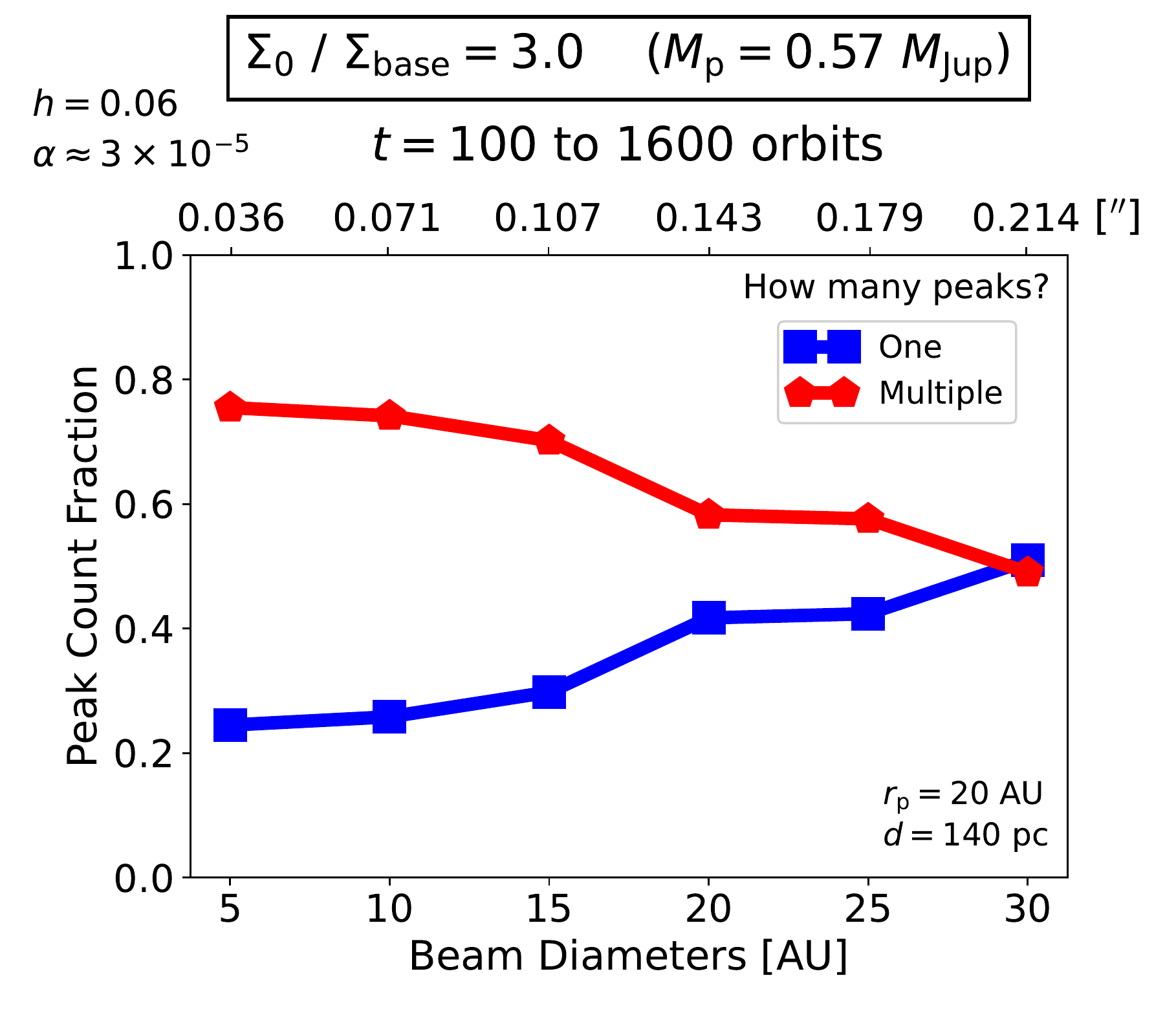}
\caption{Whether there is one peak or multiple peaks depending on the beam diameter for the featured simulation. Multiple peaks are very common with high resolution, but the fraction steadily decreases as the beam diameter increases.} 
\label{fig:peaks_h06_s3472}
\end{figure}

\subsection{High-resolution 2-D with a static planet} \label{sec:synthetic-results}

\begin{figure*} 
\centering
\includegraphics[width=0.88\textwidth]{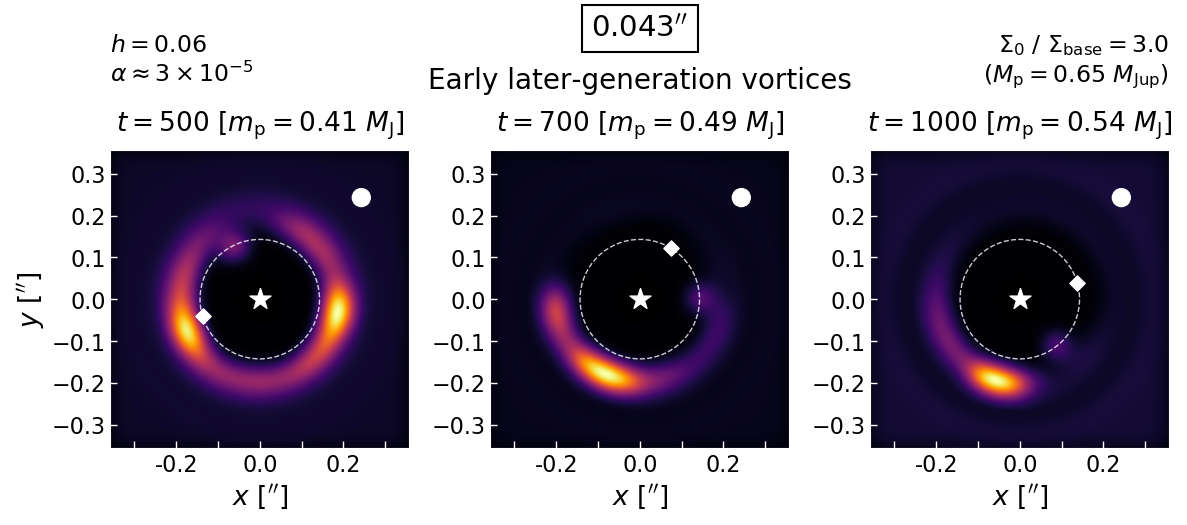}
\includegraphics[width=0.096\textwidth]{figures/intensity_colorbar.png} \\
\vspace*{-1.5em}
\includegraphics[width=0.88\textwidth]{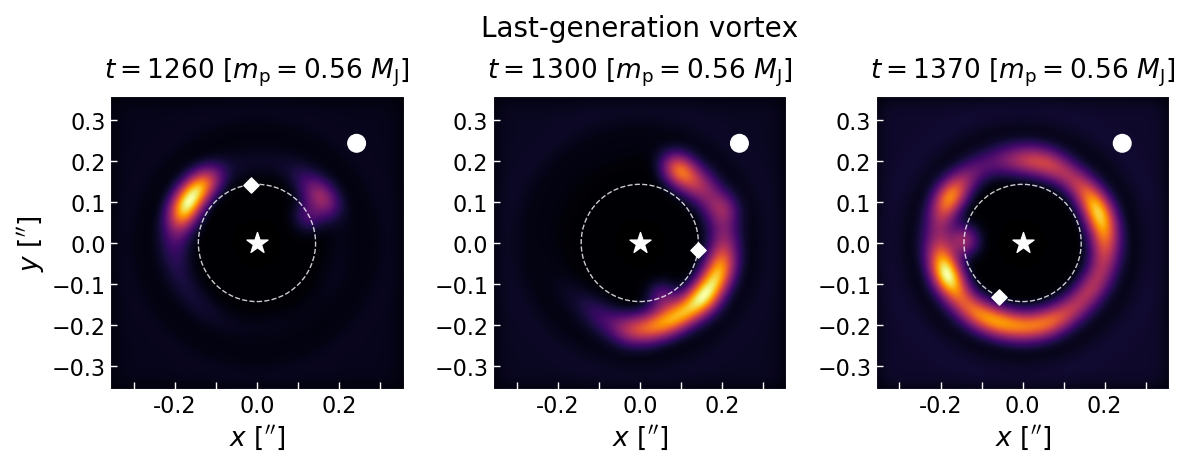}
\includegraphics[width=0.096\textwidth]{figures/intensity_colorbar.png} \\
\vspace*{-1.5em}
\includegraphics[width=0.88\textwidth]{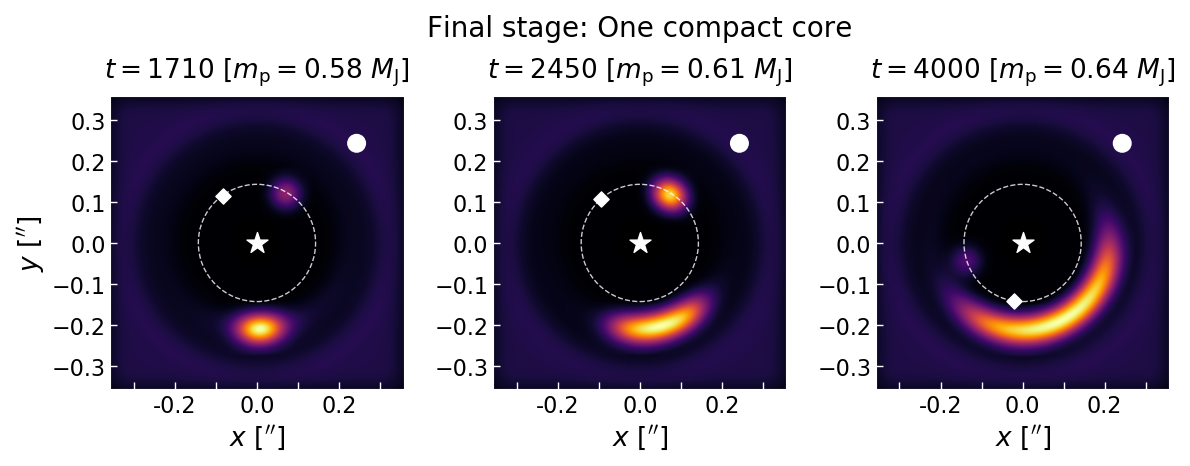}
\includegraphics[width=0.096\textwidth]{figures/intensity_colorbar.png}
\caption{Synthetic images for the featured simulation that harbours a super-thermal mass planet ($M_\mathrm{p} = 0.65~M_\mathrm{Jup}$) grown from a very massive disc with $\Sigma_0 / \Sigma_\mathrm{base} = 3.0$, including various configurations with the later-generation vortices (\textit{top panels}), the formation of the last later-generation vortices with three peaks (\textit{middle panels}), and the final stage when only a single compact core remains and the dust slowly spreads out and becomes more elongated (\textit{bottom panels}). The top two rows provide some examples of how the dust asymmetry can display multiple peaks, except for the top right snapshot at $t = 1000$ that illustrates the dust collecting in a compact core not at the center of the vortex or the observed asymmetry.} 
\label{fig:images_h06_s3472}
\end{figure*}

With better resolution in our hydrodynamic simulations, we find that the appearance of the vortex in observations can be much more intricate, particularly in the high-mass planet cases with $M_\mathrm{p} = 0.45$ to $0.77~M_\mathrm{Jup}$ (i.e. $1.5 \le \Sigma_0 / \Sigma_\mathrm{base} \le 4$) in which the vortex develops compact cores and the dust asymmetry may survive indefinitely. The azimuthal extent of the dust asymmetry can be much more elongated than our previous work, commonly exceeding $180^{\circ}$ and on occasion even approaching a full $360^{\circ}$, which we present in more detail in Appendix~\ref{sec:synthetic-extents}. The azimuthal extents can also vary on much quicker timescales on the order of tens of orbits. With larger azimuthal extents, the peak offsets can in turn also be much larger, as illustrated in Figure~\ref{fig:offsets_h06_s3472}. It is also much more common for the dust asymmetry to exhibit multiple peaks, as shown in Figure~\ref{fig:peaks_h06_s3472}. These more intricate patterns occur in the early chaotic phase of the dust asymmetry lifetime, up to and a little bit past when the last-generation gas vortex decays, as shown in the top two rows of Figure~\ref{fig:images_h06_s3472}. Once the asymmetry in the gas is gone, the dust asymmetry reverts to a more simplistic appearance, as shown in the bottom row of Figure~\ref{fig:images_h06_s3472}, once again only characterized by slowly-varying azimuthal extents and peak offsets like in our previous work.

There are several different reasons for these larger azimuthal extents, larger peak offsets, and the more frequent appearance of multiple peaks. The dust develops its most elongated extents around when the gas vortex decays or re-forms, including (i) a little bit before a vortex decays, (ii) a little bit after a vortex re-forms, and (iii) most times in-between when there is no gas vortex. Any time the asymmetry becomes more elongated also tends to coincide with the appearance of more peaks. We did not see either of these outcomes in our previous work because the gas vortex did not decay and re-form early in the simulation like with the high resolution in this work. Another factor that can be responsible for more elongated extents is the compact cores. When a compact core develops, some of the dust tends to circulate with the compact core while the rest circulates with the broader elongated vortex. Naturally, these multiple circulation paths create more elongated extents, larger peak offsets, and multiple peaks in the synthetic images.

Figure~\ref{fig:offsets_h06_s3472} shows how much the peak offsets can vary over time, in particular during the early chaotic phase. They can reach as high as about $120^{\circ}$. More typically though, they lie between $-30^{\circ}$ and $45^{\circ}$ a little over half the time. Peak shifts towards the front of the vortex ($>0^{\circ}$) are only slightly more common. The peak offset drops to zero when all the dust collects in a compact \cbf{vortex} at the end of the chaotic phase. From then on, it slowly increases over time. By $t = 4000~T_\mathrm{p}$ though, the offset still does not exceed $10^{\circ}$, leaving the peak only somewhat off-center. As such, a very large peak offset in an observed asymmetry could be a signature that the gas vortex is still present.

In the early chaotic phase, we find that it is more common for the dust asymmetry to exhibit multiple peaks than just a single peak for any size beam diameters up to even larger than the planet's separation from its star. Multiple peaks are more common with smaller beam sizes, but it is not until a beam diameter one-and-a-half times the planet's separation that a single peak becomes more common, as illustrated by Figure~\ref{fig:peaks_h06_s3472}. Whereas the multiple peaks with a high-resolution beam are only due to the dust dynamics associated with the vortex, the multiple peaks with a low-resolution beam can also be due to dust from the vortex overlapping with dust at the Lagrange points that smears further out because of the large beam diameter. With this prevalence of multiple peaks at any beam size, we do not think it would be unusual to find an asymmetry with multiple peaks in observations.\mklrc{MH:Does the prevalence of multiple peaks still happen with a migrating planet?}





\section{Discussion} \label{sec:discussion}

\subsection{When are planet-induced vortices compact?} \label{sec:compact}

\mklrc{maybe answer this question with a bullet point list of possibilities. MH: Done.}

\noindent Mechanisms for vortices that appear \textbf{compact}:
\begin{enumerate}
  \item Compact cores from elongated re-triggers \textit{(this work)}
  \item Elongated vortices from high-mass planets \textit{(this work)}
  \item Compact vortices from low-mass planets in the presence of the VSI \textit{(this work)} \\
  \item Compact vortices in thin discs with $h = 0.04$ \citep{hammer21}
  \item Elongated vortices that become compact in high-mass thick discs with $h = 0.08$ \citep{hammer21}
  \item Cooling times of order unity \citep{fung21}
\end{enumerate}

\noindent Mechanisms for vortices that appear \textbf{elongated}:
\begin{enumerate}
  \item Elongated vortices w/ dust feedback \textit{(this work)}
  \item Typical elongated planet-induced vortices \citep{hammer21}
\end{enumerate} 

Despite our previous finding that the RWI preferentially generates elongated planet-induced vortices, there are a few different ways to produce vortices that appear compact in observations as well.\footnote{\cbf{Beyond the mechanisms listed for planet-induced vortices at gap edges, another way that has been proposed to generate a compact vortex is through the accretion luminosity from the planet, which yields such a vortex co-orbital with the planet itself \citep{owen17, cummins22}.}} We had previously found in H21 that vortices in the thinnest discs ($h = 0.04$) can still be compact, albeit these vortices are also the shortest-lived. In this work, we find that elongated vortices with compact cores can sometimes appear compact, in particular when the encompassing elongated vortex has decayed and all the trapped dust has enough time to be trapped in a single remaining core. Besides that mechanism, planets around Jupiter-mass and above in a $h = 0.06$ disc are massive enough to generate a vortex with sharp enough contours to appear compact even though it may still have a weaker Rossby number in the elongated range\mklrc{observers won't relate to Rossby numbers}. Lastly, lower-mass planets around the mass of Saturn can produce a compact vortex in a $h = 0.06$ disc if the vortex is seeded by tiny vortices from the VSI. Although we had expected dust feedback to produce compact vortices, we find that the compact clumps that form through feedback in 3-D simulations with sufficient vertical resolution do not trap all of the dust in the vortex like they do in previous 2-D work \citep{fu14b} and in less-resolved 3-D simulations.

Along the lines of compact cores or the VSI's tiny seed vortices, other mechanisms that yield compact Rossby numbers may also be able to produce compact vortices. In our previous work, we found that planets in \cbf{thicker regions of a disc with high aspect ratios} ($h = 0.08$) are capable of lowering the vorticity in a vortex from elongated to compact even after a single $m = 1$ vortex arises \citep{hammer21}, an outcome we can only see in thinner discs ($h = 0.06$) if the vortex decays and re-forms like in this work. Future work should test if this mechanism is still feasible if the planet is migrating or if the disc is not isothermal. Beyond that mechanism, \cite{fung21} found that cooling timescales of order unity are also capable of lowering the vorticity in a vortex from elongated to compact in 2-D shearing box simulations of non-isothermal discs. Cooling may still not be prone to producing observable compact vortices because they find such vortices disappear relatively quickly with this mechanism. Besides that limitation, future work should test the efficacy of this cooling pathway with more of a focus on global simulations like those of \cite{lobogomes15} or \cite{les15}. It also should be tested in 3-D since an initial exploration of cooling in 3-D by \cite{rometsch21} hinted that its effects on vortices may not be the same as in 2-D. 

Overall, the most robust mechanism for producing vortices that appear compact may still be more-massive planets. Although there might be more potential mechanisms for vortices that appear compact, the simplest setups or conditions still tend to result in vortices that typically appear elongated.


\subsection{Why are there so few dust asymmetries in observations of protoplanetary discs?} \label{sec:dearth}

There are relatively few dust asymmetries in observed in protoplanetary discs in mm/sub-mm, the wavelength at which they should be easiest to detect, given the high numbers of gaps in observed discs and the presumed young ages of these systems (see H21 for a quantitative demonstration with a sample from Taurus). 

The results from our previous work suggested it was already strange to see so few vortices if observed gaps are indeed preferentially created by low-mass gap-opening planets between about half a Saturn to two Saturns in mass and the corresponding dust asymmetries survive at least one or two thousand orbits. 

Our current work exacerbates this issue in two different ways. First, it is no longer just those planets that can create long-lived vortex, but any planet above about half a Saturn-mass, including the ones around Jupiter-mass. Second, the cases where the dust asymmetry stops decaying altogether, particularly those with migrating planets, would suggest an even higher occurrence rate of dust asymmetries than what we had calculated from assuming the dust asymmetry lifetimes were just a few thousand orbits.


\subsubsection{Types of observed morphologies} \label{sec:types}

Among the discs with observed dust asymmetries, only three clear asymmetries appear to be located at the edge of a two-sided gap: one in HD 135344 B \citep{vanDerMarel16b, cazzoletti18}, another in V 1247 Ori \citep{kraus17}, \cbf{and an only recently-resolved one in SR 21 \citep{yang23}}. Almost all of the rest of the observed asymmetries are located just outside of an inner cavity, including those in Oph IRS 48 \citep{vanDerMarel13}, HD 142527 \citep{boehler17, boehler21}, RY Lupus \citep{ansdell16}, and T4 \citep{pascucci16}. Each of these asymmetries is also the sole prominent dust feature in the entire disc. Similarly, MWC 758 \citep{boehler18} and ISO-Oph 2 \citep{gonzalez20} also both have an asymmetry at the outer edge of a cavity as well as a second asymmetry at the outer edge of the dust disc itself, and the two asymmetries in ISO-Oph 2 are likewise the only dust features in the disc. 

While it is certainly possible for a planet-induced vortex to be the only dust feature(s) in a disc, it would be much more natural to expect a planet-induced vortex to be one of many features in a disc. Many observed discs have multiple sets of rings and gaps, yet none of those discs have a clear vortex candidate asymmetry at the edge of any of those gaps\mklrc{not sure i understand this statement}. It would also be more likely to see such a multi-gap structure in discs with planet-induced vortices if such vortices are only triggered by gap-opening planets that only recently made it to the gap-opening stage, as our work tends to suggest.

\subsubsection{Alternate mechanisms for dust asymmetries} \label{sec:other-ways}

Even with the dearth of observed dust asymmetries, it is still possible that these asymmetries are not vortices or are vortices not produced by planets. Recent work has shown the an inner binary can also yield an asymmetry that is not a vortex as the disc develops eccentricity at the outer edge of the cavity \citep{ragusa17, calcino19}. This explanation may be favorable for discs where the only feature or innermost feature is an asymmetry, in particular for HD 142527 which is already known to have an inner binary. More recently, \cite{kuznetsova22} demonstrated that infall from a disc's external star-forming environment can also create radial pressure bumps that trigger vortices through the RWI if the infall is focused in a limited radial range of the disc. Such a configuration could mimic a two-sided planetary gap if the range of infall has the right radial width. Another mechanism for creating RWI vortices that has been studied in-depth is the edge of a dead zone because of the pressure bumps expected to develop at these locations \citep[e.g.][]{lyra12}. More recent work, however, has suggested \cbf{such pressure bumps are not necessarily sustained unless they have a sufficient amount of small dust grains to dominate the recombination process in the ionization chemistry \citep{delage22, delage23}}.

\subsubsection{Limiting lifetimes of dust asymmetries} \label{sec:other-ways}

There have been various mechanisms proposed to shorten vortex lifetimes or inhibit their formation altogether.

\begin{enumerate}
  \item
\textit{Dust feedback} has been shown to limit gas vortex lifetimes in 2-D to around 1000 orbits or less \citep{fu14b}, but both our work and \cite{lyra18} show that feedback has little to no effect on vortex lifetimes in 3-D simulations that better resemble actual discs. 
  \item
\textit{Self-gravity} can theoretically inhibit the RWI in general for disc masses as low as $Q < h^{-1}$ \citep{lovelace13}, but in simulated discs close to that mass limit with more realistic gap profiles it may only marginally weaken the vortices after they form \citep{hammer21}. Only very massive discs with $Q < 4$ can severely alter vortices or the RWI \citep{mkl11}. Most if not all observed discs with asymmetries have too low of a local surface density to be significantly limited by self-gravity \citep{vanDerMarel21}. If self-gravity were to play a strong role of inhibiting vortex formation for many observed gaps, it could be a sign that the associated planets formed very early on in the disc lifetime.
  \item
\textit{Cooling}, specifically non-zero cooling times of order unity, has been shown in 2-D simulations to yield vortices that spin up and shrink until they fade \citep{fung21}, but \cite{rometsch21} find that such an effect may not necessarily occur in 3-D.
  \item
\textit{Migration} increases the planet mass needed to trigger a vortex because the gap-opening mass is also a bit higher \citep{kanagawa21}. As we show in this work though, migration does not shorten dust asymmetry lifetimes and only helps sustain vortices. As such, migration would only inhibit a narrow range of low-mass planets from triggering vortices.
  \item
\textit{Aspect ratios} of a disc that are too thin ($h < 0.06$) could leave any planet-induced vortices that form too short-lived to be observed. Because of flaring and the outer disc being the preferred location for most observed vortex candidates though, these candidates are more likely located in a \cbf{ thicker region of the disc with a higher aspect ratio} that helps sustain the dust asymmetry.
  \item
\textit{Viscosity} is the most straightforward way to shorten vortex lifetimes, as even a relatively low $\alpha \sim 10^{-4}$ is already strong enough to limit the lifetime of the dust asymmetry to around 1000 orbits or less. Modern expectations for the disc viscosity are that discs no longer require high values of $\alpha$ because angular momentum transport can instead be driven by magnetic disc winds \citep[e.g.][]{bai13a, bai16}. Nonetheless, these magnetic effects may still drive a small $\alpha$ \citep[e.g.][]{bethune17} that could still be large enough to inhibit or weaken planet-induced vortices. \mklrc{viscosity kills the gas vortex, but does it affect the dust asymmetry? what about dust diffusion?}
\end{enumerate}

\cbf{Overall, viscosity may still be the best physical explanation for why there are so few observed dust asymmetries and why there are so few at two-sided gaps in particular. Dust feedback and cooling may not necessarily limit vortices, while migration and self-gravity may only weaken planet-induced vortices for very low mass planets or very high mass discs respectively. Furthermore, actual observed discs may not have the very wide gaps we find in our simulations of migrating planets with a low disc viscosity of $\nu = 10^{-7}$. That discrepancy, if it holds, could also be evidence that actual discs indeed have a higher viscosity than the value used in this work.
}

As dust feedback and cooling may not necessarily limit vortices while migration and self-gravity may only weaken planet-induced vortices for very low mass planets or very high mass discs respectively, viscosity may still be the best physical explanation for why there are so few observed dust asymmetries and why there are so few at two-sided gaps in particular. Furthermore, if gaps too wide to be consistent with observations are a preferential outcome of migrating planets inducing vortices with the disc viscosity $\nu = 10^{-7}$, then even the actual discs with observed asymmetries may still have a higher viscosity than the value we use in this work or else the asymmetries may not be planet-induced vortices.

\subsection{Applications to observed dust asymmetries} \label{sec:real}

Our findings have prospects to be able to fit a set of multiple features in a few different discs. We focus on four discs in particular: MWC 758, V 1247 Ori, \cbf{SR 21}, and Oph IRS 48.

The presence of a compact core in an encompassing elongated vortex frequently creates dust configurations in which some of the dust is trapped in the core as the core circulates, while the rest of the dust also circulates around the elongated vortex but separately. Even with good image resolution, these two distinct dust signatures can frequently appear to overlap. One example of the type of combined signature it can create is the top right panel of Figure~\ref{fig:images_h06_s3472}, which resembles the inner dust asymmetry in MWC 758 \citep{boehler18}. That feature of the real disc has a concentrated clump with a tail of dust in one direction. It has been suggested that the asymmetry coincides with both a vortex \citep{baruteau19} and a spiral wave from an unseen planet, where evidence for the latter comes from the scattered light \citep{ren20}. Such a spiral likely would not explain the dust tail at mm wavelengths if it originated from the vortex itself \citep{huang19}. Although a planet's spiral may also explain the dust tail, our results show a vortex could have such a tail even without any dust from the spiral.

The outer dust asymmetry in V 1247 Ori with an azimuthal extent of about $120^{\circ}$ appears like it may be divided into two equally-elongated components with a narrow deficit in-between \citep{kraus17}. The inner dust asymmetry, which includes a full background ring, more clearly has two separate components. Although it was suggested that the deficit itself in the outer asymmetry could be an artifact of the image, the propensity to find vortices with multiple peaks in our study suggests it may be possible that the deficit is real. It could be the case that both of the asymmetries in the disc have two components, supporting our finding that multiple peaks could be more common than just a single peak for some system properties.

\cbf{The primary ring in SR 21 was already known to be asymmetric \citep{perez14} and there was some hint of an inner ring in the system \citep{muroarena20}, but only recently were the features well-resolved into two clear partially-asymmetric rings that resemble opposite sides of a planetary gap \citep{yang23}. The inner ring itself is finely split radially into a full ring and an asymmetry. It is possible that the two radial structures are a ring of dust that collected at a pressure bump and a closely-adjacent vortex, analogous to the pattern shown in the right panels of Figure~\ref{fig:co-orbital}, which is common in other types of cases as well. The outer ring is a full circle, but also contains a clear asymmetry spanning about $60^{\circ}$ on one side and a more compact clump on the complete opposite side, as well as a few other candidate clumps that may be observational artifacts. Such a complete ring with multiple clumps could be produced by a decayed or decaying elongated vortex with multiple compact cores, analogous to what is shown in the middle-right panel of Figure~\ref{fig:images_h06_s3472}.}

The dust asymmetry in Oph IRS 48 is located at 61 AU \cite{vanDerMarel15b}, much further out than the suspected location of the planet at around 20 AU based on CO observations \citep{vanDerMarel13, bruderer14}. \cite{lobogomes15} had suggested this discrepancy could be explained by the planet triggering a secondary vortex, a phenomenon they found in non-isothermal simulations. We had previously suggested the large gap between the planet and the asymmetry could be explained by the original vortex migrating outwards away from the planet, as we found could happen in discs with very thick aspect ratios of $h~\ge~0.08$ \citep{hammer21}, like Oph IRS 48. In this work, we find that the opposite is also possible and the planet may have migrated inwards away from the vortex, leaving the vortex behind.




\section{Conclusions} \label{sec:conclusions}

Building off our previous work on planet-induced vortices \citep{hammer21}, we again studied vortices in hydrodynamic simulations with a planetary core accreting the bulk of its mass from the disc, except this time at higher resolution, in 3-D, with the VSI, with feedback, or with more than one of those facets\mklrc{we only do a representative sims in 3D/VSI/feedback; this statement can be misleading if one only reads the conclusions}. We also study a range of planet masses by varying the disc mass instead of the dimensionless accretion rate. While our hope was to address the discrepancy between the longevity of dust asymmetries associated with planet-induced vortices in simulations and the paucity of dust asymmetries in observations of actual protoplanetary discs, we instead found various other ways to prolong vortex lifetimes not seen in our past work, suggesting that planet-induced vortices should be even more likely to appear in observations. We also found new ways to make planet-induced vortices compact and clarified what they should look like in such situations. These new results show the importance of using a very high resolution ($\approx$ 29 cells $/~H$ in the radial direction) to study planet-induced vortices in simulations.

Contrary to expectation from all our previous works, we found that slowly-growing planets can still generate compact vortices, or in other cases elongated vortices with compact cores. The most straightforward way we found to form compact planet-induced vortices with stronger Rossby numbers in the compact range was with the help of the VSI, using the countless tiny compact vortices from the VSI. These tiny vortices seed larger compact vortices that arise later on when a low-mass planet triggers the RWI. Without the VSI, high-mass planets can form vortices with compact cores when they re-trigger later-generation vortices. The lower vorticity in these compact cores originates from the lower vorticity just interior to the vortex circulating into the vortex as it re-forms. \cbf{These cores make it much more common for the associated asymmetry to exhibit multiple co-orbital peaks in observations.} The overall vortices, however, are still elongated, and our synthetic images show that the dust they trap should still typically appear elongated in observations. On the other hand, very high-mass planets around Jupiter-mass and above can still generate vortices that appear compact, even if the vortex itself may still have a weaker Rossby number in the elongated range. Another surprising outcome was that dust feedback, which always causes both compact and elongated vortices to appear compact in 2-D, does not make elongated vortices appear compact in 3-D with sufficient vertical resolution. These vortices still appear elongated because even though clumps still arise due to feedback, these clumps do not collect all the dust trapped in the vortex. The rest of it still spreads around almost the whole azimuthal extent, similar to what happens without feedback.

Beyond the finding from our previous work that low-mass planets can generate very long-lived dust asymmetries, in this work, we find that high-mass planets can also generate long-lived vortices with longer lifetimes that make them even more likely to be observed. In a few cases, the dust asymmetry stops decaying altogether. These longer lifetimes result from the same factors that can make the vortices appear compact. Moreover, we find that allowing the planet to migrate like it should in a real disc not only extends dust asymmetry lifetimes, but also makes them even more prone to reach a steady state with an asymmetry that no longer decays. Such a steady state may be caused by the planet migrating away from the asymmetry. In the non-migrating cases, this steady state may be inhibited by cutting off the dust supply to the asymmetry. One set of cases we did find that actually shortens dust asymmetry lifetimes compared to our last work is low-mass planets in low-mass discs. Nonetheless, such planets in those discs can still yield long-lived dust asymmetries if the VSI is present.

It is difficult to test if our results in 2-D at high resolution still hold in 3-D because the high resolution needed to allow vortices to decay and re-form are also high enough to allow the VSI to be active in our locally isothermal discs. Since the VSI only occurs in 3-D\mklrc{requires full 3D disc models}, it is difficult to do a direct comparison at high resolution. At low resolution though, the same general trends we find in 2-D for the dust asymmetry lifetimes vs. planet mass still hold in 3-D for both the parameter study in which we vary the disc mass as well as the parameter study in which we vary the dimensionless accretion rate. To do a proper comparison of 2-D vs. 3-D at high resolution, it may be more interesting to suppress the VSI in 3-D, such as by incorporating a sufficiently slow cooling rate. We are already carrying out this kind of comparison for a future work. \mklrc{say this is the subject of a future/ongoing work, or cite paper in prep}

The longevity of vortices in our simulations is inconsistent with the lack of vortex candidates in observed protoplanetary discs, particularly the lack of those located at the edge of a two-sided gap that could harbour a planet. This discrepancy may just be due to these discs having too high of an effective viscosity, although the source of that viscosity would still be unclear. If viscosity is indeed the reason there are so few vortex candidates, that may be a sign that the mechanism for angular momentum transport may be associated with at least a mildly significant viscosity, even if the mechanism is primarily laminar like with disc winds.


\mklrc{***MKL EDITING BOOKMARK***}

\section*{Acknowledgements}

We would like to thank Marcelo Alfaro-Barraza, Phil Armitage, Can Cui, He-Feng Hsieh, Lizxandra Flores-Rivera, Hubert Klahr, Marius Lehmann, Hui Li, Yaping Li, Mordecai-Mark Mac Low, and Chiara Scardoni for helpful discussions, and the referee for helpful comments. MH and MKL are supported by the National Science and Technology Council (grants 107-2112-M-001-043-MY3, 110-2112-M-001-034-, 111-2112-M-001-062-, 110-2124-M-002-012-, 111-2124-M-002-013-, 112-2112-M-001-064-, 112-2124-M-002-003-) and an Academia Sinica Career Development Award (AS-CDA-110-M06). Computations were performed on the xl cluster at ASIAA and the TWCC cluster at the National Center for High-performance Computing (NCHC). We thank NCHC for providing computational and storage resources.

\mklrc{Remember to update the acknowledgements.}

\section*{Data Availability}

The data underlying this article will be shared on reasonable request to the corresponding author.



\bibliography{vortex_bibliography}

\appendix

\section{Accretion prescriptions} \label{sec:prescriptions}

\subsection{Choice of fiducial scheme} \label{sec:choice}

Unlike in H21 in which we varied the accretion parameters to simulate a range of planet masses, we needed to choose fixed accretion parameters for this work because we wanted study the effect of varying the disc mass instead. We had avoided choosing fixed accretion parameters in our previous work because the actual accretion process is more complex than just a pair of accretion parameters and in any case not well-resolved by our global simulations. Properly incorporating the planetary accretion process would require 3-D in every simulation, radiation hydrodynamics, and a very refined grid in the vicinity of the planet that resolves the planet's atmosphere \citep[e.g.][]{moldenhauer21} and allows a circumplanetary disc to form \citep[e.g.][]{szulagyi17a}, far beyond the scope of this project. Although we cannot capture that much detail, we can base our accretion scheme on previous work that better resolves the accretion process as well as our own 3-D simulations.

\begin{figure} 
\centering
\includegraphics[width=0.40\textwidth]{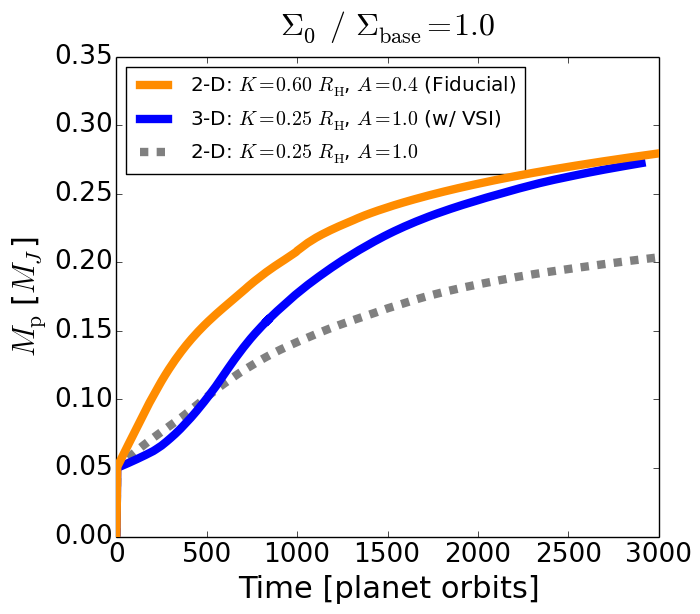}
\caption{Planet mass as a function of time with the VSI and 2-D cases with the same $\Sigma_0$. The values of $K$ and $A$ in our fiducial 2-D simulations were chosen to match the high-resolution 3-D case with the VSI. The 3-D case yields a much more massive planet than the corresponding 2-D case with the same $K$ and $A$ because the VSI drives additional angular momentum transport through the disc.} 
\label{fig:generic-mass}
\end{figure}

Past work has found planets should only accrete from a fraction of their Hill sphere within a radius of $0.25~R_\mathrm{H}$ \citep{lissauer09}. With such a low value of $K$ though, our low-resolution simulations would not resolve the accretion zone and not accrete any material onto the planet at all. To keep the accretion parameters consistent between our lower-resolution and fiducial higher-resolution parameter study, we instead set the accretion radius to $K = 0.6~R_\mathrm{H}$, a large enough value to safely allow accretion. We then choose $A = 0.4$ so that the low-resolution simulations approximate the planet's growth track in our fiducial high-resolution 3-D simulation with the VSI, as shown in Figure~\ref{fig:generic-mass}. We note that this simulation has an amplified accretion rate due to the additional Reynolds stress from the VSI compared to a 2-D simulation with the same parameters but no VSI. Even though our other simulations do not have the VSI, we still aim to mimic this accelerated growth track because actual discs may have disc winds or other additional sources of angular momentum transport even without the VSI or a high viscosity \cbf{(see \citealp{nelson23} for a recent work on this topic)}. Those other sources may not necessarily affect vortex properties. 

Since it is possible that the planet may indeed have a slower growth track without any other sources of angular momentum transport, we complement our high-resolution parameter study with an additional study that instead uses $K = 0.25~R_\mathrm{H}$ and $A = 0.9$. As discussed in the following section of the Appendix, we find similar qualitative results.

\subsection{Other accretion schemes} \label{sec:different-2D}


Beyond our fiducial accretion prescription of $K = 0.6~R_\mathrm{H}$ and $A = 0.4$, we tested out a more-massive prescription with a larger accretion rate of $A = 0.6$ and a simpler prescription in which the accretion radius is the expected $K = 0.25~R_\mathrm{H}$ and the accretion rate is near unity at $A = 0.9$. Both the more-massive prescription and the simpler prescription grow the planets faster, but the simpler prescription still results in less-massive planets. With these different prescriptions, we find that the two main regimes of low-mass and high-mass and their corresponding behavior still exist. Like the default prescription, the low-mass cases have short-lived dust asymmetries and the high-mass cases have dust asymmetries that survive the entire simulation. The only main difference we find is that the mass ranges for these regimes shift.

The simpler prescription is intended to model the accretion rate assuming there is no added accretion above the natural rate from the prescribed viscosity. We choose $A = 0.9$ instead of $A = 1.0$ to account for 2-D cylindrical accretion slightly overestimating 3-D spherical accretion, albeit this choice does not change the planet's growth trajectory by much. Although the high-mass regime is similar to that from the fiducial prescription, the highest-mass case differs in that it does have later-generation vortices. Nonetheless, the dust configuration for this case is still more compact throughout due to its steeper density contours in the vortex. Overall, the mass ranges for the regimes shift to lower values. The cases with indefinite dust asymmetry lifetimes begin with planet masses as low as $M_\mathrm{p} = 0.23~M_\mathrm{Jup}$, almost half the value of $M_\mathrm{p} = 0.42~M_\mathrm{Jup}$ at which this regime begins with our fiducial prescription. If the growth tracks of actual planets better resemble this simpler prescription, it would broaden the applicability of our fiducial results for high-mass planets to a wider range of planet masses.

The more-massive prescription is more similar to the fiducial one because it has the same accretion radius. The regime with indefinite dust starts at the only slightly higher mass of $M_\mathrm{p} = 0.45~M_\mathrm{Jup}$ with the same lowest disc mass of $\Sigma_0 / \Sigma_\mathrm{base} = 1.5$ as with our fiducial prescription. In the low-mass regime, none of the dust asymmetries survive more than 2000 orbits, a slight difference from the fiducial prescription in which the highest-mass case in this regime had a dust asymmetry lifetime of nearly 3000 orbits. Overall, these cases require a slightly lower planet mass to initially trigger the RWI.

\section{Comparing planet growth in 2-D and 3-D} \label{sec:growth-2D-3D}

\begin{figure*} 
\centering
\includegraphics[width=0.40\textwidth]{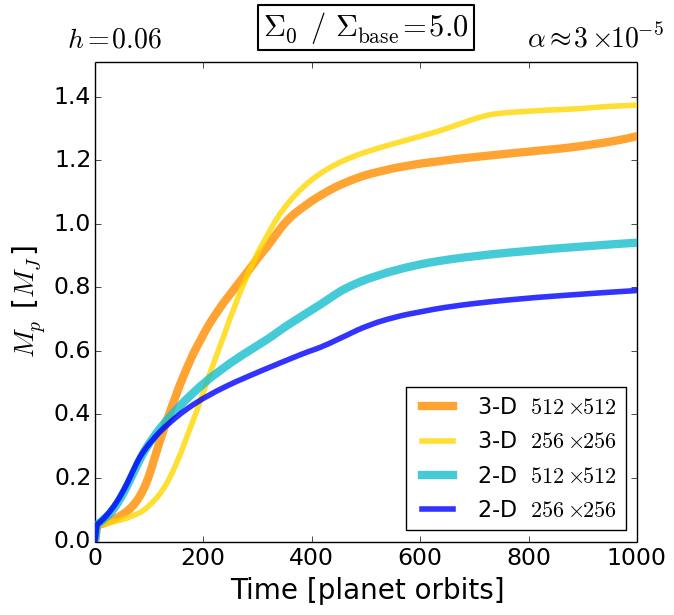}
\hspace*{2.5em}
\includegraphics[width=0.41\textwidth]{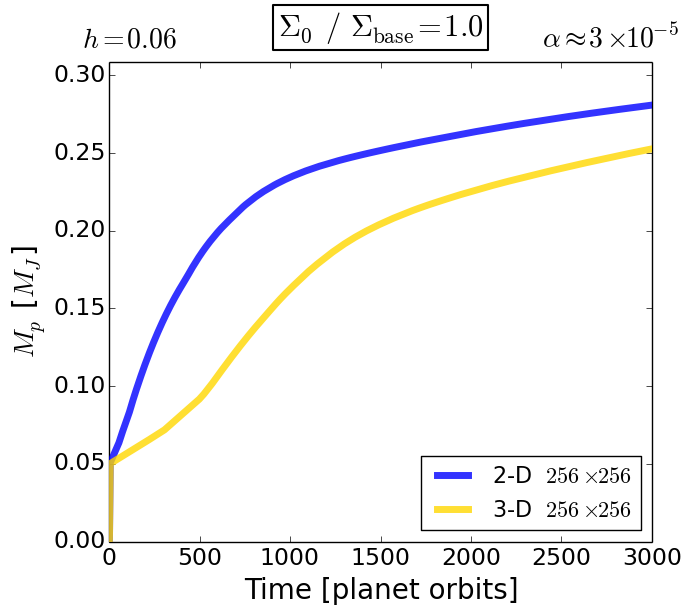}
\caption{Comparison of planet mass as a function of time for a high-mass planet in the highest-mass disc with $\Sigma_0 / \Sigma_\mathrm{base} = 5.0$ (\textit{left panel}) and a low-mass planet with the default disc mass of $\Sigma_0 / \Sigma_\mathrm{base} = 1.0$ (\textit{right panel}) for 2-D and 3-D and different resolutions. At lower planet masses, there is not a significant difference between 2-D and 3-D; however, at higher planet masses, the different smoothing lengths used creates a much larger discrepancy. Resolution has only a small effect on the planet's growth track in both 2-D and 3-D.} 
\label{fig:growth-comparison}
\end{figure*}

\begin{figure} 
\centering
\includegraphics[width=0.45\textwidth]{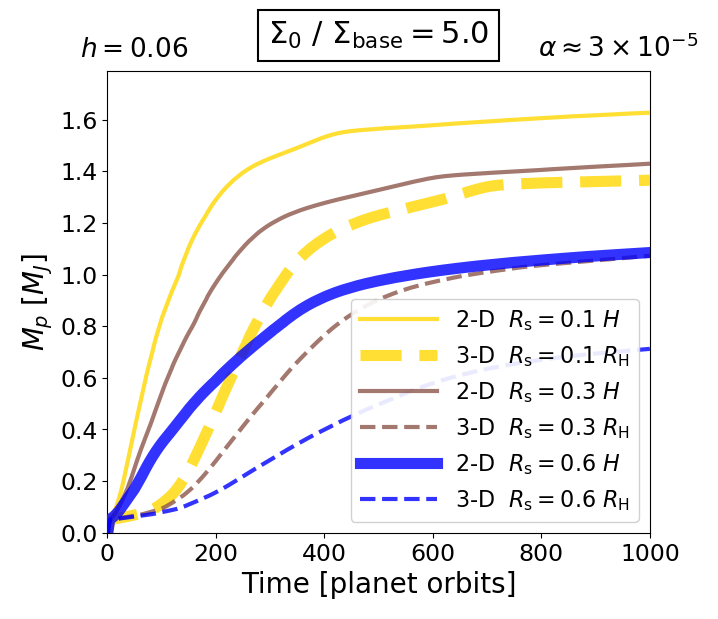}
\caption{Comparison of planet mass as a function of time for different smoothing lengths (\textit{different colors}), as well as 2-D (\textit{solid lines}) vs 3-D (\textit{dashed lines}), for the highest-mass disc. The default smoothing lengths in 2-D and 3-D are highlighted with thicker lines. Planet growth tracks are be very dependent on the choice of smoothing length with higher-mass planets.} 
\label{fig:smoothing-growth}
\end{figure}

The differences we do observe between 2-D and 3-D may in part be due to the planet's different growth tracks. We cannot use the exact same accretion prescription in 2-D and 3-D because the prescription depends on the grid cells, which are inherently different depending on the number of dimensions. Figure~\ref{fig:growth-comparison} compares the planet's growth tracks between 2-D and 3-D for two different disc masses: $\Sigma_0 / \Sigma_\mathrm{base} = 5.0$, which was the most different; and $\Sigma_0 / \Sigma_\mathrm{base} = 1.0$, which like any of the lower-mass cases yielded planets with similar mass between both numbers of dimensions. 

The most obvious difference is that the planet accretes from a sphere in 3-D compared to what is effectively a cylinder in 2-D. As such, we expect the planets in the 3-D simulations to grow slower, at least at first, which is what we do observe in every case. Naturally, we also expect that the planet's final mass would be lower in 3-D than 2-D; however, that only happens for the lower disc masses when $\Sigma_0 / \Sigma_\mathrm{base} \le 1.0$. For the higher-mass cases, the 3-D planet grows more massive than the 2-D planet. For the highest-mass case in particular, there is a huge difference. 

Some of the difference in planet masses between 2-D and 3-D, which increases with planet mass, is due to the different smoothing lengths used. To test the effect of having a $R_\mathrm{s} = 0.1 R_\mathrm{H}$ in the 3-D simulations compared to the much larger $R_\mathrm{s} = 0.6 H$ used in 2-D, we ran additional low-resolution simulations that tested the planet growth with three different coefficients: 0.1, 0.3, and 0.6. As Figure~\ref{fig:smoothing-growth} shows, we found that planet masses increase as we lower the smoothing length in both 2-D and 3-D. Had we used the same smoothing length, the planet in a 3-D disc would still grow less than the corresponding 2-D case, just like what happens with less massive planets. Since actual discs don't have a smoothing length ($R_\mathrm{s} = 0$), we suspect the planet masses attained in the 3-D simulations in which it is sensible to have a much smaller smoothing length are more realistic and the 2-D simulations may underestimate the final planet masses, particularly for the higher-mass planets. While a smoothing length of $R_\mathrm{s} = 0.6 H$ may be optimal for producing a more realistic gap structure with a fixed-mass planet \citep{muller12}, we find that it may not be realistic for modelling planetary accretion, particularly for higher-mass planets.

Another smaller difference we observe in the planet growth trajectories is the general shapes of the curves. The 2-D curves all follow a sort of logarithmic or inverse parabolic growth track with negative concavity \cbf{when the bulk of the planet's growth takes place}. Meanwhile, the 3-D curves start out with a positive concavity before switching to a negative concavity at a point of inflection somewhat early on, but not right away. 

\section{Steady-state dust asymmetry} \label{sec:steady}


In our 2-D parameter studies with $h = 0.06$ for both static and migrating planets, we find that dust asymmetries can reach a steady state and stop decaying altogether, prolonging the dust asymmetry lifetimes indefinitely, at minimum to at least through the end of our simulations. With a static planet, the lack of decay originated from later-generation vortices developing compact cores. With a migrating planet, the lack of decay happened after the planet migrated away from the dust asymmetry. We tested whether the steady-state outcome was physical in several different ways.

Our first test was to see if the indefinite lifetimes were due to the vortex receiving an unlimited supply of dust. Actual discs may indeed have an unlimited supply of dust for a significant amount of time; however, many observed asymmetries are the outermost dust feature in their discs, leaving them with no additional dust supply. To test this idea, we ran two additional simulations restarted midway from the highlighted case with $\Sigma_0 / \Sigma_\mathrm{base} = 3.0$. Specifically, we restarted the gas from the default case at $t = 1600~T_\mathrm{p}$, the point when the last elongated vortex has just decayed and all that remains is its compact core. Instead of using the dust from that timestep, we experimented with two new sets of dust profiles. The first restarted case used the initial dust profiles from $t = 0$. The second case again used the initial profiles, but with the initial surface density profile multiplied by a steep exponential cutoff. The cutoff is given by $e^{-(r/r_\mathrm{cut})^5}$, where the cutoff radius is at $r_\mathrm{cut} = 2.0~r_\mathrm{p}$. These kinds of tests are analogous to how we measured the vortex lifetimes in H21. With the regular initial profile, the dust asymmetry still survives well beyond $t = 7000~T_\mathrm{p}$. With the dust cutoff, however, we find that the dust asymmetry does not even survive until $t = 3200~T_\mathrm{p}$. Such a significant difference shows that the continual supply of dust to the vortex does indeed help extend the dust asymmetry lifetime, even after the primary series of elongated vortices has decayed.

Beyond limiting the dust supply, we also checked if other factors were responsible for prolonging the dust vortex lifetimes to help assess why the dust asymmetry outlives the gas vortex. To test whether the boundary\mklrc{testing for boundary effects is a numerical consideration, maybe more suitable for appendix} was artificially extending the lifetime, we re-ran the default case with a grid that was same resolution per scale height, but double the grid size in the radial direction. We found that the dust asymmetry still decays at about the same rate, and in fact even slower, suggesting that the proximity of the boundary to the asymmetry is not responsible for sustaining it. Lastly, we checked whether there was similar behavior in our additional tests with a lower aspect ratio that likewise form compact vortices (see Section~\ref{sec:aspect-ratios}). Based on the fact that the dust asymmetries in these cases do not have dust asymmetries that outlive the gas vortex by anywhere near as much as in the $h = 0.06$ cases, we suspect that the gas not perfectly decaying into a ring also plays a role in extending the dust asymmetry lifetime. Regardless of the aspect ratio, we observe that the extent by which the dust asymmetry outlives the gas vortex is significantly longer with compact vortices than elongated ones. We interpret this behavior as evidence that the compact cores help sustain the dust asymmetry by making the dust configuration more compact. With such a low level of dust diffusion, this compact configuration takes a significant amount of time to decay. Overall, we conclude that the compact dust configuration, the continuous dust supply, and the not perfectly-symmetric ring structure all play a role in the dust asymmetry surviving the entire simulation in our fiducial parameter studies.

\subsection{Dust asymmetry lifetimes with dust cutoff} \label{sec:cutoff}

Because we could not probe the trend of vortex lifetimes with the fiducial initial dust density profiles, we ran additional simulations that largely repeated the high-mass cases, but with the dust cutoff. We applied the cutoff from the very beginning at $t = 0$, in effect completely re-running each simulation. With the dust cutoff, the dust asymmetry decays in all but the highest-mass case. We find that the dust asymmetry lifetimes increase monotonically with the planet's final mass across the wide mass range in which the lifetimes were indefinite without any cutoff. The lifetimes with the cutoff are shown in Figure~\ref{fig:lifetimes} in comparison to the main parameter study. Even with the cutoff, the lifetimes become rather long at above 4000 orbits by around a mass of $M_\mathrm{p} = 0.50~M_\mathrm{Jup}$. The highest-mass case differs from the others because it was the only one in which the longevity of its dust asymmetry was due to the initial vortex and not compact later-generation vortices. As such, the increasing trend for the vortex lifetimes extends from around $M_\mathrm{p} = 0.25~M_\mathrm{Jup}$ to at least $M_\mathrm{p} = 1.0~M_\mathrm{Jup}$, and likely beyond.

\section{Planet mass dependence in low-resolution 2-D and 3-D} \label{sec:planet-mass-3D}

\begin{table}
\caption{3-D parameter study varying the accretion rate. The final mass $M_\mathrm{p}$ is recorded at the end of the vortex lifetime.}
\begin{tabular}{ c c c c }
  $h$ & $\Sigma$ & $A$ & $M_\mathrm{p} / M_\mathrm{J} $ \\
   \hline
        \hline
  0.06 & $5.787 \times 10^{-4}$ & 0.7 & 1.51 \\
  & & 0.3 & 1.18 \\
  & & 0.167 & 0.89 \\
  & & 0.1 & 0.69 \\
  & & 0.04 & 0.44 \\
  & & 0.02 & 0.32 \\
  & & 0.01 & 0.22 \\
\end{tabular}
\label{table:3D-simulations}
\end{table}

\begin{figure} 
\centering
\includegraphics[width=0.47\textwidth]{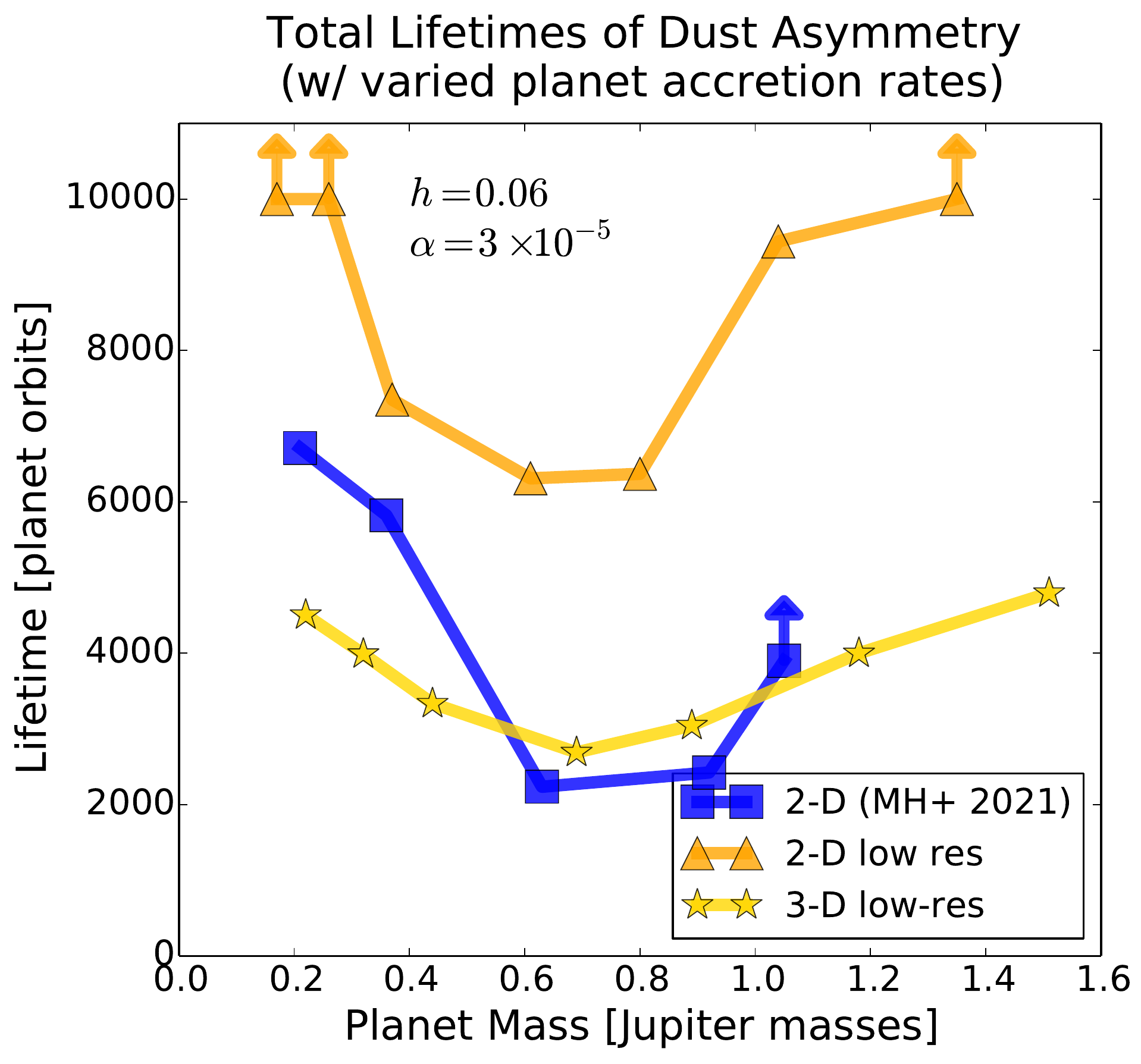}
\caption{Total dust asymmetry lifetimes as a function of planet mass, varying the planet mass with accretion parameter instead of the disc mass. The 2-D results are taken from \citealp{hammer21} (MH+ 2021), except for the most-massive planet, which was added for this study. In both 2-D and 3-D, the low-resolution simulations reproduce the U-shaped trend from MH+ 2021 in which dust asymmetry lifetimes are the shortest with intermediate-mass planets, increasing at either lower or higher planet masses. Some of the dust asymmetries survive past the end of each simulation in all cases (as indicated by the arrows).} 
\label{fig:lifetimes-p3}
\end{figure}


Besides varying the disc mass to set the planet's growth trajectory, we also ran 2-D and 3-D low-resolution simulations in which we varied the planet's accretion rate in order to test whether we could reproduce our results from H21. In that work, we had found that low-mass planets generate longer-lived dust asymmetries than high-mass planets in discs with $h = 0.06$. More generally, we found a ``U-shaped" trend in which the lifetimes increase again above a certain mass, which we only tested at $h = 0.04$ but had not fully explored at $h = 0.06$. 

In both 2-D and 3-D low-resolution simulations, we find that planets in discs with $h = 0.06$ induce vortices with dust asymmetry lifetimes that qualitatively follow the U-shaped trend, as depicted in Figure~\ref{fig:lifetimes-p3}. There is some quantitative disagreement in that the 2-D low-resolution simulations appear to overestimate the lifetimes by a significant margin, while the 3-D low-resolution simulations largely underestimate the lifetimes \cbf{from H21}, although by a lesser margin. Nevertheless, the qualitative trend still holds in both cases. Like the dust asymmetry lifetime dependence on disc mass, the trend at high-resolution from our previous work was due to re-triggered vortices. Also like that dependence, we are able to reproduce the U-shaped trend even though the low-resolution simulations do not have any re-triggered vortices. Lastly, we also ran an additional higher-resolution 2-D simulation following the fiducial parameters from our previous work but with $A = 1.0$. We find that this extra simulation confirms that the ``U-shaped" trend also occurs at $h = 0.06$ at higher resolution, just as it did at $h = 0.04$ in our previous work.

\section{Extents in synthetic images} \label{sec:synthetic-extents}

\begin{figure*} 
\centering
\includegraphics[width=0.48\textwidth]{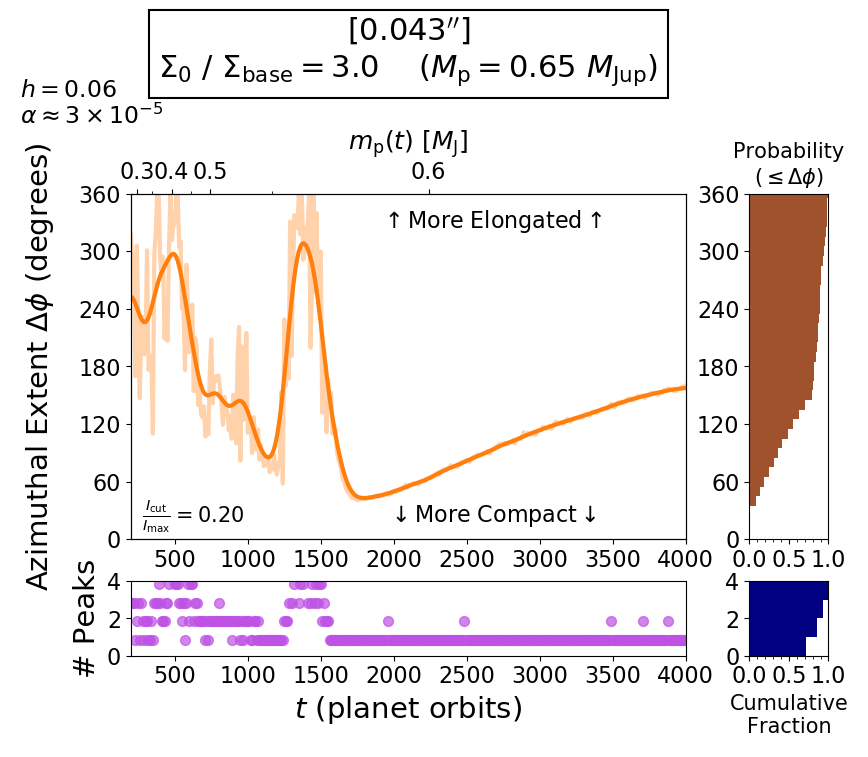}
\hspace*{2em}
\includegraphics[width=0.48\textwidth]{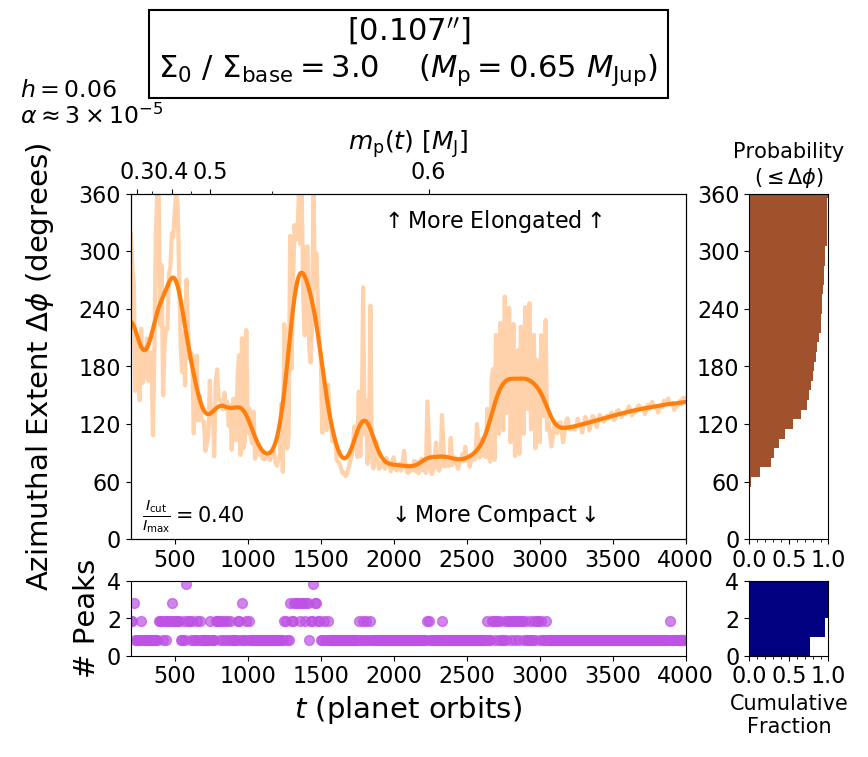} \\
\caption{Evolution of the azimuthal extents and number of peaks for the featured high-mass planet that grows to $M_\mathrm{p} = 0.65~M_\mathrm{Jup}$ from a disc with $\Sigma_0 / \Sigma_\mathrm{base} = 3.0$, analyzed with beam diameters of 0.043$^{\prime \prime}$ = 6 AU (\textit{left panel}) and 0.107$^{\prime \prime}$ = 15 AU (\textit{right panel}). The probabilities the vortex will have a certain extent or number of peaks are shown. Both the azimuthal extents and numbers of peaks vary widely in the early chaotic stage before settling into a steady-state in the later stage when the vortex has decayed. With a poor beam diameter, the dust at the Lagrange point can widen the azimuthal extent and add another peak, particularly in the later stage.} 
\label{fig:extents_h06_s3472}
\end{figure*}

\begin{figure*} 
\centering
\includegraphics[width=0.48\textwidth]{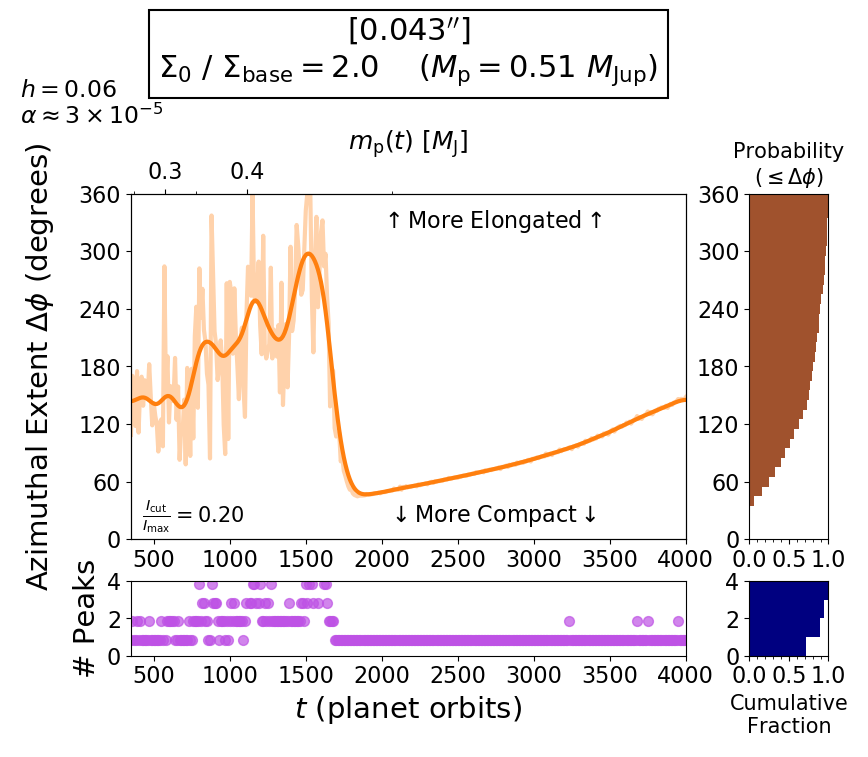}
\hspace*{1em}
\includegraphics[width=0.48\textwidth]{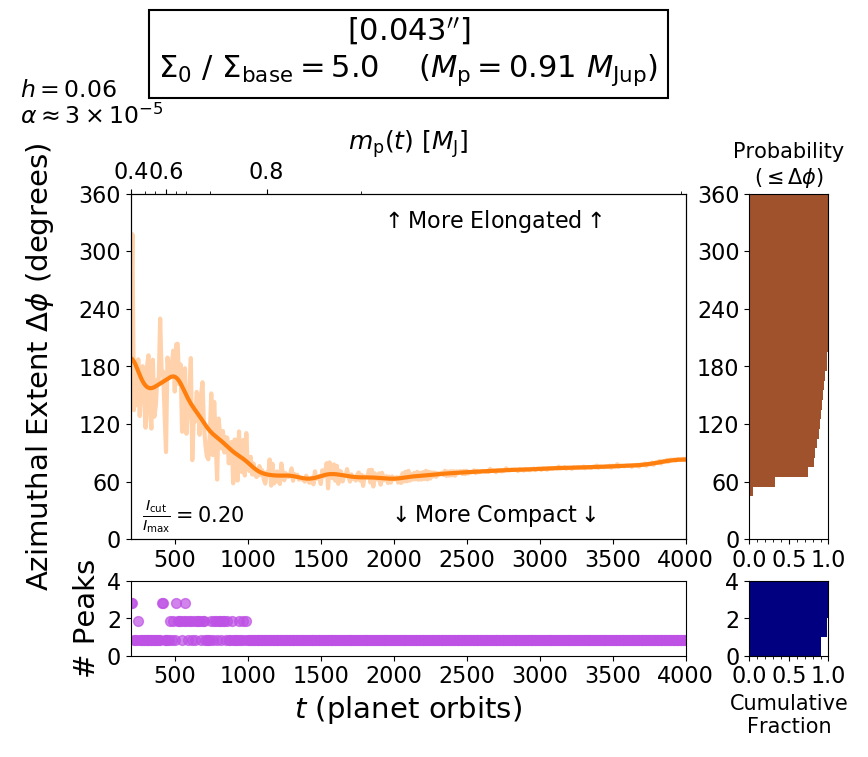}
\caption{Evolution of the azimuthal extents and number of peaks for two high-mass planets, analyzed with a beam diameter of 0.043$^{\prime \prime}$ = 6 AU. The first case (\textit{left panel}) also has two distinct stages like the previous figure, while the second case (\textit{right panel}) has a much simpler evolution because the initial vortex never decays or re-forms.} 
\label{fig:extents_h06_other}
\end{figure*}

Figures~\ref{fig:extents_h06_s3472} and~\ref{fig:extents_h06_other} show how much the azimuthal extents and the number of peaks can vary over time. The former figure highlights the featured simulation that grows to $M_\mathrm{p} = 0.65~M_\mathrm{Jup}$ ($\Sigma_0 / \Sigma_\mathrm{base} = 3.0$) using two different beam sizes, and features a clear demarcation between its early chaotic stage and simpler later stage. The latter figure shows two other high-mass planet cases ($\Sigma_0 / \Sigma_\mathrm{base} = 2.0$ and $5.0$) with indefinite dust asymmetry lifetimes, the first of which likewise has an early chaotic stage because of the compact cores that develop in the re-triggered vortices and the second of which has just a simple first-generation vortex that never decays.

In our previous work, the expected range of observed azimuthal extents was typically $100^{\circ}$ to $180^{\circ}$. During the early chaotic stage, the default case ($\Sigma_0 / \Sigma_\mathrm{base} = 3.0$) only has such an extent for a few hundred orbits from about $t = 650$ to $900~T_\mathrm{p}$, and the other chaotic case ($\Sigma_0 / \Sigma_\mathrm{base} = 2.0$) only has such an extent for several hundred orbits at the beginning. Both of these cases also have extents in this range in the simpler later stage for around 1000 orbits, even longer than in the chaotic, as the remaining dust from the last compact core slowly spreads out towards a full ring. For the final case with no early chaotic stage, the vortex begins with an azimuthal extent in the expected range before settling at around $70^{\circ}$ and decaying at a much slower rate than the compact cores do in the more chaotic cases.

A loose pattern that emerges is that vortices tend to have more peaks when they have a more elongated azimuthal extent. These additional peaks are typically due to the dust dynamics in the vortex. With a larger beam size, however, a second peak can emerge due to the dust away from the vortex and co-orbital with the planet at a Lagrange point. In the simpler later stage, that is the sole cause for additional peaks, as shown in the right panel of Figure~\ref{fig:extents_h06_s3472}.

 \end{document}